\documentstyle[12pt]{book}
\include{psfig}

\newcommand{\mi}[1]{#1\index{#1}}

\newfont{\bftitle}{cmbx12 scaled\magstep5}

\newcommand{\be}{\begin{equation}}
\newcommand{\bea}{\begin{eqnarray}}
\newcommand{\eea}{\end{eqnarray}}
\newcommand{\ee}{\end{equation}}

\begin{document}


\bibliographystyle{unsrt}


\begin{titlepage}

\begin{center}

\vspace*{2.0cm}

{\bftitle Relativistic constituent quark \\[0.4cm]
model for baryons }
\vspace{4.5cm}

{\large
{\Large\bf Inaugural-Dissertation}\\
zur\\
Erlangung der Philosophischen Doktorw\"{u}rde\\
vorgelegt der\\
Philosophischen Fakult\"{a}t II\\
der\\
{\Large\bf Universit\"{a}t Z\"{u}rich}\\[2.0cm]

von\\
{\Large\bf Felix Schlumpf}\\[0.15cm]

aus Z\"{u}rich ZH\\[1.5cm]

Begutachtet von den Herren\\
Prof. Dr. G. Rasche und Prof. Dr. W. Jaus\\[1.0cm]

Z\"{u}rich 1992}

\end{center}
\end{titlepage}

\pagenumbering{roman}

\thispagestyle{empty}

\centerline{  }\vspace{2in}
\centerline{\bf Dedicated to Priska}

\vspace{4in}
{\noindent\it
\hspace*{2.5in}Now we are seeing a dim reflection in a mirror;\\
\hspace*{2.5in}but then we shall be seeing face to face.\\
\hspace*{2.5in}The knowledge that I have now is imperfect;\\
\hspace*{2.5in}but then I shall know as fully as I am known.\\}
\hspace*{5in}{1. Cor. 13.12}

\setcounter{page}{1}


\chapter*{Abstract}
\addcontentsline{toc}{chapter}{Abstract}

The electroweak properties of nucleons and hyperons are calculated
in a relativistic constituent quark model. The baryons are treated as
three quark bound states, and the diagrams of perturbation theory are
considered on the light front. The electroweak properties of the baryons
are of nonperturbative nature and can be represented by one-loop diagrams.
We consider different extensions of the simplest model:
\begin{itemize}
\item  Quark form factors.
\item  Configuration mixing of the wave function.
\item  Asymmetric wave function.
\item  Wave function different from the one of a harmonic oscillator valid
up to energies of more than 30 GeV$^2$.
\end{itemize}
A comprehensive study of various baryonic properties is given:
\begin{itemize}
\item Elastic form factors of the nucleon.
\item Magnetic moments of the baryon octet.
\item Semileptonic weak form factors.
\end{itemize}

This analysis also gives the Kobayashi-Maskawa matrix element $V_{us}$
and a sound symmetry breaking scheme for the Cabibbo theory (see
Sec.~\ref{sec:cabibbo}).

A consistent physical picture appears in this work. The nucleon consists
of an unmixed, symmetric three quark state, the wave function of the
hyperons is however asymmetric with a spin-isospin-0 diquark. Only for
the strangeness-changing weak decay do we need nontrivial form factors.

\tableofcontents
\listoffigures
\listoftables
\cleardoublepage
\pagestyle{headings}

\pagenumbering{arabic}
\chapter{Introduction}\label{ch:intro}

The purpose of this thesis is to present the results of comprehensive
calculations of electromagnetic and weak form factors of the baryon octet
in a relativistic constituent quark model. Ever since the proposal of the
quark model in the early sixties by \mi{Gell-Mann} \cite{gell64}, the
modeling of the hadrons has been a very active area of theoretical
research. There are many interesting current questions, which can only be
answered by
quark model calculations. New high-statistics data from hyperon beta decay
\index{hyperon decay}
raise questions about SU(3) breaking \cite{roos90,ratc90} and the related
controversy about the meson and baryon derived values for the
\mi{Kobayashi-Maskawa
matrix} element $V_{us}$ has not been solved yet \cite{garc92}. This value
is important for studying the unitarity of the Kobayashi-Maskawa matrix.
There are also many recent papers on the \mi{magnetic moments} of the baryon
octet, which are a hard testing ground for various hadronic models.

In this thesis we consider a relativistic constituent quark model on the
\mi{light front} that was first formulated by \mi{Terent'ev} and
\mi{Berestetskii}
\cite{bere76,bere77}. It has been applied to various hadronic processes by
Aznauryan et~al.
\cite{azna82,azna82b,azna84}. Recently, new studies have been carried out
by \mi{Jaus} in the meson sector \cite{jaus90,jaus90b,jaus91} and by
\mi{Chung} and
\mi{Coester} on the electromagnetic form factors of the nucleons
\cite{chun91}. \mi{Dziembowski} et~al. \cite{dzie88,dzie88b,dzie88c}
and \mi{Weber}
et~al. \cite{webe87,kone90} treat the Melosh transformation
\index{Melosh rotation} of the quark spin in a ``weak-binding
approximation'' of questionable validity.

Nonrelativistic constituent quark models are successful in describing the
mass spectrum of baryons (for a review, see
Refs.~\cite{godf89,luch91,rich92}). The dominant effects of the gluonic
degrees of freedom are absorbed into constituent quark masses and into an
effective confining potential. In addition, effective dynamics inspired by
QCD has been considered in Refs.~\cite{caps86,caps87}; but this is
inconsistent for light quarks. A review in Ref.~\cite{thom84} gives an
estimate of the r.m.s. radius of the nucleon from 0.8~fm (derived from the
charge radius) to 0.4~fm (obtained from hyperon decays). Considering the
\mi{uncertainty principle} these values of the r.m.s. radius imply a quark
momentum in the range 250--500~MeV, which has to be compared with the light
constituent quark mass in the range 210--360~MeV. The use of
nonrelativistic quantum mechanics is therefore inconsistent, even for
static properties of the hadrons, because the relativistic corrections are
of the order $\langle p^2\rangle /m^2$ and must be expected to be large.

In a relativistic theory the Poincar\'{e} invariance has to be
respected; this means, on the quantum level, the fulfillment of the
commutation relations between the generators of the \mi{Poincar\'{e} group}.
Dirac \cite{dira49} has given a general formulation of methods to
simultaneously satisfy the requirements of special relativity and
Hamiltonian quantum mechanics. An extension of the Dirac classes of dynamics
can be found in Ref.~\cite{leut78}. The \mi{light front} scheme is in
particular distinguished from the other Dirac classes. Among the ten
generators of the Poincar\'{e} group, there are in the light front
approach seven (the maximal number) generators of kinematical character, and
only the remaining three generators contain
interaction, which is the minimal possible number.
The \mi{light front} dynamics is
therefore the most economical scheme for dealing with a relativistic
system. If we introduce the light front variables $p^\pm\equiv p^0\pm p^3$,
the Einstein mass relation $p_\mu p^\mu = m^2$ is linear in $p^-$ and
linear in $p^+$, in contrast to the quadratic form in $p^0$ and $\vec p$ in
the usual dynamical scheme. A consequence is a single solution of the mass
shell relation in terms of $p^-$, in contrast to two solutions for $p^0$:
\begin{equation}
p^- = (p_\perp^2 + m^2)/p^+\;, \qquad p^0 = \pm \sqrt{\vec p\,^2 + m^2}\;.
\label{eq:primo}
\end{equation}
The quadratic relation of $p^-$ and $p_\perp \equiv (p^1,p^2)$
in Eq.~(\ref{eq:primo})
resembles the nonrelativistic scheme \cite{suss68}, and the variable $p^+$
plays the role of ``mass'' in this nonrelativistic analogy. It is therefore
a good idea to introduce relative variables like the Jacobi momenta
when dealing with several particles. As in the nonrelativistic scheme
such variables allow us to decouple the \mi{center of mass motion} from the
internal dynamics. Hence we do not have the problems with the center of
mass motion which occur in the \mi{bag model}. The light front scheme shows
another
attractive feature that it has in common with the infinite momentum
technique \cite{wein66}. In terms of the old fashioned (\mi{Heitler} type, time
ordered, pre-Feynman) perturbation theory, the diagrams with quarks
created out of or annihilated into the vacuum do not contribute. The
usual $qqq$ quark structure is therefore conserved as in the
nonrelativistic theory. It is, however, harder to get the hadron states to
be eigenfunctions of the spin operator \cite{melo74}.

In this thesis we describe the baryon as a three quark \mi{bound state} with
 relativistic Faddeev equations \index{Faddeev equation}
and a Bethe-Salpeter interaction \index{Bethe-Salpeter equation}
kernel. Using the quasiparticle method these equations can be reduced to a
relativistic Schr\"{o}dinger equation with an effective potential. We
could either start with an Ansatz for the effective potential and derive
the \mi{wave function} or else start with the wave function itself.
For simplicity we take the latter approach and work with an Ansatz for
the wave function. In addition to the standard Gauss shaped wave
function we also investigate a Lorentz shaped one that fits the nucleon
form factors up to more than 30 GeV$^2$. In addition to the minimal
quark model (MQM), which uses the smallest number of parameters, we also
consider some extensions to the MQM: structure of the constituent
quark, several different radial wave functions, and asymmetry in the
wave function. For the MQM we have five parameters: the quark
masses $m_u=m_d\equiv m_{u/d}$ and $m_s$,
and the wave function parameters $\beta$ for
the nucleons, the $\Sigma$s and the $\Xi$s respectively, which
essentially determine the confinement scale. For the asymmetric model
we have three additional parameters: instead of $\beta$ we have
$\beta_q$ and $\beta_Q$. In the configuration mixing case there are some
more parameters that determine the mixing and the confinement scale of
the various radial wave functions. For the structure of the constituent
quarks there are seven additional parameters, i.e. the anomalous quark
moments $f_{2u}$, $f_{2d}$, $f_{2s}$, and the vector and axial weak form
factors $f_{1ud}$, $f_{1us}$, $g_{1ud}$, and $g_{1us}$ at zero momentum
transfer. Our goal is to
keep the number of parameters as small as possible in order to have a
model with predictive power. On the other hand the MQM is not able to
fit all electroweak properties of the baryons. The parametric dependence
on the data is highly nonlinear, so it is not obvious that even 15
parameters may fit two experimental data. Our analysis gives a physical
picture of a baryon, which is an asymmetric three quark state with quark
form factors in the weak sector.

This thesis is organized as follows:
\begin{description}

\item[Chapter \ref{ch:lff}] In this Chapter we present the \mi{light front}
formalism for bound states of three quarks. We start with the light front
variables, elaborate on the quasipotential reduction of the Bethe-Salpeter
\index{Bethe-Salpeter equation}
equation, and calculate the one loop diagram. We also study the wave
function and compare it with the one given by Chernyak and Zhitnisky
\cite{cher84}, and Lepage and Brodsky \cite{lepa80}.

\item[Chapter \ref{ch:electro}] In Sec.~\ref{sec:nucleon} we give the
details of our calculation of the electromagnetic form factors. We fit the
mass $m_{u/d}$ and the scale parameter $\beta$ to the data of the
magnetic moments of the proton and neutron, and the weak axial form factor
$g_1(n\rightarrow p)$. We not only examine the Gauss shaped wave
function, but also a Lorentz shaped one. With the latter it is possible to
fit the data in the whole experimentally accessible energy region up
to more than 30
GeV$^2$. In Sec.~\ref{sec:mm} we derive the explicit formulae for the
magnetic moments. The MQM is not able to give a reasonable fit, even with
non-zero quark anomalous moments. Only an asymmetric wave function can fit
all magnetic moments of the baryon octet as is shown in
Chapter~\ref{ch:Asymmetric}. The mass $m_s$ and the range parameters
$\beta_{\Sigma / \Lambda}$ and $\beta_\Xi$ can be fixed in
Chapter~\ref{ch:electro}.

\item[Chapter \ref{ch:decay}] Semileptonic beta decay\index{hyperon decay}
 of hyperons is an
active area of current research, both experimentally with high-statistic data
and theoretically with the problem of the \mi{Kobayashi-Maskawa matrix}
element $V_{us}$ \cite{roos90,ratc90}. In addition to the calculation of
the weak form factors and their derivatives we present a Cabibbo fit that
gives almost the same value for $V_{us}$ as the one recently published
\cite{garc92}.

\item[Chapter \ref{ch:Asymmetric}] The smallest extension to the MQM is to
use an asymmetric wave function, in which the scale parameter $\beta$ is
replaced by two scales $\beta_Q$ and $\beta_q$ corresponding to the
quark-diquark binding and the quark-quark binding within the \mi{diquark}.
In the limit of two particles this extended MQM reduces to a MQM of the
meson, which is
successfully used in Refs. \cite{jaus90,jaus91}.

\item[Chapter \ref{ch:conclusion}] summarizes our investigation. We
discuss the physical picture derived from our analysis and draw some
conclusions.

\item[Appendix \ref{ch:methods}] describes in detail the symbolic and
numerical methods used in this thesis.

\item[Appendix \ref{ch:formulae}] contains all formulae for the spin matrix
elements to all orders of the momentum transfer.

\end{description}

The index at the end of this thesis should give quick access to
special topics.

\chapter{Light-front formalism for baryons}\label{ch:lff}
\section{Relativistic three-body equation}
To specify the dynamics of a \mi{many-particle system} one
has to express the ten generators of the \mi{Poincar\'{e} group} $P_\mu$ and
$M_{\mu\nu}$ in terms of dynamical variables. The kinematic subgroup
is the set of generators that are independent of the interaction. There
are five ways to choose these subgroups \cite{leut84}.
Usually a physical state is defined
at fixed $x_0$, and the corresponding hypersurface is left
invariant under the \mi{kinematic subgroup}.

\index{vector!four-vector}
\index{vector!light-front vectors}
We shall use the light-front formalism which is specified by the invariant
hypersurface $x^+ = x^0+x^3 =$ constant. The following notation is used: The
four-vector is given by $x = (x^+,x^-,x_\perp)$, where $x^\pm = x^0
\pm x^3$ and $x_\perp=(x^1,x^2)$.
Light-front vectors are denoted by an arrow $\vec x =
(x^+,x_\perp)$, and they are covariant under kinematic Lorentz
transformations \cite{chun88}. The three momenta $\vec p_i$ of the quarks
can be transformed to the total and relative momenta to facilitate
the separation of the \mi{center of mass motion}
\cite{bakk79}.

\begin{eqnarray}
\vec P&=&\vec p_1+\vec p_2+\vec p_3, \quad \xi={p_1^+\over p_1^++p_2^+}\;,
\quad
\eta={p_1^++p_2^+\over P^+}\;,\nonumber\\
&&\\
q_\perp&=&(1-\xi)p_{1\perp}-\xi p_{2\perp}\;, \quad
Q_\perp =(1-\eta)(p_{1\perp}+p_{2\perp})-\eta p_{3\perp}\;.\nonumber
\end{eqnarray}
Note that the four-vectors are not conserved, i.e. $p_1+p_2+p_3\not= P$.
In the light-front dynamics the Hamiltonian takes the form
\begin{equation}
H={P^2_\perp +\hat M^2 \over 2P^+}\;,
\end{equation}
where $\hat M$ is the \mi{mass operator} with the interaction term $W$
\begin{eqnarray}
\hat M &=&M+W\;, \nonumber\\
M^2&=&{Q_\perp^2\over \eta(1-\eta)}+{M_3^2\over \eta}+{m_3^2\over 1-\eta},
\label{eq:2.3} \\
M_3^2&=&{q_\perp^2\over \xi (1-\xi)}+{m_1^2 \over \xi}+{m_2^2\over 1-\xi}\;,
\nonumber
\end{eqnarray}
with $m_i$ being the masses of the constituent quarks. To get a clearer
picture of $M$ we transform to $q_3$ and $Q_3$ by
\begin{eqnarray}
\xi&=&{E_1+q_3\over E_1+E_2}\;, \quad \eta={E_{12}+Q_3\over E_{12}+E_3}\;,
\nonumber\\
&&\\
E_{1/2}&=&({\bf q}^2+m_{1/2}^2)^{1/2}\;,\quad
E_{3}=({\bf Q}^2+m_{3}^2)^{1/2}\;,\quad
E_{12}=({\bf Q}^2+M_{3}^2)^{1/2}\;,\nonumber
\end{eqnarray}
where ${\bf q}=(q_1,q_2,q_3)$, and ${\bf Q}=(Q_1,Q_2,Q_3)$.
The expression for the mass operator is now simply
\begin{equation}
M=E_{12}+E_3\;, \quad M_3=E_1+E_2\;.
\end{equation}
We shall assume only two-particle forces interacting in a ladder-type
\index{ladder approximation}
pattern so that the dynamics of the three-body system is governed
by the Bethe-Salpeter (BS) interaction kernel \index{Bethe-Salpeter equation}
for the two body system and
the relativistic Faddeev equations.\index{Faddeev equation}

Using the Faddeev decomposition for the \mi{vertex function} $\Gamma =
\Gamma^{(1)} +\Gamma^{(2)} +\Gamma^{(3)}$, we can write down a BS equation for
the various components in operator notation
\begin{equation}
\Gamma^{(1)}=T^{(1)}G_2G_3(\Gamma^{(2)}+\Gamma^{(3)})
\end{equation}
with
\begin{equation}
G_i=\not\! p_i - m_i, \quad T^{(1)}=(1-VG_2G_3)^{-1}V\;,
\end{equation}
and similarly for $\Gamma^{(2)}$ and $\Gamma^{(3)}$. $V$ is the one
gluon exchange kernel between two quarks, and $T$ is already the ladder
sum to all orders. It is useful to consider the second iteration of the
vertex equation, which is given by:
\begin{equation}
{\bf \Gamma} = U G_1 G_2 G_3 {\bf \Gamma}\;,
\label{eq:2.8}
\end{equation}
where ${\bf \Gamma}=(\Gamma^{(1)},\Gamma^{(2)},\Gamma^{(3)})$
and $U$ is the matrix
\begin{equation}
U_{ij}=\cases{T^{(i)}G_jT^{(k)}&for $i\neq j$ with $k\neq i,j$ ,\cr
              T^{(i)}(G_kT^{(l)}+G_lT^{(k)})&for $i=j$ with $k\neq l\neq
			  i$ .}
\end{equation}
The four-dimensional Eq.~(\ref{eq:2.8}) can be reduced to a three-dimensional
equation
\begin{equation}
{\bf \Gamma}=W g_3 {\bf \Gamma}\;, \quad W=(1-U R_3)^{-1}U
\end{equation}
by writing $G_1 G_2 G_3=g_3+R_3$ where $g_3$ has only three-particle
singularities.  We choose a $g_3$ which puts the quarks on
their mass shells:
\begin{equation}
g_3=(2\pi i)^2\int ds{1 \over P^2-s}\prod_{i=1}^3\delta^+(p_i^2-m_i^2)
(\not\! p_i + m_i)\;,
\end{equation}
where $P$ is the total momentum of the \mi{bound state}, $s=(p_1+p_2+p_3)^2$
and $p_i$ are restricted by $p_i^+ \ge 0$. We get
\begin{equation}
g_3=(2\pi i)^2 \delta(p_2^2-m_2^2)\delta(p_3^2-m_3^2)\Theta(\xi)\Theta
(1-\xi)\Theta(\eta)\Theta(1-\eta){\Lambda^+(p_1)\Lambda^+(p_2)
\Lambda^+(p_3)\over \xi\eta (P^2-M^2)}
\label{eq:green}
\end{equation}
with the spin projection operator
\begin{equation}
\Lambda^+(p_i)=\sum_\lambda u(p_i,\lambda)\bar u(p_i,\lambda)\;.
\end{equation}
Writing
\begin{eqnarray}
\hat g_3&=&{1\over P^2-M^2}\;,\nonumber\\
\hat \Gamma^{(i)}&=&\left( {M_3M\over E_1E_2E_3E_{12}}\right)^{1/2}
\Gamma^{(i)}u(p_1\lambda_1)u(p_2\lambda_2)u(p_3\lambda_3)\;,\\
\hat W_{ij}&=&\left( {M_3M'_3MM'\over E_1E'_1E_2E'_2E_3E'_3E_{12}E'_{12}}
\right)^{1/2}u(p_1\lambda_1)u(p_2\lambda_2)u(p_3\lambda_3)W_{ij}
\bar u(p'_1\lambda'_1)\bar u(p'_2\lambda'_2)\bar u(p'_3\lambda'_3)
\nonumber
\end{eqnarray}
we are led to the integral equation
\begin{eqnarray}
\hat \Gamma^{(i)}({\bf q},{\bf Q},\lambda_1,\lambda_2,\lambda_3)&=
&{1\over (2\pi)^6}\sum_{\lambda'_1\lambda'_2\lambda'_3j}\int d^3\!q'd^3\!Q'
\hat W^{ij}({\bf q},{\bf q}',{\bf Q},{\bf Q}',\lambda_1,\lambda'_1,
\lambda_2,\lambda'_2,\lambda_3,\lambda'_3)\nonumber\\
&&\times \hat g_3({\bf q}',{\bf Q}') \hat \Gamma^{(j)}
({\bf q}',{\bf Q}',\lambda'_1,\lambda'_2,\lambda'_3)\;.
\end{eqnarray}
We can write this equation in terms of the \mi{wave function} $\Psi$. The
Faddeev decomposition \index{Faddeev equation}
is $\Psi=\Psi^{(1)}+\Psi^{(2)}+\Psi^{(3)}$,
the relation to the \mi{vertex function} is $\Psi^{(i)}=\hat g_3
\hat \Gamma^{(i)}$, and writing ${\bf\Psi}=(\Psi^{(1)}, \Psi^{(2)},
\Psi^{(3)})$ we get
\begin{equation}
(M_B^2-M^2){\bf\Psi}=\hat W {\bf\Psi}
\end{equation}
with $M_B$ being the mass of the baryon. If we put $\hat W=MW+WM+W^2$
we see that the wave function is an eigenfunction of the \mi{mass operator}
$\hat M^2$, given in Eq.~(\ref{eq:2.3}):
\begin{equation}
\hat M^2 \Psi = M^2_B \Psi
\end{equation}
which is equivalent to the equation usually used in constituent quark
models \cite{godf89}
\begin{equation}
(E_{12}+E_3+W)\Psi = M_B \Psi\;.
\label{eq:2.18}
\end{equation}
This last equation is the starting point for an explicit calculation of the
wave function, which has been done for the meson sector
\cite{jaco90,hill90}.

\newpage
\section{Current matrix element}
\begin{figure}[bhtp]
\centerline{\psfig{figure=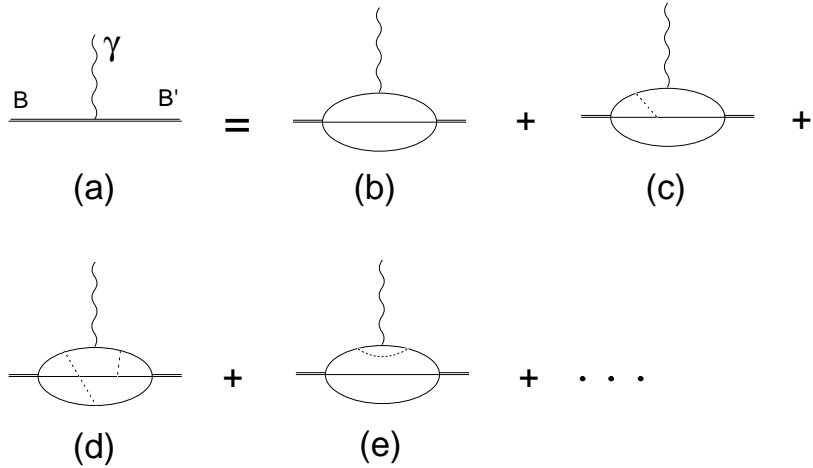,height=180pt}}
\caption[Feynman diagrams for the elastic form factor of baryons.]
{Feynman diagrams for the elastic form factor of baryons. Only
the three quark core of the baryon is considered.}
\label{fig:1}
\end{figure}

We would like to calculate the current matrix element $M^\mu=
\left< B' | \bar q\gamma^\mu q | B\right>$ corresponding to the
Feynman diagram of Fig. \ref{fig:1} (a).

We restrict ourselves to the three quark core of the baryon. The diagrams of
Fig. \ref{fig:1} (c+d), and (e) can be absorbed into the wave function and
quark form factors, respectively. We are left with the diagram Fig.
\ref{fig:1} (b).

\begin{figure}[bhtp]
\centerline{\psfig{figure=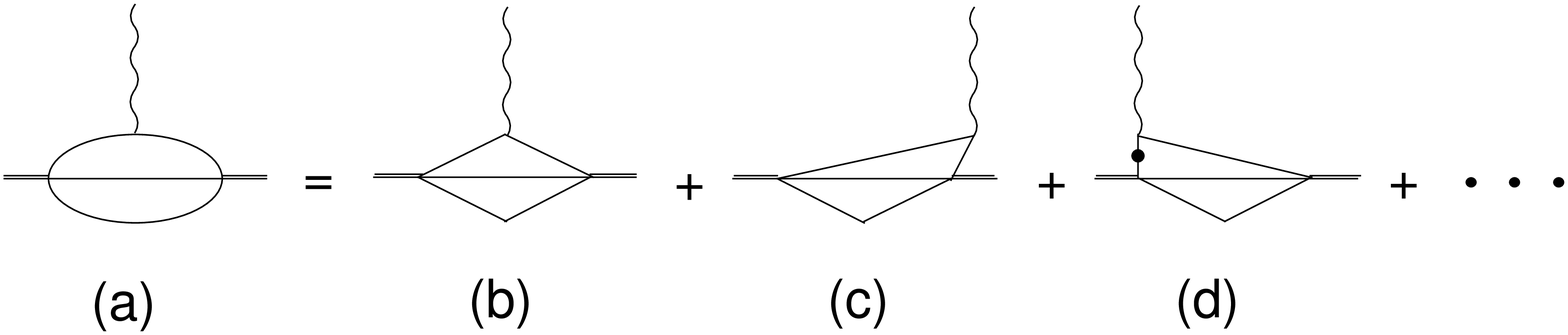,height=86.4pt}}
\caption[Feynman diagrams for the elastic form factor.]
{(a) Feynman diagram for the elastic form factor.
(b)--(d) Time $x^+$-ordered diagrams corresponding to (a).
Pointed lines represent instantaneous quark propagators.}
\label{fig:2}
\end{figure}

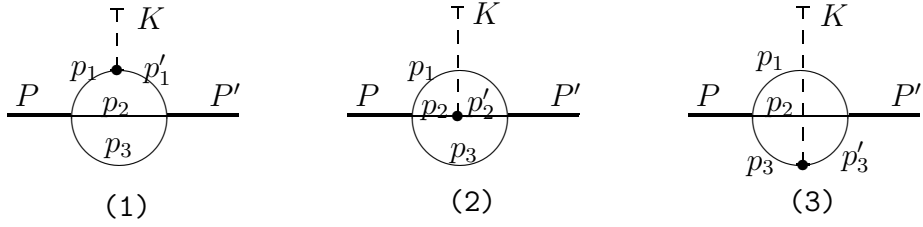
\begin{figure}
\centerline{{\tt    \setlength{\unitlength}{0.92pt}
\begin{picture}(411,111)
\thicklines   \multiput(10,32)(140,0){3}{\begin{picture}(111,40)
\thinlines        \put(83,25){$P'$}
                  \put(3,25){$P$}
                  \put(66,22){\line(1,0){29}}
                  \put(0,22){\line(1,0){26}}
                  \put(0,21){\line(1,0){95}}
                  \put(46,20){\circle{40}}
                  \end{picture}}
\thinlines    \put(49,12){(1)}
              \put(191,14){(2)}
              \put(331,14){(3)}
              \put(55,72){\dashbox{5}(0,25){}}
              \put(195,53){\dashbox{5}(0,45){}}
              \put(337,33){\dashbox{5}(0,65){}}
              \put(63,89){$K$}
              \put(201,91){$K$}
              \put(344,90){$K$}
              \put(50,38){$p_3$}
              \put(192,36){$p_3$}
              \put(49,55){$p_2$}
              \put(175,70){$p_1$}
              \put(36,70){$p_1$}
              \put(66,69){$p_1'$}
              \put(180,54){$p_2$}
              \put(199,55){$p_2'$}
              \put(318,74){$p_1$}
              \put(322,55){$p_2$}
              \put(314,31){$p_3$}
              \put(353,33){$p_3'$}
              \put(55,72){\circle*{4}}
              \put(337,33){\circle*{4}}
              \put(195,53){\circle*{4}}
\end{picture}}}
\caption[Feynman diagrams that represent the transition of the baryon
state.]{Feynman diagrams that represent the transition of the baryon
state with four-momentum $P$ to the baryon state with four-momentum $P'$.
$K = P - P'$. The photon or the $W$ boson is coupled either to the first,
second or third quark line, corresponding to the diagrams (1),(2) and (3),
respectively.}
\label{fig:3diagram}
\end{figure}

In Fig \ref{fig:2}, the Feynman diagram (a) is
equivalent to the sum of diagrams (b) -- (d) in the old-fashioned
perturbation theory. If we consider $x^+$-ordering and put $K^+=0$,
diagram (c) drops out because of conservation of $+$-momentum.
The $x^+$-instantaneous propagator in diagram (d) is proportional
to $\gamma^+$, and gives no contribution for $K^+=0$ since
$(\gamma^+)^2=0$. We are left with diagram (b) which can be expressed
with the help of the \mi{vertex function} $\Gamma$ ($N_c$ is the
number of colors):
\index{color}
\begin{eqnarray}
M^+&=&{N_c\over (2\pi)^8}\int d^4\!k\,d^4l\sum_{\rm indices}
\Gamma_{ijk}{u^i(p_1\lambda_1)\bar u^l(p_1\lambda_1)\over (p_1^2-m_1^2+i
\epsilon)}\gamma^+_{lm}{u^m(p'_1\lambda'_1)\bar u^n(p'_1\lambda'_1)\over
(p_1^{'2}-m_1^{'2}+i\epsilon)}\nonumber\\
&&\times\Gamma^\dagger_{nop}{u^j(p_2\lambda_2)\bar u^o
(p_2\lambda_2)u^k(p_3\lambda_3)\bar u^p(p_3\lambda_3)\over
(p_2^2-m_2^2+i\epsilon)(p_3^2-m_3^2+i\epsilon)}+{\rm
permutations}\;,\nonumber\\
\vec p_1&=&{1\over 3}\vec P+\vec k+{1\over 2}\vec l\;,\quad
\vec p_2={1\over 3}\vec P-\vec k+{1\over 2}\vec l\;,\quad
\vec p_3={1\over 3}\vec P-\vec l\;,\\
\vec p_1\,'&=&\vec p_1-\vec K\;,\quad K=P-P'\;.\nonumber
\end{eqnarray}
On the \mi{light front}, we have exact correspondence with the choice of the
Greens function in Eq.~(\ref{eq:green}).
If the vertex function $\Gamma$ is assumed to be independent of the
components $k^-$ and $l^-$, we can calculate $M^+$ by contour methods in the
$k^-$ and $l^-$ planes. $M^+$ is given by the residua of the two
noninteracting quark poles. Replacing vertex functions by wave functions
\index{wave function!from vertex functions to}
we get
\begin{eqnarray}
M^+&=& {N_c\over (2\pi)^6}\int d^3qd^3Q\left({E'_1E'_2E'_3E'_{12}
M_3M\over E_1E_2E_3E_{12}M'_3M'}\right)^{1/2}
\sum_{\rm spin}\Psi^\dagger({\bf q}',{\bf Q}',\lambda'_1,\lambda'_2,\lambda'_3)
\nonumber\\
&&\times(O_1+O_2+O_3)\Psi({\bf q},{\bf Q},\lambda_1,
\lambda_2,\lambda_3)\;,\nonumber\\
O_1&=&{1\over\xi\eta}\bar u(\vec p_1\,'\lambda'_1)\gamma^+u(\vec p_1\lambda_1)
\;,\\
O_2&=&{1\over (1-\xi)\eta}\bar u(\vec p_2\,'\lambda'_2)\gamma^
+u(\vec p_2\lambda_2)\;,\nonumber\\
O_3&=&{1\over (1-\eta)}\bar u(\vec p_3\,'\lambda'_3)\gamma^+
u(\vec p_3\lambda_3)\;.\nonumber
\end{eqnarray}
The $O_i$s correspond to the $i$th diagram in Fig. \ref{fig:3diagram}.
For $O_1$ the primed variables are
\begin{equation}
q'_\perp = q_\perp -(1-\xi)K_\perp,\quad Q'_\perp = Q_\perp -(1-\eta)
K_\perp\;,\nonumber
\end{equation}
for $O_2$
\begin{equation}
q'_\perp = q_\perp +\xi K_\perp\;,\quad Q'_\perp = Q_\perp -(1-\eta)
K_\perp\;, \nonumber
\end{equation}
and for $O_3$
\begin{equation}
q'_\perp =q_\perp\;, \quad Q'_\perp = Q_\perp +\eta K_\perp\;.\nonumber
\end{equation}

For pointlike quarks the matrix element of the current is
\begin{eqnarray}
\bar u(\vec p_1\,'\lambda'_1)\gamma^+u(\vec p_1\lambda_1)&=&2\xi\eta P^+\delta_
{\lambda_1\lambda'_1}\;,\nonumber\\
\bar u(\vec p_2\,'\lambda'_2)\gamma^+u(\vec p_2\lambda_2)&=&2(1-\xi)\eta P^+
\delta_{\lambda_2\lambda'_2}\;,\\
\bar u(\vec p_3\,'\lambda'_3)\gamma^+u(\vec p_3\lambda_3)&=&2(1-\eta)P^+\delta_
{\lambda_3\lambda'_3}\;.\nonumber
\end{eqnarray}
After factoring out \mi{color} the wave function is totally symmetric and we
have $\sum_i\left< O_i \right> = 3\left< O_j \right>$ for any $j$.
Since primed variables take a simple form for $O_3$ we choose
$3\left< O_3 \right>$. We arrive at the form
\begin{equation}
M^+= 2P^+{N_c\over (2\pi)^6}\int d^3qd^3Q\left({E'_3E'_{12}M\over E_3E_{12}M'}
\right)^{1/2}\sum_{\rm spin}\Psi^\dagger({\bf q}',{\bf Q}',\lambda'_3)
3\delta_{\lambda_3\lambda'_3}\Psi({\bf q},{\bf Q},\lambda_3)\;.
\label{eq:2.25}
\end{equation}

\newpage
\section{Wave function} \label{sec:wavefunction}
In light front variables\index{light front!variables}
one can separate the \mi{center of mass motion} from the
internal motion. The wave function $\Psi$ is therefore a function of the
relative momenta ${\bf q}$ and ${\bf Q}$. The product $\Psi=
\Phi\chi\phi$, with $\Phi=$ flavor, $\chi=$ spin, and $\phi=$
momentum distribution, is a symmetric function. This is consistent with
Fermi statistics since the color wave function is totally antisymmetric.

The angular momentum ${\bf j}$ can be expressed as a sum of orbital and
spin contributions
\begin{eqnarray}
{\bf j}=i\nabla_{\bf p}\times {\bf p}+\sum_{j=1}^3 {\cal R}_{Mj}{\bf s}_j \;,
\end{eqnarray}
where ${\cal R}_M$ is a Melosh rotation \index{Melosh rotation}
acting on the quark spins
${\bf s}_j$, which has the matrix representation (for two particles)
\begin{equation}
\left< \lambda' |{\cal R}_M(\xi,q_\perp,m,M)|\lambda\right> =
\left[ {m+\xi M-i{\bf \sigma}\cdot({\bf n}\times {\bf q})\over
\sqrt{(m+\xi M)^2+q_\perp^2}}\right]_{\lambda'\lambda}
\end{equation}
with ${\bf n}=(0,0,1)$. In previous works
\cite{dzie88,dzie88b,dzie88c,webe87,kone90} this
rotation has been approximated by putting $M=M_B$. This corresponds to a
weak-binding limit which cannot be justified for a \mi{bound state} in QCD.
In this limit our model has a close connection to many other
relativistic quark models as shown by Koerner et al. \cite{korn91}.

The operator ${\bf j}$ commutes with the \mi{mass operator} $\hat M$; this is
necessary and sufficient for Poincar\'e-invariance of the bound state.

In terms of the relative momenta the angular momentum takes the
form
\begin{eqnarray}
{\bf j}&=& i\nabla_{\bf Q}\times {\bf Q}+{\cal R}_M(\eta,Q_\perp,M_3,M)
{\bf j}_{12}+{\cal R}_M(1-\eta,-Q_\perp,m_3,M){\bf s}_3\;,\nonumber\\
&&\\
{\bf j}_{12}&=& i\nabla_{\bf q}\times {\bf q}+{\cal R}_M(\xi,q_\perp,m_1,M_3)
{\bf s}_1+{\cal R}_M(1-\xi,-q_\perp,m_2,M_3){\bf s}_2\;.\nonumber
\end{eqnarray}
We can drop the orbital contribution.
\begin{eqnarray}
{\bf j}&=&\sum {\cal R}_i{\bf s}_i\;,\nonumber\\
{\cal R}_1&=&{1\over \sqrt{a^2+Q_\perp^2}\sqrt{c^2+q_\perp^2}}
\pmatrix{ac-q_RQ_L&-aq_L-cQ_L\cr
         cQ_R+aq_R&ac-q_LQ_R}\;,\nonumber\\
{\cal R}_2&=&{1\over \sqrt{a^2+Q_\perp^2}\sqrt{d^2+q_\perp^2}}
\pmatrix{ad+q_RQ_L&aq_L-dQ_L\cr
         dQ_R-aq_R&ad+q_LQ_R}\;,\label{eq:melosh}\\
{\cal R}_3&=&{1\over \sqrt{b^2+Q_\perp^2}}\pmatrix{b&Q_L\cr
         -Q_R&b}\;,\nonumber
\end{eqnarray}
with
\begin{eqnarray}
a&=&M_3+\eta M\;,\quad b=m_3+(1-\eta)M\;,\nonumber\\
c&=&m_1+\xi M_3\;, \quad d=m_2+(1-\xi)M_3\;,\nonumber\\
q_R&=&q_1+iq_2\;,\quad q_L=q_1-iq_2\;,\\
Q_R&=&Q_1+iQ_2\;,\quad Q_L=Q_1-iQ_2\;.\nonumber
\end{eqnarray}
The momentum wave function $\phi$ is normalized according to \cite{mich82}
\begin{equation}
{N_c\over (2\pi)^6} \int d^3qd^3Q|\phi |^2=1\;,
\label{eq:2.32}
\end{equation}
and can be chosen as a function of $M$ to fulfill the spherical and
permutation symmetry. That is the same as to express it in terms of the
off-shell energy ${\cal E}$ since
\begin{equation}
{\cal E}=P^+(P^--p^-_1-p^-_2-p^-_3)=M_B^2-M^2\;.
\end{equation}
The $S$-state orbital function $\phi(M)$ is approximated by either
\begin{equation}
\phi(M)=N\exp\left[-{M^2\over 2\beta^2}\right] \qquad\hbox{or}\qquad
\phi(M)={N'\over (M^2+\beta^2)^{3.5}}\;,
\label{eq:2.34}
\end{equation}
which depend on two free parameters, the constituent quark mass and the
\mi{confinement scale} parameter $\beta$. The first
function is the conventional choice used in spectroscopy,
but it has a too strong falloff for
high $K^2$. Both functions give nearly the same result for low values
of $K^2$, the second one performs obviously better for high $K^2$.
This independence of the wave function $\phi$ for low $K^2$
suggests that the static properties are mainly given by the flavor and
spin part of the wave function.

\index{wave function!baryon octet}
The total wave functions for the baryon octet are
\footnote{The overall sign for $\Sigma^+$,$\Lambda$ and $\Xi^0$ has to be
changed in Ref. \cite[p. 46]{clos79}.}
\begin{eqnarray}
p&=&{-1\over\sqrt 3}\left(uud\chi^{\lambda3}+udu\chi^{\lambda2}
+duu\chi^{\lambda1}\right)\phi\;,\nonumber\\
n&=&{1\over\sqrt 3}\left(ddu\chi^{\lambda3}+dud\chi^{\lambda2}
+udd\chi^{\lambda1}\right)\phi\;,\nonumber\\
\Lambda&=&{-1\over\sqrt 6}\left[(uds-dus)\chi^{\rho3}+(usd-dsu)
\chi^{\rho2}+(sud-sdu)\chi^{\rho1}\right]\phi\;,\nonumber\\
\Sigma^+&=&{1\over\sqrt 3}\left(uus\chi^{\lambda3}+usu\chi^{\lambda2}
+suu\chi^{\lambda1}\right)\phi\;,\label{eq:wavefunction}\\
\Sigma^0&=&{-1\over\sqrt 6}\left[(uds+dus)\chi^{\lambda3}+(usd+dsu)
\chi^{\lambda2}+(sud+sdu)\chi^{\lambda1}\right]\phi\;,\nonumber\\
\Sigma^-&=&{-1\over\sqrt 3}\left(dds\chi^{\lambda3}+dsd\chi^{\lambda2}
+sdd\chi^{\lambda1}\right)\phi\;,\nonumber\\
\Xi^0&=&{-1\over\sqrt 3}\left(ssu\chi^{\lambda3}+sus\chi^{\lambda2}
+uss\chi^{\lambda1}\right)\phi\;,\nonumber\\
\Xi^-&=&{1\over\sqrt 3}\left(ssd\chi^{\lambda3}+sds\chi^{\lambda2}
+dss\chi^{\lambda1}\right)\phi\;,\nonumber
\end{eqnarray}
with
\begin{eqnarray}
\chi^{\lambda3}_\uparrow&=&{1\over\sqrt 6}(\downarrow\uparrow\uparrow+
\uparrow\downarrow\uparrow-2\uparrow\uparrow\downarrow),\nonumber\\
\chi^{\lambda3}_\downarrow&=&{1\over\sqrt 6}(2\downarrow\downarrow\uparrow-
\downarrow\uparrow\downarrow-\uparrow\downarrow\downarrow)\;,\nonumber\\
&&\label{eq:spinfunction}\\
\chi^{\rho3}_\uparrow&=&{1\over\sqrt 2}(\uparrow\downarrow\uparrow-
\downarrow\uparrow\uparrow)\;,\nonumber\\
\chi^{\rho3}_\downarrow&=&{1\over\sqrt 2}(\uparrow\downarrow\downarrow
-\downarrow\uparrow\downarrow)\;.\nonumber
\end{eqnarray}
The spin wave functions $\chi^{\lambda2}$ and $\chi^{\lambda1}$ are the
appropriate permutations of $\chi^{\lambda3}$, and $\chi^{\rho2}$ and
$\chi^{\rho1}$ are the appropriate permutations of $\chi^{\rho3}$.
The spin-wave function of the $i$th quark is given by
\begin{equation}
\uparrow={\cal R}_i\pmatrix{1\cr 0} \hbox{  and  }
\downarrow={\cal R}_i\pmatrix{0\cr 1}\;.
\end{equation}
\index{wave function|(}

\newpage
\section{Extensions of the model}\label{sec:extension}

As already mentioned in the introduction we are also considering some
extensions to the minimal quark model (MQM), because the MQM is not able
to fit experimental data for both the electromagnetic and weak sector
with the same parameters. We give an overview of the different models in
this Section. Questions concerning particular experimental data
are discussed in the appropriate Chapters.

\subsection{Minimal quark model}
The MQM uses the wave function presented in Eq.~(\ref{eq:2.34}) with
structureless quarks. It is called minimal because it uses the smallest
number of parameters. These are the mass of the up and down quark
$m_u=m_d\equiv m_{u/d}$, the mass of the strange quark $m_s$, and the wave
function parameter $\beta$ for the nucleons, the $\Sigma$s and the
$\Xi$s; this $\beta$ essentially determines the \mi{confinement scale}.
With these five parameters we either fit the electromagnetic properties
(parameter set~4 of Table \ref{tab:para} on page~\pageref{tab:para}) or the
semileptonic weak decays (parameter set~6). The contrary  statement in
Ref.~\cite{azna84} has to be questioned, since their numerical results for
the \mi{magnetic moments} of the baryon octet are wrong.

\subsection{Quark structure}
\index{quark structure}
One extension to the MQM is the introduction of nontrivial quark form
factors. This would give us three new parameters in the electromagnetic
sector (quark anomalous magnetic moments), and four new parameters in the
weak sector (axial and vector quark form factors at zero momentum transfer).
\mi{Chung} and
\mi{Coester} \cite{chun91} investigate the nucleon sector of the same
model, and favor quark anomalous magnetic moments and a modified axial
coupling for quarks. There are two reasons why we think that the quarks
should be structureless in the electromagnetic sector:
\begin{enumerate}
\item Whatever the nature of this form factor is, whether it be due to a
composite model or to radiative corrections, one expects the quark form
factors to fall off for large $K^2$ in a different way from that used in
Ref.~\cite{chun91}.
\item There exists no parameter set of the anomalous magnetic moments that
can improve the magnetic moments of the baryon octet (see
Sec.~\ref{sec:mm}).
\end{enumerate}

\subsection{Configuration mixing}
Another extension of the MQM is the admixture of different radial wave
functions. The configuration mixing suggested by spectroscopy reads:
\be
| {\rm Baryon} \rangle = {\rm A}\; [56,0^+][8\times 2] + {\rm B}\;
[56,0^+]^*[8\times 2] + {\rm C}\; [70,0^+] [8\times 2],\label{eq:mixing}
\ee
in the notation [SU(6),L$^p$][SU(3)$_{\rm flavour}\times$SU(2)$_{\rm spin}$],
where
$
{\rm A}^2+{\rm B}^2+{\rm C}^2=1\;,
$
L denotes the angular momentum, and p is the parity of the nucleon.
The values for A, B, C are listed in Table \ref{tab:mixing} for different
references.

\begin{table}
\caption[Parameters for the configuration mixing of the baryon octet.]
{Parameters for the configuration mixing of the baryon octet
given in Eq. (\protect\ref{eq:mixing}) for two different references.}
\begin{center}\begin{tabular}{cccc}
\hline\hline
&A&B&C\\
\hline
Ref. \cite{isgu82}&0.93&--0.29&--0.23\\
Ref. \cite{caps86}&0.90&--0.34&--0.27\\
\hline\hline
\end{tabular}\end{center}
\label{tab:mixing}
\end{table}

In principle no additional parameters are required, if we use the same
scale parameter $\beta$ for every wave function, but in practice,
it is convenient
to choose different $\beta$s or admixture parameters. The wave functions
in Eq.~(\ref{eq:mixing}) are as follows:
\bea
\lbrack 56,0^+\rbrack \lbrack 8\times 2\rbrack &=& \frac{1}{\sqrt{2}}
\left(\chi^\rho\Phi^\rho +
\chi^\lambda\Phi^\lambda\right)\phi_s\;, \nonumber\\
\lbrack 56,0^+\rbrack ^*\lbrack 8\times 2\rbrack &=& \frac{1}{\sqrt{2}}
\left(\chi^\rho\Phi^\rho +
\chi^\lambda\Phi^\lambda\right)\phi^*_s\;, \\
\lbrack 70,0^+\rbrack \lbrack 8\times 2\rbrack &=& \frac{1}{2}
\left(\chi^\rho\Phi^\lambda +
\chi^\lambda\Phi^\rho\right)\phi_\rho + \frac{1}{2}
\left(\chi^\rho\Phi^\rho -
\chi^\lambda\Phi^\lambda\right)\phi_\lambda\;. \nonumber
\eea
The spin functions $\chi^\rho$ and $\chi^\lambda$ are the same as in
Eq.~(\ref{eq:spinfunction}), the flavor wave functions $\Phi^\rho$ and
$\Phi^\lambda$ correspond to $\chi^\rho$ and $\chi^\lambda$ with spin up
and down exchanged with the appropriate flavors of the baryon. For the
momentum wave functions $\phi_s$, $\phi^*_s$, $\phi_\rho$, and
$\phi_\lambda$ we first define
\be
M_i = \frac{k^2_{i\perp}+m^2_i}{x_i}\;, \quad x_i=p^+_i/P^+\;, \quad
k_{i\perp}=p_{i\perp} -x_iP_\perp\;.
\ee
With these functions $M_i$ it is easier to build wave functions with
\index{wave function!configuration mixing}
special symmetries. Note that $M=M_1+M_2+M_3$, the combination
$M_1+M_2-2M_3$ is symmetric in the particles (12), and $M_1-M_2$ is
antisymmetric in the same particles. We therefore write:
\bea
\phi_s &=& N_s e^{-M^2/2\beta^2}\;, \nonumber\\
\phi^*_s &=& N^*_s (M^2/\beta^2-c)\phi_s\;, \nonumber\\
\phi_\lambda &=& N_\lambda (M_1+M_2-2M_3)\phi_s\;, \nonumber\\
\phi_\rho &=& N_\rho (M_1-M_2)\phi_s\;.
\label{eq:wavemix}
\eea
The constant $c$ in $\phi^*_s$ is evaluated from the orthogonality of
$\phi_s$ and $\phi^*_s$:
\be
c=\frac{\int \phi^2_s M^2/\beta^2}{\int \phi^2_s}\;,
\ee
and the constants $N_s$, $N^*_s$, $N_\lambda$, and $N_\rho$ are given by
the normalization in Eq.~(\ref{eq:2.32}).
These wave functions in Eq.~(\ref{eq:wavemix}) go over into the
nonrelativistic ones \cite{bhad88,clos79} in the limit, where the masses
$m_i$ go to zero and the $\xi$ and $\eta$ to their nonrelativistic values.

Unfortunately, the mixing configuration does not improve the fit, it
is even worse for the crucial ratio in Eq.~(\ref{eq:ratio}). A rough
estimate gives
\begin{equation}
{g_1/f_1(\Lambda\to pe^-\bar\nu_e) \over g_1/f_1(\Sigma^-\to ne^-\bar\nu_e)}
\simeq -3\left( 1+ \frac{8}{3}{\rm C}^2\right) = -3.5\pm 0.1\;,
\end{equation}
to be compared with the MQM value $-3$, and the experimental data $-2.11\pm
0.15$. Other values like the ratio $\mu (p)/\mu (n)$ also get worse. We
therefore do not consider this extension any further.

\subsection{Asymmetric wave function}
\index{wave function!asymmetric}
The extension to the MQM with a two quark clustering in the
\mi{light front} wave function is minimal in the sense that there
is no difference between the extension and MQM in the meson sector. Instead
of one scale parameter $\beta$, we have two of them, $\beta_q$
for the scale between the two spin-isospin-zero quarks and $\beta_Q$ for
the scale between the third quark and the \mi{diquark}. We devote the
entire Chapter~\ref{ch:Asymmetric} to the asymmetric wave function, because
it improves the fit dramatically in many details and it provides a
comprehensive fit of both the electromagnetic and weak sectors.

\newpage
\section{Discussion of the wave function}\label{sec:disswavefun}

An analysis of the baryon spectrum based on Eq.~(\ref{eq:2.18}) could in
principle determine the wave function, but we restrict ourselves to the
two approximations in Eq.~(\ref{eq:2.34}). It is therefore important to
compare our Ansatz with the wave function of other authors. Usually the
transverse momenta are integrated out up to a scale $\mu$, and the wave
function is expressed in \mi{light front} fractions $x_i=p_i^+/P^+$, written in
our variables as
\bea
x_1 &=& \xi\eta\;, \nonumber\\
x_2 &=& \eta (1-\xi)\;, \\
x_3 &=& 1-\eta\;. \nonumber
\eea
The valence quark distribution amplitude $\phi(x_1,x_2,x_3,\mu^2)$ is
\be
\phi(x_i,\mu^2)=\int^{|q_\perp^2| <\mu^2}\int^{|Q_\perp^2| <\mu^2}
\phi(x_i,q_\perp,Q_\perp) d^2q_\perp d^2Q_\perp\;.
\ee
This amplitude is well known in two limits, which are unfortunately not
interesting. In the static, symmetric SU(6) quark model, the
variables $x_i$ take on only discrete values:
\be
\phi_{\rm NR}(x_i)=\delta\left(x_1-\frac 1 3\right)\delta\left(x_2-\frac 1
3\right)
\delta\left(x_3-\frac 1 3\right)=\delta\left(\xi-\frac 1
2\right)\delta\left(\eta-\frac 2 3\right)
\ee
and the asymptotic amplitude $\phi_{\rm as}$ for large $K^2$ is known as
\cite{lepa80}
\be
\phi_{\rm as}(x_i)=\phi(x_i,\mu^2\rightarrow \infty)=120x_1x_2x_3\;,
\label{eq:asym}
\ee
with a normalization such that
\be
\int \phi(x_i,\mu^2) \delta(\sum x_i -1) dx_1 dx_2 dx_3 =1\;.
\ee
Notice that this is not the same normalization
as the one used in Eq. (\ref{eq:2.32}).
Unfortunately the knowledge of both forms is not very useful
since they contradict experimental data \cite{isgu84}.

Using the \mi{QCD sum rule technique}, \mi{Chernyak} and \mi{Zhitnitsky}
 \cite{cher84}
suggest the quark distribution amplitudes for the proton as follows:
\bea
\lefteqn{\phi_{\rm CZ}(x_i,\mu\approx 1\hbox{ GeV}) =}\\
& & 120x_1x_2x_3\left[
11.35(x_1^2+x_2^2) + 8.82x_3^2 - 1.68x_3 - 2.94 - 6.72(x_2^2-x_1^2)\right]\;.
\nonumber
\eea
For the Gauss shaped wave function $\phi_{\rm G}$ and the Lorentz shaped
one $\phi_{\rm L}$ in Eq.~(\ref{eq:2.34}) we can write
\bea
\lefteqn{\phi_{\rm G} = N e^{-M^2/2\beta^2}\;,\qquad
\phi_{\rm L} = \frac{N'}{(M^2/\beta^2+1)^n}\;,} \nonumber\\
\lefteqn{\int_{-\infty}^\infty \int_{-\infty}^\infty \phi_{\rm G}\; d^2q_\perp
d^2Q_\perp =}
\nonumber\\
 & &\tilde N \beta^4
\xi\eta^2(1-\eta)(1-\xi)\exp{\left(-\frac{m_1^2}{2\beta^2\eta\xi}
-\frac{m_2^2}{2\beta^2(1-\xi)}-\frac{m_3^2}{2\beta^2(1-\eta)}\right)}\;,\\
\lefteqn{\int_{-\infty}^\infty \int_{-\infty}^\infty \phi_{\rm L}\; d^2q_\perp
d^2Q_\perp =}
\nonumber\\
& &\frac{\tilde{N'}\beta^{2n}(1-\eta)^{n-1}\eta^n(1-\xi)^{n-1}\xi^{n-1}}
{\left[m_1^2(1-\xi)(1-\eta)+m_2^2\xi(1-\eta)+m_3^2\xi\eta(1-\xi)
+\beta^2\eta\xi(1-\xi)(1-\eta)\right]^{n-2}}\;.\nonumber
\eea
Letting the quark masses $m_i$ go to zero the amplitudes both converge
to the asymptotic form (\ref{eq:asym}):
\bea
\phi_{\rm G}(x_i) &\buildrel m_i\rightarrow 0\over {\hbox to 1in{
\rightarrowfill} }& \tilde N \beta^4\xi\eta^2(1-\eta)(1-\xi) =
\tilde N \beta^4 x_1x_2x_3\;,\nonumber\\
\phi_{\rm L}(x_i) &\buildrel m_i\rightarrow 0\over {\hbox to 1in{
\rightarrowfill} }& \tilde N' \beta^4\xi\eta^2(1-\eta)(1-\xi) =
\tilde N' \beta^4 x_1x_2x_3\;.
\eea

\begin{figure}
\centerline{\psfig{figure=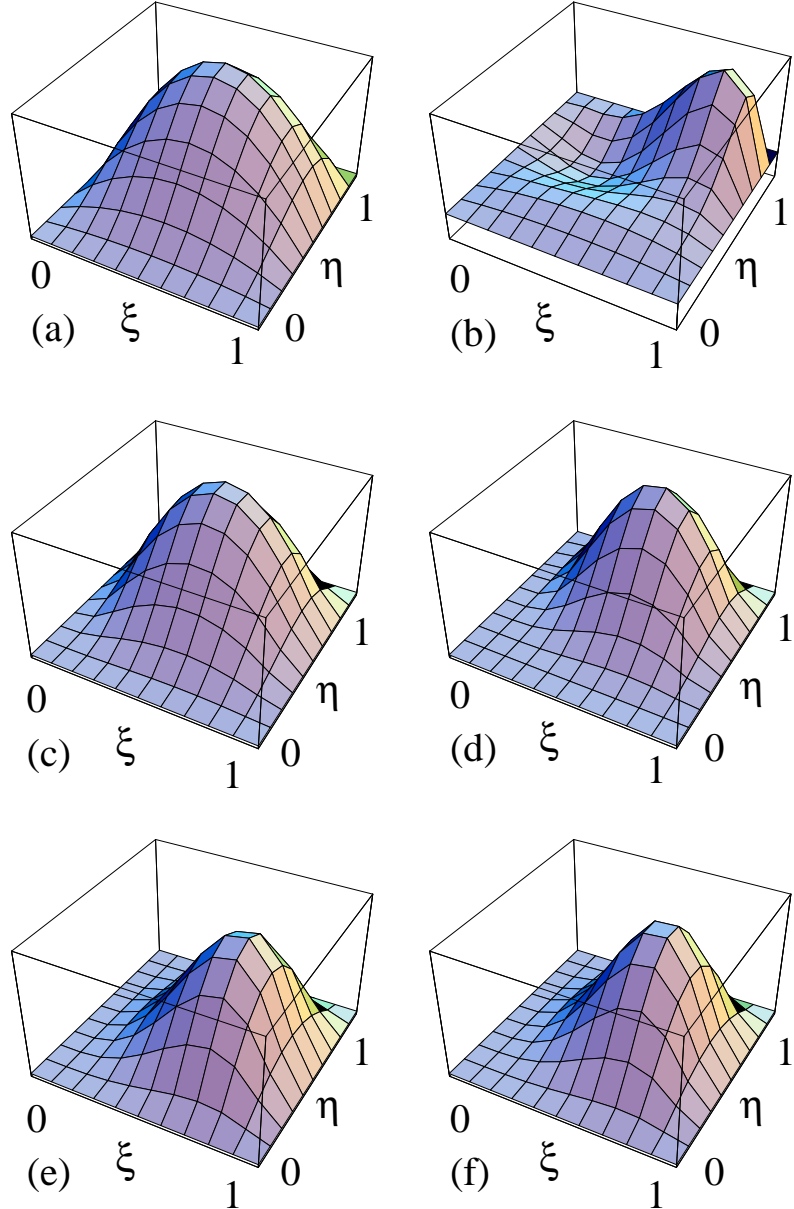,height=500pt}}
\caption[Different wave functions used by other authors and used in this
work.]{Different wave functions used by other authors and used in this
work. (a) asymptotic quark distribution \protect\cite{lepa80}; (b)
amplitude derived by QCD sum-rule technique \protect\cite{cher84}; Gauss
shaped wave function with parameter set~6 (c) and set~8 for hyperons (d);
Lorentz shaped wave function with parameter set~6 (e) and set~8 for
hyperons (f). A trend to more structured, asymmetric wave functions can be
seen.}
\label{fig:wavefun}
\end{figure}

The differences between these various wave functions are best seen in a
plot. In Fig.~\ref{fig:wavefun} we show $\phi_{\rm as}$ (a),
$\phi_{\rm CZ}$ (b), $\phi_{\rm G}$ (c+d), and $\phi_{\rm L}$ (e+f).
The plots (c) and (e) are the symmetric wave functions (parameter set~6),
(d) and (f) are the asymmetric ones (parameter set~8 for hyperons). The broad,
unstructured distribution in the asymptotic limit gets sharper and more
structured for the phenomenological amplitudes. The wave function in (c)
usually used in quark models is close to the asymptotic function (a).

\begin{figure}
\centerline{\psfig{figure=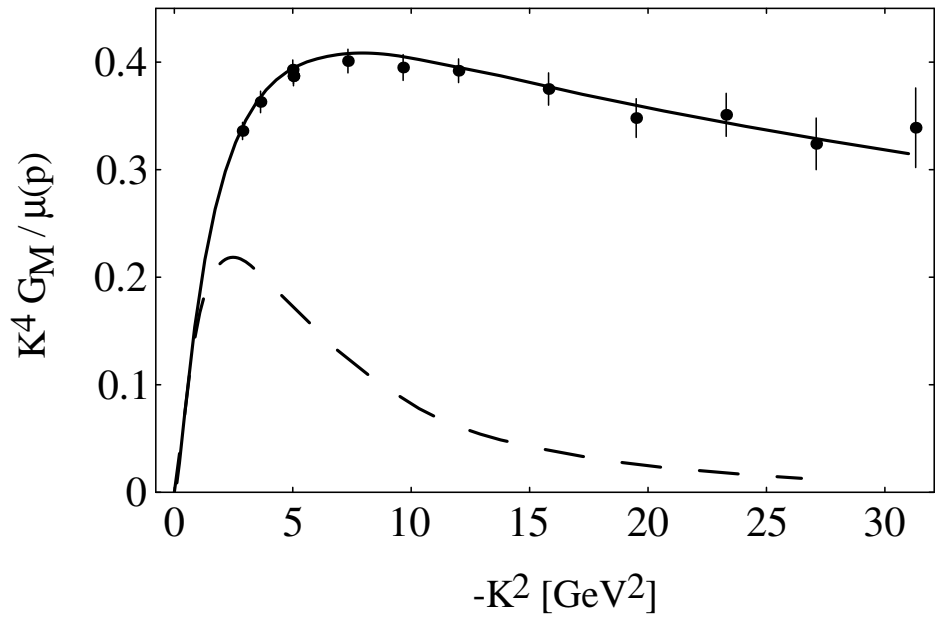,height=240pt}}
\caption[Proton form factor $G_{Mp}(K^2)$.]{Proton form factor $G_{Mp}(K^2)$
for both wave functions in
Eq.~\protect\ref{eq:2.34}. Continuous line, Lorentz shaped wave function
with parameter set~3; broken line; Gauss shaped wave function with
parameter set~2. The experimental data are taken from
Ref.~\protect\cite{arno86}.}
\label{fig:gm}
\end{figure}

The important difference between $\phi_G$ and $\phi_L$ is their large
momentum behavior. For $| K^2| \to \infty$ the wave functions behave as:
\be
\phi_G \to e^{-| K^2 |/2\beta^2} \;, \qquad
\phi_L \to \left[\frac{| K^2 |}{\beta^2} \right]^{-n} \;.
\ee
The exponential falloff for $\phi_G$ becomes too strong at a momentum
scale of about 2 GeV$^2$ (see Fig.~\ref{fig:gm}).

\index{wave function|)}

\chapter{Electromagnetic properties}\label{ch:electro}

\section{Nucleon electromagnetic form factors}\label{sec:nucleon}
\index{nucleons!electromagnetic form factors|(}
\subsection{Calculation of nucleon form factors}
The electromagnetic current matrix element can be written
in terms of two form factors taking into account current and parity
conservation:

\begin{equation}
\left< N,\lambda ' p' \left| J^\mu \right| N,\lambda p\right> =
\bar u_{\lambda '}(p') \left[ F_1(K^2)\gamma^\mu + {F_2(K^2) \over
2 M_N}i\sigma^{\mu\nu}K_\nu \right] u_\lambda (p)
\label{eq:3.1}
\end{equation}
with momentum transfer $K = p' - p$ and $J^\mu = \bar q \gamma^\mu q$.
For $K^2 = 0$ the form factors $F_1$ and $F_2$ are respectively equal
to the charge and the anomalous magnetic moment in units $e$ and
$e/M_N$, and the magnetic moment is $\mu = F_1(0) + F_2(0)$.
The \mi{Sachs form factors} are defined as
\begin{equation}
G_E = F_1 + {K^2 \over 4M_N^2}F_2\;, \quad\hbox{and}\quad
G_M = F_1 + F_2\;,
\end{equation}
and the charge radii \index{charge radius} of the nucleons are
\begin{equation}
\left< r_i^2 \right> = 6{dF_i(K^2) \over dK^2}\Bigg\vert_{K^2=0}\;,
\quad\hbox{and}\quad
\left< r_{E/M}^2 \right> = {6 \over G_{E/M}(0)}{dG_{E/M}(K^2) \over dK^2}
\Bigg\vert_{K^2=0}\;.
\end{equation}
The form factors can be expressed in terms of the $+$ component of
the current:
\begin{eqnarray}
F_1(K^2) &=&{1 \over 2P^+}\left< N,\uparrow\left| J^+\right| N,
\uparrow\right>\;,\nonumber\\
&&\\
K_\perp F_2(K^2) &=&-{2M_N \over 2P^+}\left< N,\uparrow\left|
J^+\right| N,\downarrow\right>\;.\nonumber
\end{eqnarray}
Therefore Eq.~(\ref{eq:2.25}) can be used to calculate the form factors.
In addition, the Dirac quark current $\bar q \gamma^\mu q$ can be
generalized to include the quark structure in the following way:
\begin{equation}
\bar q \left(F_{1q}\gamma^\mu + {F_{2q}\over 2m_q}i\sigma^{\mu\nu}
K_\nu \right) q\;.
\label{eq:quarkf}
\end{equation}
We get
\begin{eqnarray}
F_1(K^2) &=& {N_c \over (2\pi)^6} \int\!d^3\!q d^3Q
\left( {E_3'E_{12}'M \over E_3E_{12}M'} \right)^{1/2}\phi^\dagger(M')\phi(M)
\nonumber\\
&&\times\sum_{i=1}^3\left[F_{1i}(K^2)\left<\chi_\uparrow^{\lambda i}|\chi_
\uparrow^{\lambda i}\right>+{K_\perp\over 2m_i}F_{2i}(K^2)
\left<\chi_\uparrow^{\lambda i}|O_3|\chi_\uparrow^{\lambda i}\right>
\right]\nonumber\\
&&\label{eq:bracket1}\\
K_\perp F_2(K^2) &=& -2M_N{N_c \over (2\pi)^6} \int\!d^3\!q d^3Q
\left( {E_3'E_{12}'M \over E_3E_{12}M'} \right)^{1/2}\phi^\dagger(M')\phi(M)
\nonumber\\
&&\times\sum_{i=1}^3\left[ F_{1i}(K^2)\left<\chi_\uparrow^{\lambda i}|\chi_
\downarrow^{\lambda i}\right>+{K_\perp\over 2m_i}F_{2i}(K^2)
\left<\chi_\uparrow^{\lambda i}|O_3|\chi_\downarrow^{\lambda i}\right>\right]
\nonumber
\end{eqnarray}
with $O_3 =\pmatrix{0&-1\cr 1&0}$, $i=(uud)$ for the proton and $i=(ddu)$
for the neutron. The form factors $F_{1u}(0)$ and $F_{1d}(0)$ are the charges,
$F_{2u}(0)$ and $F_{2d}(0)$ are the  anomalous magnetic moments
\index{quark structure} of the
$u$ and $d$ quarks, respectively. Only the $F_{1i}$-terms contribute
for $K^2=0$, and $\left< \chi_\uparrow^{\lambda i}|\chi_\uparrow^{\lambda i}
\right> =1$. The matrix elements $\left< \chi_\uparrow^{\lambda i}|\chi_
\downarrow^{\lambda i} \right> $ and $\left< \chi_\uparrow^{\lambda i}|O_3|
\chi_\downarrow^{\lambda i} \right> $ are given in the next Subsection. The
expressions for $K^2\neq 0$ are quite long, we therefore give only
$F_1$ for the proton with vanishing quark anomalous moment and
$F_{1u}={2 \over 3}$, $F_{1d}=-{1 \over 3}$:
\begin{equation}
\sum_{i=1}^3F_{1i}\left<\chi_\uparrow^{\lambda i}|\chi_\uparrow^
{\lambda i}\right>={{\rm Num}\over (a^{'2}+Q^{'2}_\perp)(a^{2}+Q^{2}_\perp)
\sqrt{(b^{'2}+Q^{'2}_\perp)}\sqrt{(b^{2}+Q^{2}_\perp)}
(c^2+q^2_\perp)(d^2+q^2_\perp)}
\label{eq:3.7}
\end{equation}

\begin{eqnarray}
{\rm Num}&=&(a^{'2}+Q^{'2}_\perp)(a^2+Q^2_\perp)
(b'b+Q'_\perp\!\!\cdot\! Q_\perp)(c^2d^2+q^4_\perp)\nonumber\\
&&
+(c^2+d^2)q^2_\perp\Bigl\lbrace a^2a^{'2}(bb'+Q'_\perp\!\!\cdot\! Q_\perp)+
(aa'+{1\over 2}Q'_\perp\!\!\cdot\! Q_\perp)\nonumber\\
&&
\times [2bb'(Q'_\perp\!\!\cdot\! Q_\perp) +Q^{'2}_\perp
Q^2_\perp+(Q'_\perp\!\!\cdot\! Q_\perp)^2]\Bigr\rbrace\nonumber\\
&&
+cdq^2_\perp\Bigl\lbrace 4aa'bb'(Q'_\perp\!\!\cdot\! Q_\perp)
 -2(a^2Q^{'2}_\perp
+a^{'2}Q^2_\perp)(bb'+Q'_\perp\!\!\cdot\! Q_\perp)\nonumber\\
&&
+2aa'[Q^{'2}_\perp Q^2_\perp+(Q'_\perp\!\!\cdot\! Q_\perp)^2]
+(2bb'+Q'_\perp\!\!\cdot\! Q_\perp)[(Q'_\perp\!\!\cdot\!
Q_\perp)^2-Q^{'2}_\perp
Q^2_\perp ]\Bigr\rbrace \nonumber
\end{eqnarray}
For $K^2=0$, Eq.~(\ref{eq:3.7}) reduces to 1 and we get $F_{1p}(0)=1$, the
charge of the proton in units of $e$.

\subsection{Results and conclusions}
\begin{figure}
\centerline{\psfig{figure=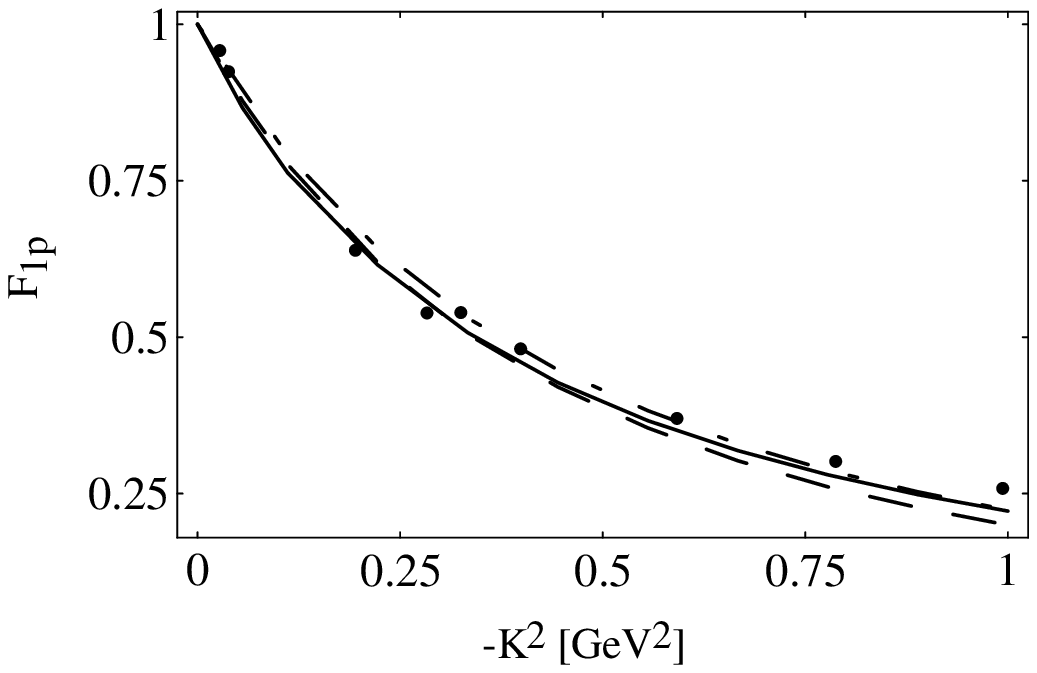,height=240pt}}
\caption[The proton form factor $F_{1p}(K^2)$.]
{The proton form factor $F_{1p}(K^2)$. Continuous line, Parameter
set~3; dashed line, parameter set~2; dashed-dotted line, parameter set~1.
The experimental points are taken from
Ref.~\protect\cite{litt70,hoel76,walk89}.}
\label{fig:lpf1}
\end{figure}

\begin{figure}
\centerline{\psfig{figure=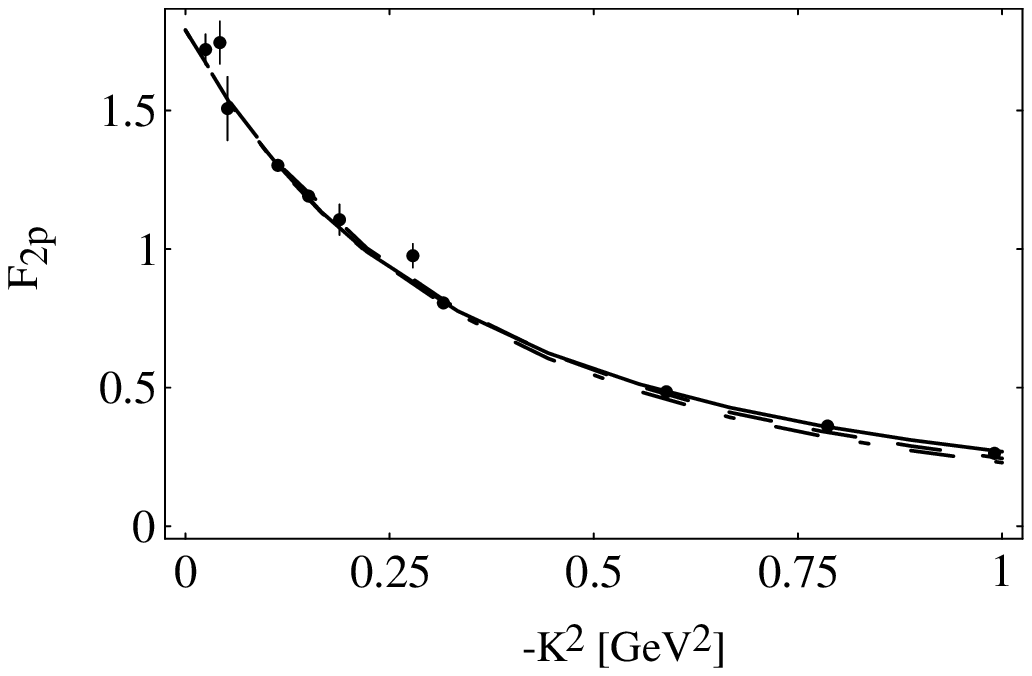,height=240pt}}
\caption[The proton form factor $F_{2p}(K^2)$.]
{The proton form factor $F_{2p}(K^2)$. Continuous line, Parameter
set~3; dashed line, parameter set~2; dashed-dotted line, parameter set~1.
The experimental data are taken from
Ref.~\protect\cite{litt70,hoel76,walk89}.}
\label{fig:lpf2}
\end{figure}

Appendix \ref{ch:methods} describes how the formulae were generated and
integrated. The exponential function $\phi (M)$ in Eq.~(\ref{eq:2.34})
falls off too fast, it can only be valid for low $K^2$. In general $\phi$
has to be just a function decreasing with $M$. We try
\begin{equation}
\phi (M) = {N\over (M^2 +\beta^2)^{3.5}}\;,
\label{eq:3.11}
\end{equation}
and $N$ is chosen so that Eq.~(\ref{eq:2.32}) is fulfilled. The wave
function $\phi (M)$ does correspond to a confining potential in the sense
that we need more energy to ionize the \mi{bound state} than to produce a
new quark pair. This guarantees that no free quark appears. Figures
\ref{fig:lpf1} and \ref{fig:lpf2} show that the low $K^2$ behavior is the
same for both wave functions.

\begin{figure}[bhtp]
\centerline{\psfig{figure=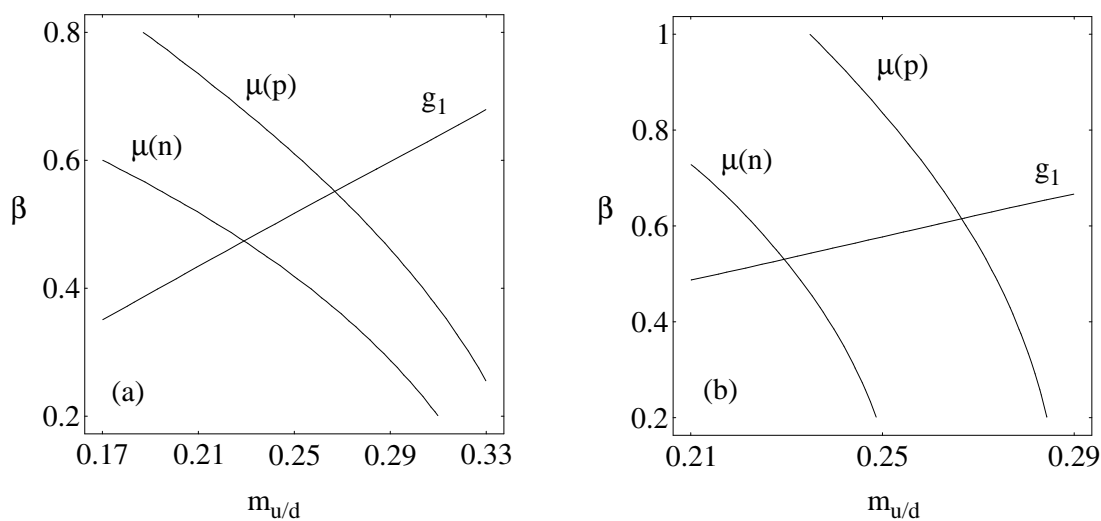,height=200pt}}
\caption[Parameter space for the nucleon sector.]
{The lines represent a set of parameters $\beta$ and $m_{u/d}$,
which reproduce respectively the experimental data for the
\mi{magnetic moments} of
the proton and neutron, and for $g_1(n \to pe^-\bar\nu)$. (a) Parameters
for the Gauss shaped wave function; (b) parameters for the Lorentz shaped
wave function.}
\label{fig:para}
\end{figure}

We plot the parameter $\beta$ against $m_{u/d}$ for the different
experimental data. In Fig.~\ref{fig:para} the fits for the Gauss shaped wave
function (a) and for the Lorentz shaped one (b) are given. We see that it
is not possible to exactly fit the three data, since the three lines
should ideally meet in one single point.
Figure \ref{fig:para} shows for case (a) that the proton and neutron magnetic
moments alone tend to a large quark mass $m_{u/d}\simeq
0.33$ GeV, whereas $g_1$ of the \mi{neutron beta decay} together with the
neutron
magnetic moment favors a small $m_{u/d}\simeq 0.23$ GeV. We
compromise and fit to the proton magnetic moment and to $g_1$ of the
neutron decay \index{neutron beta decay},
 which yields $m_{u/d}\simeq 0.267$ GeV. The anomalous \index{quark structure}
moments of the $u$ and $d$ quarks are fitted to $F'_1(0)$ of the neutron.
To analyze the results we have chosen two sets of parameters given in
Table~\ref{tab:para} on page~\pageref{tab:para}, set~1 with
quark anomalous magnetic moments and set~2 without them. The situation
for case (b) in Fig~\ref{fig:para} is similar, with the exception that the
proton and neutron magnetic moments do not favor a large mass. We
use parameter set~3 for the Lorentz shaped wave function.

\begin{table}
\caption[The parameters of the constituent quark model.]
{The parameters of the constituent quark model. All
numbers are given in units of GeV. Note that only set~3 is used for the
Lorentz shaped wave function.}
\begin{center}\begin{tabular}{cccccccccc}
\hline\hline
&$m_u$&$m_d$&$m_s$&$\beta_N$&$\beta_{\Sigma/\Lambda}$&$\beta_\Xi$&
$F_{2u}$&$F_{2d}$&$F_{2s}$\\
\hline
Set 1&0.267&0.267&--&0.56&--&--&--0.0069&--0.028&--\\
Set 2&0.267&0.267&--&0.56&--&--&0.0&0.0&--\\
Set 3&0.263&0.263&--&0.607&--&--&0.0&0.0&--\\
Set 4&0.33&0.33&0.55&0.16&1.00&1.08&--0.0086&--0.034&0.077\\
Set 5&0.267&0.267&0.33&0.56&0.63&0.70&0.0&0.0&0.0\\
Set 6&0.267&0.267&0.40&0.56&0.60&0.62&0.0&0.0&0.0\\
Set 7&0.267&0.267&0.40&0.56&0.60&0.62&--0.0069&--0.028&0.056\\
\hline\hline
\end{tabular}\end{center}
\label{tab:para}
\end{table}

\begin{table}
\caption[The quantity $F'(0)$ for the nucleons.]
{The quantity $F'(0)$ for the nucleons with parameter sets 1 and 2 of
Table~\protect\ref{tab:para}. The values for the Lorentz shaped wave
function (set 3) are almost the same as the one from set 2.}
\begin{center}\begin{tabular}{cccc}
\hline\hline
Form factor&Set 1 [fm$^2$]&Set 2 [fm$^2$]&Expt. [fm$^2$]\\
\hline
$F'_{1p}$&0.0874&0.0924&0.0966\\
$F'_{2p}$&0.179&0.177&0.234\\
$F'_{1n}$&0.0027&0.012&0.0017\\
$F'_{2n}$&--0.186&--0.170&--0.236\\
\hline\hline
\end{tabular}\end{center}
\label{tab:static}
\end{table}


\begin{figure}
\centerline{\psfig{figure=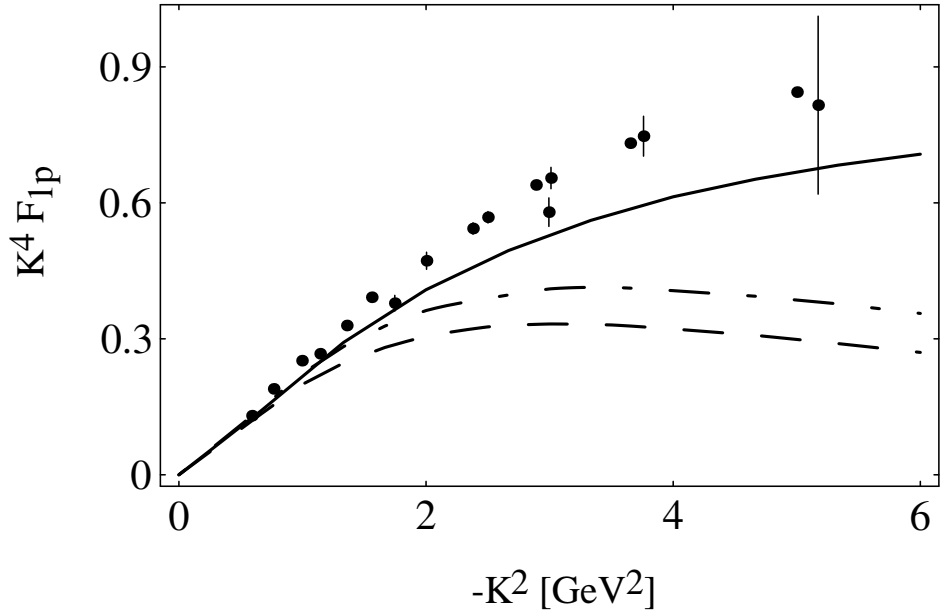,height=240pt}}
\caption[The proton form factor $F_{1p}(K^2)$.]
{The proton form factor $F_{1p}(K^2)$. Continuous line, Parameter
set~3; dashed line, parameter set~2; dashed-dotted line, parameter set~1.
The experimental data are taken from
Ref.~\protect\cite{litt70,hoel76,walk89}.}
\label{fig:pf1}
\end{figure}

\begin{figure}
\centerline{\psfig{figure=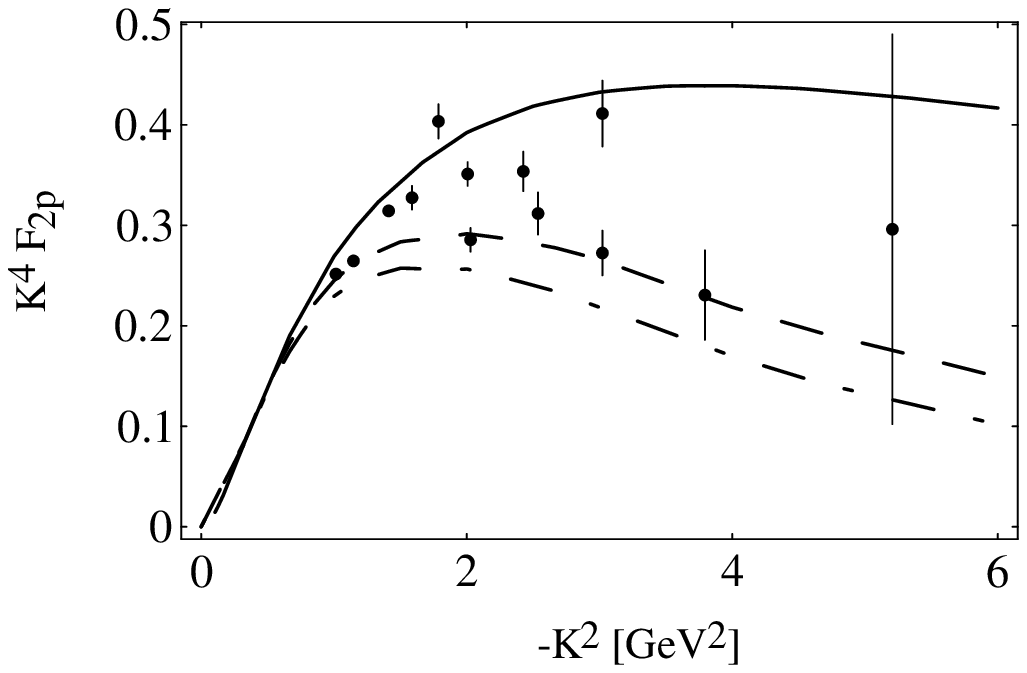,height=240pt}}
\caption[The proton form factor $F_{2p}(K^2)$.]
{The proton form factor $F_{2p}(K^2)$. Continuous line, Parameter
set~3; dashed line, parameter set~2; dashed-dotted line, parameter set~1.
The experimental data are taken from
Ref.~\protect\cite{litt70,hoel76,walk89}.}
\label{fig:pf2}
\end{figure}

\begin{figure}
\centerline{\psfig{figure=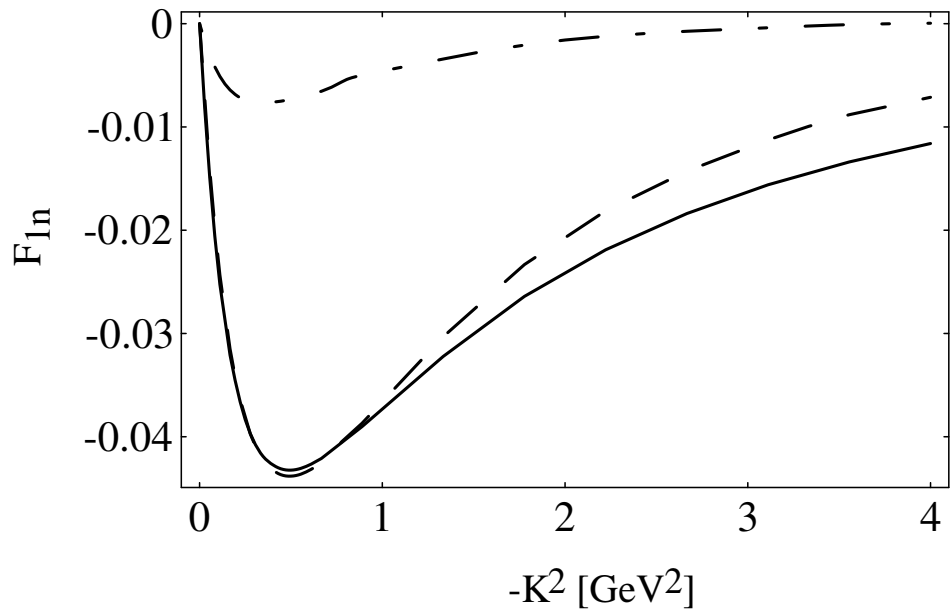,height=240pt}}
\caption[The neutron form factor $F_{1n}(K^2)$.]
{The neutron form factor $F_{1n}(K^2)$. Continuous line, Parameter
set~3; dashed line, parameter set~2; dashed-dotted line, parameter set~1.}
\label{fig:nf1}
\end{figure}

\begin{figure}
\centerline{\psfig{figure=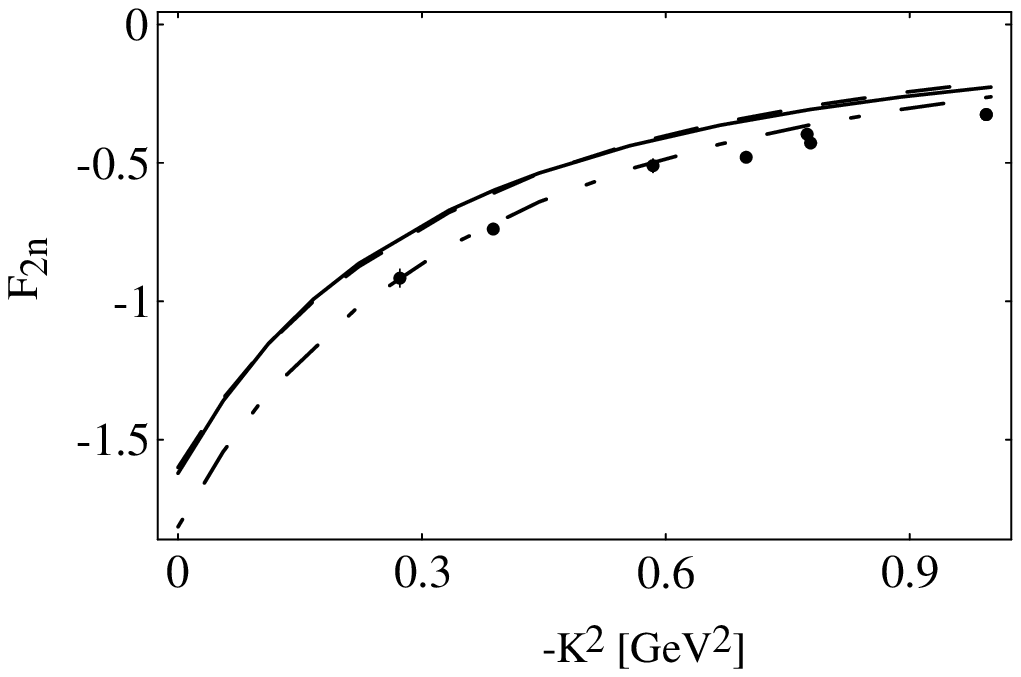,height=240pt}}
\caption[The neutron form factor $F_{2n}(K^2)$.]
{The neutron form factor $F_{2n}(K^2)$. Continuous line, Parameter
set~3; dashed line, parameter set~2; dashed-dotted line, parameter set~1.
The experimental data are taken from
Ref.~\protect\cite{albr68,bart73,hans73}.}
\label{fig:nf2}
\end{figure}

Figures \ref{fig:lpf1} -- \ref{fig:nf2} show $F_1$ and $F_2$ for the proton and
neutron for the various versions of the model. If we neglect the quark
anomalous magnetic moments, only the form factor $F_{1n}$ (Fig~\ref{fig:nf1})
changes and $F_{2n}$ (Fig~\ref{fig:nf2}) gets shifted by a small amount.
Both changes are welcome but only of minor importance. $F_{1n}$ is very
small and therefore sensitive to any corrections.

All data for form factors can be reproduced  well
up to 2 GeV$^2$ for both wave functions in Eq.~(\ref{eq:2.34}). But only
the Lorentz shaped wave function has a good high energy behavior up to
more than 30 GeV$^2$. In this region we already have QCD predictions for
$G_M$ \cite{ji88}. Note that this is an extremely large $K^2$ region, since
other models can only fit the data up to 2.5 GeV$^2$ \cite{warn90} and
6 GeV$^2$ \cite{chun91}. Our fit is even better than the well known
dipole formula
\be
\frac{G_{Mp}}{\mu(p)} = \left(1-\frac{K^2}{M_V^2}\right)^{-2} \;,\quad
M_V=0.84\; \hbox{ GeV }\;.
\ee

One can approximate our form factors with the parameterization of
Eq.~(\ref{eq:pol}) on page~\pageref{eq:pol}. It is valid up to 3 GeV$^2$
with a deviation of less than $5\%$. This justifies the use of that
parameterization in Chapter~\ref{ch:decay}.

The values for magnetic moments derived with the parameter sets 1 and 2
have been collected in Table \ref{tab:moment} and will be discussed
further in Sec.~\ref{sec:mm}. We do not consider the Lorentz shaped wave
function (set~3) any further in this thesis, because the results for
small momentum transfer or even static properties are almost the same for
both types of wave functions.



\index{nucleons!electromagnetic form factors|)}

\clearpage
\section{Magnetic moment of the baryon octet}\label{sec:mm}
\index{magnetic moments|(}
\subsection{$F_2(0)$ in the quark model}
According to Eq.~(\ref{eq:3.1}) the magnetic moment of a baryon is
$\mu=F_1(0)+F_2(0)$. The form factor $F_1(0)$ is equal to the charge, and
$F_2(0)$ to the anomalous magnetic moment $\kappa$ of the particle.
We have [see Eqs.~(\ref{eq:2.25}) and (\ref{eq:quarkf})]
\begin{equation}
M^+_{\lambda'\lambda} =
2P^+{N_c\over (2\pi)^6}\int d^3qd^3Q\left({E'_3E'_{12}M
\over E_3E_{12}M'}\right)^{1/2}\Psi^\dagger({\bf q}',{\bf Q}',
\lambda')O\Psi({\bf q},{\bf Q},\lambda)\;,\label{eq:4.1}
\end{equation}
where $O$ is given by:
\[
O=3\bar u_3\left( F_{13}\gamma^++{1\over 2m_3}F_{23}i\sigma^{+\nu}
K_\nu \right)u_3\;.
\]
We get for the baryon octet
\begin{eqnarray}
\Psi^\dagger_\uparrow O \Psi_\uparrow (K^2=0)&=&\sum_{i=1}^3
F_{1i}\left<\chi^{bi}_\uparrow |\chi^{ai}_\uparrow\right>|\phi|^2\;,
\nonumber\\ &&\label{eq:4.2}\\
\Psi^\dagger_\uparrow O\Psi_\downarrow &=&\sum_{i=1}^3\phi^\dagger\phi
\left( F_{1i} \left<\chi^{bi}_\uparrow |\chi^{ai}_\downarrow\right>
+{K_\perp\over 2m_i}F_{2i}\left<\chi^{bi}_\uparrow |O_3|\chi^{ai}
_\downarrow\right> \right)\;,\nonumber
\end{eqnarray}
with $O_3 =\pmatrix{0&-1\cr 1&0}$, and with

\begin{center}\begin{tabular}{cccccccccc}
\hline\hline
&$p$&$n$&$\Sigma^+$&$\Sigma^-$&$\Sigma^0$&$\Lambda$&$\Xi^-$&$\Xi^0$&$\Sigma^0
\Lambda$\\
\hline
a/b&$\lambda/\lambda$&$\lambda/\lambda$&$\lambda/\lambda$&$\lambda/\lambda$&
$\lambda/\lambda$&$\rho/\rho$&$\lambda/\lambda$&$\lambda/\lambda$&
$\rho/\lambda$\\
i=1&u&d&u&d&(u+d)/2&(u+d)/2&s&s&(d--u)/2\\
i=2&u&d&u&d&(u+d)/2&(u+d)/2&s&s&(d--u)/2\\
i=3&d&u&s&s&s&s&d&u&--\\
\hline\hline
\end{tabular}\end{center}

In order to get $F_1(0)$ and $F_2(0)$ the matrix elements
$\left<\chi_\uparrow | \chi_\uparrow
\right>$ and $\left<\chi_\uparrow | O_3 | \chi_\downarrow\right>$ have
to be calculated to order 1, and $\left<\chi_\uparrow | \chi_\downarrow
\right>$ to order $K_\perp$:
\begin{eqnarray}
\left<\chi^{\lambda}_\uparrow |\chi^{\lambda}_\uparrow\right>&=&
\left<\chi^{\rho}_\uparrow |\chi^{\rho}_\uparrow\right>=1\;,\nonumber\\
\left<\chi^{\lambda1}_\uparrow |O_3|\chi^{\lambda1}_\downarrow\right>&=&
\left<\chi^{\lambda2}_\uparrow |O_3|\chi^{\lambda2}_\downarrow\right>=
-{2\over 3}{b^2\over b^2+Q^2_\perp}\;,\nonumber\\
\left<\chi^{\lambda3}_\uparrow |O_3|\chi^{\lambda3}_\downarrow\right>&=&
{1\over 3}{b^2\over b^2+Q^2_\perp}\;,\nonumber\\
\left<\chi^{\rho1}_\uparrow |O_3|\chi^{\lambda1}_\downarrow\right>&=&
\left<\chi^{\rho2}_\uparrow |O_3|\chi^{\lambda2}_\downarrow\right>=
{1\over\sqrt 3}{b^2\over b^2+Q^2_\perp}\;,\nonumber\\
\left<\chi^{\rho1}_\uparrow |O_3|\chi^{\rho1}_\downarrow\right>&=&
\left<\chi^{\rho2}_\uparrow |O_3|\chi^{\rho2}_\downarrow\right>=
0\;,\nonumber\\
\left<\chi^{\rho3}_\uparrow |O_3|\chi^{\rho3}_\downarrow\right>&=&
-{b^2\over b^2+Q^2_\perp}\;,\nonumber\\
\left<\chi^{\lambda1}_\uparrow |\chi^{\lambda1}_\downarrow\right>&=&
{1\over 3}K_\perp (2A_1+2A_2-A_3)\;,\nonumber\\
\left<\chi^{\lambda2}_\uparrow |\chi^{\lambda2}_\downarrow\right>&=&
{1\over 3}K_\perp (2A_1+2A_3-A_2)\;,\label{eq:4.3}\\
\left<\chi^{\lambda3}_\uparrow |\chi^{\lambda3}_\downarrow\right>&=&
{1\over 3}K_\perp (2A_2+2A_3-A_1)\;,\nonumber\\
\left<\chi^{\rho1}_\uparrow |\chi^{\rho1}_\downarrow\right>&=&
K_\perp A_3\;,\nonumber\\
\left<\chi^{\rho2}_\uparrow |\chi^{\rho2}_\downarrow\right>&=&
K_\perp A_2\;,\nonumber\\
\left<\chi^{\rho3}_\uparrow |\chi^{\rho3}_\downarrow\right>&=&
K_\perp A_1\;,\nonumber\\
\left<\chi^{\rho1}_\uparrow |\chi^{\lambda1}_\downarrow\right>&=&
{1\over\sqrt 3}K_\perp (A_2-A_1)\;,\nonumber\\
\left<\chi^{\rho2}_\uparrow |\chi^{\lambda2}_\downarrow\right>&=&
{1\over\sqrt 3}K_\perp (A_2-A_1)\;,\nonumber\\
\hbox{with}\nonumber\\
A_1&=&{{Q_\perp^2\over 2M}-\eta b\over b^2+Q_\perp^2}\;,\nonumber\\
A_2&=&{\eta\left( a-{Q_\perp^2\over 2(1-\eta)M}\right)\over a^2+Q_
\perp^2}{d^2\over d^2+q^2_\perp}\;,\nonumber\\
A_3&=&{\eta\left( a-{Q_\perp^2\over 2(1-\eta)M}\right)\over a^2+Q_
\perp^2}{c^2\over c^2+q^2_\perp}\;.\nonumber
\end{eqnarray}
Eq.~(\ref{eq:4.1}) together with Eqs.~(\ref{eq:4.2}) and (\ref{eq:4.3})
is the most general
formula for $F_2(0)$. Putting $m_u=m_d$ for the hyperons there is an
equality $A_2=A_3$ under the integral, which reduces the number of
integrations.
The simplified formulae read:
\begin{eqnarray}
\kappa(p)&=&-{1\over 3}S_N^{(3)}+{4\over 3}S_N^{(2)}+
{1\over 3}\left( {4F_{2u}\over 2m_u}-{F_{2d}\over 2m_d}\right)Z_N\;,
\nonumber\\
\kappa(n)&=&{2\over 3}S_N^{(3)}-{2\over 3}S_N^{(2)}-
{1\over 3}\left( {F_{2u}\over 2m_u}-{4F_{2d}\over 2m_d}\right)Z_N\;,
\nonumber\\
\kappa(\Sigma^+)&=&-{1\over 3}S_\Sigma^{(3)}+{4\over 3}S_\Sigma^{(2)}+
{4\over 3}{F_{2u}\over 2m_u}Z_\Sigma^{(2)}-{1\over 3}
{F_{2s}\over 2m_s}Z_\Sigma^{(3)}\;,\nonumber\\
\kappa(\Sigma^-)&=&-{1\over 3}S_\Sigma^{(3)}-{2\over 3}S_\Sigma^{(2)}-
{1\over 3}{F_{2s}\over 2m_s}Z_\Sigma^{(3)}+{4\over 3}
{F_{2d}\over 2m_d}Z_\Sigma^{(2)}\;,\nonumber\\
\kappa(\Lambda)&=&-{1\over 3}A_\Sigma^{(3)}+{1\over 3}A_\Sigma^{(2)}+
{F_{2s}\over 2m_s}Z_\Sigma^{(3)}\;,\nonumber\\
\kappa(\Xi^-)&=&-{1\over 3}S_\Xi^{(3)}-{2\over 3}S_\Xi^{(2)}-
{1\over 3}{F_{2d}\over 2m_d}Z_\Xi^{(3)}+{4\over 3}
{F_{2s}\over 2m_s}Z_\Xi^{(2)}\;,\nonumber\\
\kappa(\Xi^0)&=&{2\over 3}S_\Xi^{(3)}-{2\over 3}S_\Xi^{(2)}-
{1\over 3}{F_{2u}\over 2m_u}Z_\Xi^{(3)}+{4\over 3}
{F_{2s}\over 2m_s}Z_\Xi^{(2)}\;,\\
\kappa(\Sigma^0\Lambda)&=&{\sqrt 3 \over 2}\left(S_\Sigma^{(2)}-
A_\Sigma^{(2)}\right)+{1\over \sqrt 3}\left({F_{2u}\over 2m_u}-
{F_{2d}\over 2m_d} \right)Z_\Sigma^{(2)}\;,\nonumber\\
\hbox{with}\nonumber\\
S_B^{(2)}&=&-2M_B{N_c\over (2\pi)^6}\int d^3qd^3Q |\phi|^2
(2A_1+A_2)/3\;,\nonumber\\
S_B^{(3)}&=&-2M_B{N_c\over (2\pi)^6}\int d^3qd^3Q |\phi|^2
(4A_2-A_1)/3\;,\nonumber\\
A_B^{(2)}&=&-2M_B{N_c\over (2\pi)^6}\int d^3qd^3Q |\phi|^2
A_2\;,\nonumber\\
A_B^{(3)}&=&-2M_B{N_c\over (2\pi)^6}\int d^3qd^3Q |\phi|^2
A_1\;,\nonumber\\
Z_B&=&2M_B{N_c\over (2\pi)^6}\int d^3qd^3Q |\phi|^2
{b^2\over b^2+Q^2_\perp}\;.\nonumber
\end{eqnarray}
The masses $m_i$ are set as follows:
\begin{eqnarray}
S_N^{(2)},S_N^{(3)},Z_N\quad :&&\quad
m_1=m_2=m_3=m\;,\nonumber\\
S_\Sigma^{(2)},A_\Sigma^{(2)},Z_\Sigma^{(2)}\quad :&&\quad
m_1=m_3=m,\quad m_2=m_s\;,\nonumber\\
S_\Sigma^{(3)},A_\Sigma^{(3)},Z_\Sigma^{(3)}\quad :&&\quad
m_1=m_2=m,\quad m_3=m_s\;,\\
S_\Xi^{(2)},Z_\Xi^{(2)}\quad :&&\quad
m_1=m_3=m_s,\quad m_2=m\;,\nonumber\\
S_\Xi^{(3)},Z_\Xi^{(3)}\quad :&&\quad
m_1=m_2=m_s,\quad m_3=m\;,\nonumber
\end{eqnarray}
where $m=m_u=m_d$. In $\kappa(\Sigma^0\Lambda)$ the mass $M_B$ is
equal to $(M_\Lambda+M_\Sigma)/2$.

\subsection{Results and conclusions}

Table \ref{tab:moment} shows the magnetic moments for all baryons and the
transition moment $\Sigma^0 \to \Lambda \gamma$. For comparison we
have chosen set~4 from Table \ref{tab:para}.

\begin{table}
\caption{Magnetic moments of the baryon octet and transition moment for
$\Sigma^0\to \Lambda\gamma$ in units of the nuclear magneton.}
\begin{center}\begin{tabular}{ccrrr}
\hline\hline
Particle&Experiment \cite{part90}&Set 4&Set 6&Set 7\\
\hline
$p$&2.79 $\pm 10^{-7}$&2.85&2.78&2.78\\
$n$&--1.91 $\pm 10^{-7}$&--1.83&--1.62&--1.73\\
$\Sigma^+$&2.42 $\pm$ 0.05 &2.59&3.23&3.15\\
$\Sigma^-$&--1.157 $\pm$ 0.025 &--1.30&--1.36&--1.56\\
$\Lambda$&--0.613 $\pm$ 0.004 &--0.48&--0.72&--0.58\\
$\Xi^0$&--1.250 $\pm$ 0.014 &--1.25&--1.87&--1.64\\
$\Xi^-$& --0.679 $\pm$ 0.031&--0.99&--0.96&--0.71\\
$\Sigma^0\Lambda$&1.61 $\pm$ 0.08&1.22&1.74&1.79\\
\hline\hline
\end{tabular}\end{center}
\label{tab:moment}
\end{table}

These parameters are fitted to the magnetic moments and good agreement
can be achieved even without anomalous quark magnetic moments.
\index{quark structure} But we have
\begin{equation}
\beta_N \ll \beta_\Sigma \approx \beta_\Xi\;,
\end{equation}
which cannot be reconciled with the weak beta decay\index{hyperon decay},
 since the small
wave function overlap would cause
too large a suppression of the $\Delta S =1$ transitions.
Therefore, we reject the parameter set~4.
In Chapter~\ref{ch:decay} it is shown that beta decay requires
\begin{equation}
\beta_N \approx \beta_\Sigma \approx \beta_\Xi\;.
\end{equation}
So another set of parameters (set~6) will be considered in
Chapter~\ref{ch:decay}, but Table \ref{tab:moment} shows that it
leads to results which are
in disagreement with experiment. Use of set~6 with non-zero quark
anomalous magnetic moments (set~7) leads only to a slight improvement.
For the quark form factor $F_{2s}$ we get 0.056 GeV by fitting
$\mu (\Lambda)$.

We did not used the Lorentz shaped wave function for the magnetic moments
because this wave function is only important for high $K^2$-values.

We conclude that, within the symmetric wave function model, either the
magnetic moments can be fitted and the weak decay parameters are poorly
fitted or vice versa. The
opposite statement in Ref.~\cite{azna84} has to be questioned, because
their numerical results for the magnetic moments are wrong. Our results
agree with Ref.~\cite{tupp88} on this point. The inconsistency just
described between the
electromagnetic and the weak sector can be resolved by using asymmetric wave
functions as described in Chapter~\ref{ch:Asymmetric}.

\index{magnetic moments|)}

\chapter{Hyperon semileptonic beta decay}\label{ch:decay}
\index{hyperon decay|(}

There is evidence for SU(3) \mi{symmetry breaking} in semileptonic
beta decay of
hyperons \cite{roos90,gens90}. Up to now one does not have any means
to calculate such a behavior from first principles. Some symmetry
breaking schemes have been proposed in the framework of bag
models \cite{dono82} \index{bag model}
and \mi{chiral perturbation theory} \cite{krau90}. Since they are
different from each other we need some support from other models.

Another uncertainty in the analysis of data is the $K^2$ dependence of the
form factors. Although it is generally small, a change of $M_V$ or
$M_A$ by $\pm 0.15$ GeV in the case $\Sigma^- \to ne\nu$ (which has the
largest $K^2_{\rm max}$) causes a relative change of $g_1/f_1$ of
$\pm 2\%$. Ignoring the $K^2$ dependence altogether would shift
$g_1/f_1$ by $17 \%$. Our quark model provides a unique scheme for
the calculation of these form factors.

\section{Hyperon semileptonic decay}
In the low energy limit the standard model for semileptonic weak decays
reduces to an effective \mi{current-current interaction Hamiltonian}
\begin{equation}
H_{\rm int} = {G \over \sqrt{2}} J_\mu L^\mu + {\rm h.c. }\;,
\end{equation}
where $G \simeq 10^{-5}/M_p^2$ is the weak coupling constant,
\begin{equation}
L^\mu = \bar{\psi_e} \gamma^\mu (1 - \gamma_5) \psi_\nu +
   \bar{\psi_\mu} \gamma^\mu (1 - \gamma_5) \psi_\nu
\end{equation}
is the lepton current, and
\begin{equation}
J_\mu = V_\mu - A_\mu\;, \quad
V_\mu = V_{ud} \bar u \gamma_\mu d + V_{us} \bar u \gamma_\mu s\;, \quad
A_\mu = V_{ud} \bar u \gamma_\mu \gamma_5 d + V_{us} \bar u \gamma_\mu
\gamma_5 s \;,
\end{equation}
is the hadronic current, and $V_{ud}, V_{us}$ are the elements of the
Kobayashi-Maskawa mixing matrix\index{Kobayashi-Maskawa matrix}.
The $\tau$-lepton current cannot
contribute since $m_\tau$ is much too large.

The matrix elements of the hadronic current between spin-${1 \over 2}$
states are
\begin{equation}
\left< B',p' \left| V^\mu\right| B,p \right> = V_{qq'} \bar u(p') \left[
  f_1(K^2) \gamma^\mu - {f_2(K^2) \over M_i} i \sigma^{\mu\nu} K_\nu
  + {f_3(K^2) \over M_i}  K^\mu \right] u(p)\;,
\end{equation}
\begin{equation}
\left< B',p' \left| A^\mu\right| B,p \right> = V_{qq'} \bar u(p') \left[
  g_1(K^2) \gamma^\mu - {g_2(K^2) \over M_i} i \sigma^{\mu\nu} K_\nu
  + {g_3(K^2) \over M_i}  K^\mu \right] \gamma_5 u(p)\;,
\end{equation}
where $K = p - p'$ and $M_i$ is the mass of the initial baryon.
The quantities $f_1$ and $g_1$ are the vector and axial-vector
form factors, $f_2$ and $g_2$ are the \mi{weak magnetism} and electric form
factors and $f_3$ and $g_3$ are the induced scalar and pseudoscalar form
factors, respectively. T invariance implies real form factors.
We do not calculate $f_3$ and $g_3$ since we put $K^+ = 0$
and their dependence on the decay spectra is of the order
\begin{equation}
\left( m_l \over M_i \right)^2 \ll 1\;,
\end{equation}
where $m_l$ is the mass of the final charged lepton.
The other form factors are
\begin{eqnarray}
2P^+ f_1 &=&\left< B',\uparrow\left| V^+\right| B,\uparrow
\right>\;,\nonumber\\
2P^+ K_\perp f_2 &=&M_i\left< B',\uparrow\left| V^+\right|
B,\downarrow \right>\;,\nonumber\\
2P^+ g_1 &=&\left< B',\uparrow \left| A^+\right|
B,\uparrow\right>\;,\nonumber\\
2P^+ K_\perp g_2 &=&-M_i\left< B',\uparrow \left| A^+\right|
B,\downarrow\right>\;.
\label{eq:formfactors}
\end{eqnarray}

What is usually measured is the total decay rate $R$, the electron-neutrino
correlation $\alpha_{e\nu}$ and the electron $\alpha_e$, neutrino
$\alpha_\nu$ and final baryon $\alpha_B$ asymmetries.
\index{angular correlation}
The $e$-$\nu$ correlation is defined as
\begin{equation}
\alpha_{e\nu} = 2{N(\Theta_{e\nu} < {1 \over 2}\pi) - N(\Theta_{e\nu} >
 {1 \over 2}\pi) \over N(\Theta_{e\nu} < {1 \over 2}\pi) +
N(\Theta_{e\nu} > {1 \over 2}\pi)}\;,
\end{equation}
where $N(\Theta_{e\nu} < {1 \over 2}\pi)$ is the number of $e$-$\nu$
pairs that form an angle $\Theta_{e\nu}$ smaller than $90^\circ$.
The correlations $\alpha_e$,$\alpha_\nu$ and $\alpha_B$ are defined
analogously with
$\Theta_e$,$\Theta_\nu$ and $\Theta_B$ now being the angles between
the $e$, $\nu$, $B$ directions and the polarization of the initial
baryon.

\index{hyperon decay!decay rate}
\index{hyperon decay!angular correlation}
Ignoring the lepton-mass one can calculate expressions for the
measured quantities. We copy the expressions for $R$, $\alpha_{e\nu}$,
$\alpha_e$, $\alpha_\nu$ and $\alpha_B$ from Ref. \cite{garc85}:
\begin{eqnarray}
R &=&G^2 {\Delta M^5 |V|^2 \over 60\pi^3}\Bigl[ (1 - {3 \over 2} \beta +
  {6 \over 7} \beta^2) f_1^2 + {4 \over 7} \beta^2 f_2^2 + (3 -
  {9 \over 2} \beta + {12 \over 7} \beta^2) g_1^2  \nonumber\\
  &&+ {12 \over 7} \beta^2 g_2^2
  + {6 \over 7} \beta^2 f_1 f_2 + (-4\beta + 6\beta^2) g_1 g_2
  + {4 \over 7} \beta^2 (f_1 \lambda_f + 5g_1\lambda_g) \Bigr]\;,
\label{eq:rate}\\
R \alpha_{e\nu} &=&G^2 {\Delta M^5 |V|^2\over 60\pi^3}\Bigl[
(1 - {5 \over 2} \beta
 + {11 \over 7} \beta^2) f_1^2 - {2 \over 7} \beta^2 f_2^2 +(-1-{3 \over 2}
\beta+{25 \over 7}\beta^2)g_1^2 \nonumber\\
  && -2\beta^2g_2^2 -{2 \over 7}\beta^2f_1f_2+(4\beta-2
\beta^2)g_1g_2
-{24 \over 7}\beta^2g_1\lambda_g \Bigr]\;,\\
R \alpha_e&=&G^2 {\Delta M^5 |V|^2 \over 60\pi^3}\Bigl[ (-{1 \over 3}\beta+
{3 \over 14}\beta^2)f_1^2-{4 \over 21}\beta^2f_2^2+(-2+{8 \over 3}\beta-{9
\over 14}
\beta^2)g_1^2-{4 \over 3}\beta^2g_2^2 \nonumber\\
&&+(-{2 \over 3}\beta+{14 \over 21}\beta^2)f_1f_2
+(2-{11 \over 3}\beta+{15 \over 7}\beta^2)f_1g_1+(-{2 \over 3}\beta+{32
\over 21}\beta^2)f_1g_2 \nonumber\\
&&+(-{2 \over 3}\beta+{32 \over 21}\beta^2)f_2g_1+{16 \over 21}
\beta^2f_2g_2+({10 \over 3}\beta-{94 \over 21}\beta^2)g_1g_2\nonumber\\
&&+{4 \over 7}\beta^2(g_1\lambda_f+f_1\lambda_g-4g_1\lambda_g)\Bigr]\;,\\
R \alpha_\nu&=&G^2 {\Delta M^5 |V|^2 \over 60\pi^3}\Bigl[ ({1 \over 3}\beta-
{3 \over 14}\beta^2)f_1^2+{4 \over 21}\beta^2f_2^2+(2-{8 \over 3}\beta+{9
\over 14}
\beta^2)g_1^2+{4 \over 3}\beta^2g_2^2 \nonumber\\
&&+({2 \over 3}\beta-{14 \over 21}\beta^2)f_1f_2
+(2-{11 \over 3}\beta+{15 \over 7}\beta^2)f_1g_1+(-{2 \over 3}\beta+{32
\over 21}\beta^2)f_1g_2 \nonumber\\
&&+(-{2 \over 3}\beta+{32 \over 21}\beta^2)f_2g_1+{16 \over 21}
\beta^2f_2g_2+(-{10 \over 3}\beta+{94 \over 21}\beta^2)g_1g_2\nonumber\\
&&+{4 \over 7}\beta^2(g_1\lambda_f+f_1\lambda_g+4g_1\lambda_g)\Bigr]\;,\\
R \alpha_B&=&G^2 {\Delta M^5 |V|^2 \over 60\pi^3}{5 \over 2}\Bigl[(-1
+{11 \over 6}\beta-\beta^2)f_1g_1+({1 \over 3}\beta-{5 \over 6}\beta^2)f_1g_2
\nonumber\\
&&+({2 \over 3}\beta
-{7 \over 6}\beta^2)f_2g_1 -{2 \over 3}\beta^2f_2g_2-{1 \over 3}
\beta^2(f_1\lambda_g+g_1\lambda_f)  \Bigr]\;,
\label{eq:asy}
\end{eqnarray}
where $\beta$ is defined as $\beta = (M_i - M_f) / M_i$, and $\Delta M
=M_i-M_f$,
$M_i$, $M_f$ being the masses of the initial and final baryon,
respectively. The $K^2$ dependence of $f_2$ and $g_2$ is ignored and
$f_1$ and $g_1$ are expanded as
\begin{equation}
f_1(K^2) = f_1(0) + {K^2 \over M_i^2} \lambda_f\;, \quad
g_1(K^2) = g_1(0) + {K^2 \over M_i^2} \lambda_g \;.
\end{equation}
We get the corresponding expression for the dipole parameterization
by putting
\begin{equation}
\lambda_f = 2 M_i^2 f_1 / M_V^2\;, \quad
\lambda_g = 2 M_i^2 g_1 / M_A^2 \;.
\end{equation}

Two corrections have to be made to these quantities: (a) the correction
due to the
nonvanishing lepton mass and (b) the \mi{radiative corrections}.
To keep the lepton mass non-zero we integrate the differential decay
rate for $f_2 = g_2 = 0$ :
\begin{equation}
R = G^2 {M_i^5 | V |^2 \over 32 \pi^3} \int_{x_l^2}^{(1-x')^2}\!
dz \left[ \left( f_1^2 + g_1^2 \right) I_1 + \left( f_1^2 - g_1^2
\right) I_2 \right]
\end{equation}
with
\begin{eqnarray}
I_1 &=&(z - x_l^2)(1 + x'^2 - z)f \;,\nonumber\\
I_2 &=&-(z - x_l^2)x'f \;,\nonumber\\
f &=&\lambda^{1/2}(1,z,x'^2) \lambda^{1/2}(z,x_l^2,0)/z \;,\nonumber\\
\lambda(x,y,z) &=&x^2 + y^2 + z^2 - 2xy - 2yz - 2xz \;,\nonumber\\
x_l &=&{m_l \over M_i},\ x' = {M_f \over M_i},\ K^2 = zM_i^2 \;.\nonumber
\end{eqnarray}
The ratio $R/R(m_l=0)$ for the various reactions is given in Table
\ref{tab:ratio} for the $e$-mode $(l=e)$ and the $\mu$-mode $(l=\mu)$.

\begin{table}
\caption{Ratio of the rate to the rate with vanishing lepton mass.}
\begin{center}\begin{tabular}{cll}
\hline\hline
&\multicolumn{2}{c}{$R/R(m_l=0)$} \\
Reaction& $e$-mode & $\mu$-mode \\
\hline
$np$ & 0.472 & -- \\
$\Sigma^+\Lambda$ & 1.000 & -- \\
$\Sigma^-\Lambda$ & 1.000 & -- \\
$\Sigma^-\Sigma^0$ & 0.955 & -- \\
$\Sigma^0\Sigma^+$ & 0.830 & -- \\
$\Xi^-\Xi^0$ & 0.971 & -- \\
$\Lambda p$ & 1.000 & 0.161 \\
$\Sigma^0 p$ & 1.000 & 0.431 \\
$\Sigma^- n$ & 1.000 & 0.443 \\
$\Xi^-\Lambda$ & 1.000 & 0.271 \\
$\Xi^-\Sigma^0$ & 1.000 & 0.0136 \\
$\Xi^0\Sigma^+$ & 1.000 & 0.00821 \\
\hline\hline
\end{tabular}\end{center}
\label{tab:ratio}
\end{table}

The \mi{radiative corrections} are well known \cite{garc85}. The rate can
be written as
\begin{equation}
R' = R(1 + \delta_a)(1 + \delta_b) = R(1 + \delta)\;,
\end{equation}
where $R$ is the rate defined in Eq.~(\ref{eq:rate}).
The term $\delta_a$ comes from the
model independent virtual corrections and the \mi{bremsstrahlung}
\cite{garc82}. For
the model dependent term $\delta_b$ one takes the value $0.021$
\cite{gail84}. The $\delta$ term is the whole radiative correction
given in Table \ref{tab:rcorr} together with $\delta_a$.

\begin{table}
\caption{Radiative corrections to the semileptonic decay rates.}
\begin{center}\begin{tabular}{ccrr}
\hline\hline
 & Reaction & $\delta_a$ & $\delta$ \\
\hline
$e$-mode & $np$ & 0.0486 & 0.0706 \\
&$ \Sigma^+\Lambda$ & 0.0015 & 0.0225 \\
&$ \Sigma^-\Lambda$ & 0.0012 & 0.0222 \\
&$ \Sigma^0\Sigma^+$ & 0.0226 & 0.0441 \\
&$ \Xi^-\Xi^0$ & 0.0104 & 0.0316 \\
&$ \Lambda p$ & 0.0207 & 0.0421 \\
&$ \Sigma^0 p$ & 0.0196 & 0.0410 \\
&$ \Sigma^- n$ & -0.0025 & 0.0184 \\
&$ \Xi^-\Lambda$ & -0.0015 & 0.0195 \\
&$ \Xi^-\Sigma^0$ & 0.0000 & 0.0210 \\
$\mu$-mode &$ \Lambda p$ & 0.0468 & 0.0688 \\
&$ \Sigma^0 p$ & 0.0336 & 0.0553 \\
&$ \Sigma^- n$ & -0.0022 & 0.0188 \\
&$ \Xi^-\Lambda $& -0.0013 & 0.0197 \\
&$ \Xi^-\Sigma^0$ & 0.0002 & 0.0212 \\
\hline\hline
\end{tabular}\end{center}
\label{tab:rcorr}
\end{table}

For the angular correlation $\alpha_{e\nu}$ and the asymmetries
$\alpha_e$ and $\alpha_\nu$ the model dependent corrections vanish
and the model independent corrections are of the order of $0.001$
\cite{garc85}. We therefore do not include these corrections.

\newpage
\section{The form factors in the quark model}
The basic formula for the matrix elements in Eq.~(\ref{eq:formfactors}) is
Eq.~(\ref{eq:2.25}). The Dirac quark current
\begin{equation}
\bar{q} \Gamma^\mu q \;, \quad \Gamma^\mu = \gamma^\mu (1 - \gamma_5)
\end{equation}
can be generalized by the Ansatz
\begin{equation}
\Gamma^\mu = f_{1q}\gamma^\mu - {f_{2q} \over m_q} i\sigma^{\mu\nu}
K_\nu + {f_{3q} \over m_q} K^\mu
+ g_{1q}\gamma^\mu\gamma_5 - {g_{2q} \over m_q} i\sigma^{\mu\nu}
K_\nu\gamma_5 + {g_{3q} \over m_q} K^\mu\gamma_5 \;.
\end{equation}
The subscript '$q$' stands for a transition on the quark level.
The form factor $f_{2q}$ is
determined by the anomalous quark moments through CVC.
\index{quark structure} But $f_{2q}$ as
well as $f_{3q}$, $g_{2q}$ and $g_{3q}$ do not contribute to
$f_1(0)$ and $g_1(0)$ because of their factor $K$.  Since $K^2$ is
small and contributions different from
$f_1(0)$  and $g_1(0)$ are of higher order in $\beta$ we put
$f_{2q} = f_{3q} = g_{2q} = g_{3q} =0$ without loss of generality.
For the form factors $f_{1q}$ and $g_{1q}$ we shall distinguish between
the transitions $d\to u$ and $s\to u$. In the last Chapter we have put the
parameters $f_{1ud}$ and $g_{1ud}$ equal to one. For the $s\to u$
transition we note that we have to put $f_{1us}=1$, if we wish to
predict $V_{us}$, since the rate is proportional to $f_{1us}^2 |V_{us}|^2$.
The determination of $g_{1us}$ is discussed in the next Section. Notice
that we would have to put $f_{1us}\sim g_{1us}\sim 5$, if we liked to
fit the weak sector with parameter set~4 (see Table~\ref{tab:diquark}).

As seen from Eq.~(\ref{eq:formfactors}) we have to calculate ${\left<\chi_
\uparrow \left| V^+\right| \chi_
\uparrow \right>}$,  ${\left<\chi_\uparrow \left| V^+\right| \chi_\downarrow
\right>}$,
${\left<\chi_\uparrow \left| A^+\right| \chi_\uparrow \right>}$ and  ${
\left<\chi_\uparrow \left| A^+\right|
\chi_\downarrow \right>}$. Table \ref{tab:matrix} gives the matrix elements
for the various reactions with $O_3 = V^+,A^+$.

\begin{table}
\caption{Matrix elements for weak beta decay.}
\label{tab:matrix}
\begin{center}\begin{tabular}{cc}
\hline\hline
Reaction& \\
\hline
$np$ & $-\left( \left<\chi^{\lambda2} \left| O_3\right| \chi^{\lambda1}
\right> +
\left<\chi^{\lambda1}
\left| O_3\right| \chi^{\lambda2} \right> \right)$ \\
$\Sigma^+\Lambda $&$ -{1 \over \sqrt{2}}\left( \left<\chi^{\rho1} \left|
O_3\right| \chi^ {\lambda1} \right> + \left<\chi^{\rho2} \left| O_3\right|
\chi^{\lambda2}
\right> \right)$ \\
$\Sigma^-\Lambda $&$ -{1 \over \sqrt{2}}\left( \left<\chi^{\rho1} \left|
O_3\right| \chi^ {\lambda1} \right> + \left<\chi^{\rho2} \left| O_3\right|
\chi^{\lambda2}
\right> \right)$ \\
$\Sigma^-\Sigma^0$ & ${1\over\sqrt{2}}\left(\left<\chi^{\lambda1} \left|
O_3\right|
\chi^{\lambda1} \right> + \left<\chi^{\lambda2} \left| O_3 \right| \chi^{
\lambda2} \right> \right)$\\
$\Sigma^0\Sigma^+$ & $-{1 \over \sqrt{2}}\left( \left<\chi^{\lambda1}
\left| O_3
\right|
\chi^{\lambda1} \right> + \left<\chi^{\lambda2} \left| O_3 \right| \chi^{
\lambda2} \right> \right)$ \\
$\Xi^-\Xi^0$ & $-\left<\chi^{\lambda3} \left| O_3 \right| \chi^{\lambda3}
\right> $\\
$\Lambda p$ & ${1 \over \sqrt{2}}\left<\chi^{\lambda2} - \chi^{\lambda1}
\left| O_3 \right| \chi^{\rho3} \right>$\\
$\Sigma^0 p$ & ${1 \over \sqrt{2}}\left<\chi^{\lambda1} +
\chi^{\lambda2} \left| O_3 \right| \chi^{\lambda3} \right>$\\
$\Sigma^- n$ & $-\left<\chi^{\lambda3} \left| O_3 \right| \chi^{\lambda3}
\right>$\\
$\Xi^-\Lambda$ & ${1 \over \sqrt{2}}\left( \left<\chi^{\rho2} \left| O_3
\right|
\chi^{\lambda1} \right> + \left<\chi^{\rho1} \left| O_3 \right|
\chi^{\lambda2} \right> \right)$\\
$\Xi^-\Sigma^0$ & $-{1\over\sqrt{2}}\left( \left<\chi^{\lambda1} \left| O_3
\right|
\chi^{\lambda2} \right> + \left<\chi^{\lambda2} \left| O_3 \right| \chi^{
\lambda1} \right> \right)$\\
$\Xi^0\Sigma^+$ & $-\left( \left<\chi^{\lambda1} \left| O_3 \right|
\chi^{\lambda2} \right> + \left<\chi^{\lambda2} \left| O_3 \right|
\chi^{\lambda1} \right> \right)$\\
\hline\hline
\end{tabular}\end{center}
\end{table}

For $K^2 =0$ we have for $\Delta S = 0$ transitions
\begin{eqnarray}
f_1 &=&A(f_1) \;,\nonumber\\
f_2 &=&{N_c \over (2\pi)^6}\int d^3q d^3Q |\Phi|^2 A(f_2) \;,\nonumber\\
&&\label{eq:5.20}  \\
g_1 &=&g_{1ud} A(g_1) {N_c \over (2\pi)^6} \int d^3q d^3Q |\Phi|^2
{b^2 - Q_\bot^2 \over b^2 + Q_\bot^2} \;,\nonumber\\
g_2 &\simeq &0\;, \nonumber
\end{eqnarray}
with $A$s given in Table \ref{tab:reaction1}.
The values $A(f_1)$ and $A(g_1)$ are the values in the
nonrelativistic quark model.

\begin{table}
\caption{Parameters in Eq.~(\protect\ref{eq:5.20}).}
\begin{center}\begin{tabular}{cccc}
\hline\hline
Reaction& $A(f_1)$ & $A(f_2)$ & $A(g_1)$ \\
\hline
$np$ & 1 &$(2A_2-5A_1)/3$ & ${5 \over 3} $\\
$\Sigma^+\Lambda$ & 0 &$(A_2+A_3-2A_1)/\sqrt 6$&$ \sqrt{{2 \over 3}}$\\
$\Sigma^-\Lambda$ & 0 &$(A_2+A_3-2A_1)/\sqrt 6$&$ \sqrt{{2 \over 3}}$\\
$\Sigma^-\Sigma^0$ & $\sqrt{2} $&$-(4A_1+A_2+A_3)/(3\sqrt 2)$
&$ {2\sqrt{2} \over 3}$\\
$\Sigma^0\Sigma^+$ &$ -\sqrt{2}$ &$(4A_1+A_2+A_3)/(3\sqrt 2)$
&$-{2\sqrt{2} \over 3}$\\
$\Xi^-\Xi^0$ & --1 &$(2A_2+2A_3-A_1)/3$  &$ {1 \over 3}$\\
\hline\hline
\end{tabular}\end{center}
\label{tab:reaction1}
\end{table}

The $\Delta S = 1$ transitions for $K^2 = 0$ are
\begin{eqnarray}
f_1 &=&{N_c \over (2\pi)^6}\int d^3\!q d^3Q \left({E_3'E_{12}'M \over
E_3 E_{12} M'} \right)^{1/2}\!{\Phi^\dagger(M')\Phi(M)
B(f_1) \over (a'^2+Q_\bot^2)(a^2+Q_\bot^2)\sqrt{b'^2+Q_\bot^2}
\sqrt{b^2+Q_\bot^2}}\;,\nonumber\\
&&\label{eq:weak1}\\
g_1 &=&{N_c \over (2\pi)^6}\int d^3\!q d^3Q \left({E_3'E_{12}'M \over
E_3 E_{12} M'} \right)^{1/2}\!{\Phi^\dagger(M')\Phi(M)
B(g_1) \over (a'^2+Q_\bot^2)(a^2+Q_\bot^2)\sqrt{b'^2+Q_\bot^2}
\sqrt{b^2+Q_\bot^2}} \;,\nonumber
\end{eqnarray}

\begin{eqnarray}
B(f_1) &=&B_1(a'a+Q_\bot^2)^2(b'b+Q_\bot^2)\nonumber\\
 &&+B_2(a'-a)^2Q_\bot^2(b'b+Q_\bot^2){(cd-q_\bot^2)^2 \over (c^2+q_\bot^2)
(d^2+q_\bot^2)}\nonumber\\
 &&+B_3(a'-a)(b'-b)Q_\bot^2(a'a+Q_\bot^2)\left({c^2 \over c^2+q_\bot^2}
+ {d^2 \over d^2+q_\bot^2} \right)\;,\nonumber\\
&&\label{eq:5.22}\\
B(g_1) &=&B_4(b'b-Q_\bot^2)\left[(a'a+Q_\bot^2)^2 + (a'-a)^2Q_\bot^2
{(cd-q_\bot^2)^2 \over (c^2+q_\bot^2)(d^2+q_\bot^2)} \right]\nonumber\\
 &&+B_5(a'-a)^2Q_\bot^2(b'b-Q_\bot^2){cdq_\bot^2 \over (c^2+q_\bot^2)
(d^2+q_\bot^2)}\nonumber\\
 &&+B_6(a'-a)Q_\bot^2(b'+b)(a'a+Q_\bot^2)\left({c^2 \over c^2+q_\bot^2}
+ {d^2 \over d^2+q_\bot^2} \right)\;.\nonumber
\end{eqnarray}
The $B_i$ for the different decays are given in Table \ref{tab:reaction2}.

\begin{table}
\caption{Parameters in Eq.~(\protect\ref{eq:5.22}).}
\label{tab:reaction2}
\begin{center}\begin{tabular}{ccccccc}
\hline\hline
Reaction& $B_1$ & $B_2$ & $B_3$ & $B_4$ & $B_5$ & $B_6$ \\
\hline
$\Lambda p$ &$ -\sqrt{{3 \over 2}}$ & $-\sqrt{{3 \over 2}}$ & 0 &$
-\sqrt{{3 \over 2}}$& 0 & 0 \\
$\Sigma^0 p$ &$ -{1 \over \sqrt{2}}$ &$ {1 \over 3\sqrt{2}}$ &
${\sqrt{2} \over 3}$ & ${1 \over 3\sqrt{2}}$ &$ {4\sqrt{2} \over 3}$ &$
{\sqrt{2} \over 3}$ \\
$\Sigma^- n $& --1 &$ {1 \over 3}$ & ${2 \over 3}$ & ${1 \over 3}$ &
${8 \over 3}$ &$ {2 \over 3}$ \\
$\Xi^-\Lambda$ & $\sqrt{{3 \over 2}} $&$ {1 \over \sqrt{6}}$ &$ -{1 \over
\sqrt{6}} $&$ {1 \over \sqrt{6}}$ & $-2\sqrt{{2 \over 3}} $&$ -{1 \over 6}$ \\
$\Xi^-\Sigma^0$ &$ {1 \over \sqrt{2}}$ &$ {5 \over 3\sqrt{2}}$ & ${1 \over
3\sqrt{2}}$ & ${5 \over 3\sqrt{2}} $&$ {4 \over 3\sqrt{2}}$ &$
{\sqrt{2} \over 6}$ \\
$\Xi^0\Sigma^+$ & 1 & ${5 \over 3} $&$ {1 \over 3} $&$ {5 \over 3}$ & $
{4 \over 3}$ &$ {1 \over 3} $\\
\hline\hline
\end{tabular}\end{center}
\end{table}

Eqs.~(\ref{eq:weak1}) and (\ref{eq:5.22}) confirm the Ademollo-Gatto theorem
\cite{adem64}.
Since $(a'-a) \sim \Delta m$ and $(b'-b) \sim \Delta m$ the symmetry
breaking for $f_1$ is of the order $(\Delta m)^2$ whereas it is of the order
$\Delta m$ for $g_1$ owing to the term containing $B_6$. In addition to
Ademollo-Gatto
\index{Ademollo-Gatto theorem!extension of}
we see that the \mi{symmetry breaking} for $g_1(\Lambda \to p)$ is
also of second order.

The full formulae for $K^2 \le 0$ are longer than the ones for
$K^2 = 0$; they are given in Appendix~\ref{ch:formulae}.

\newpage
\section{Results}

The form factors can be determined by the generalization of
Eqs.~(\ref{eq:5.20}) and (\ref{eq:weak1}). With the parameterization
of the form factor $f(K^2)$:
\begin{equation}
f(K^2) \simeq {f(0) \over 1-K^2 / \Lambda_1^2+K^4 / \Lambda_2^4} \;,
\label{eq:pol}
\end{equation}
we get the result shown in Tables \ref{tab:nstrange} and \ref{tab:strange}
together
with the rates, angular correlation and asymmetries from
Eqs.~(\ref{eq:rate})--(\ref{eq:asy}). The parameters $\Lambda_n$ are determined
by the
calculation of the appropriate derivatives of $f(K^2)$ at
$K^2=0$. The rates have been corrected taking into account
the non-vanishing lepton mass and radiative corrections.

\begin{table}
\caption[Results for $\Delta S = 0$ weak beta decay.]
{Results for $\Delta S = 0$ weak beta decay with parameter set~6.
Experimental data are from PDG \protect\cite{part90}.}
\label{tab:nstrange}
\begin{center}\small\begin{tabular}{cccccccc}
\hline\hline
 & & $np$ & $\Sigma^+\Lambda $& $\Sigma^-\Lambda $& $\Sigma^-\Sigma^0$ &
$\Sigma^0\Sigma^+$ & $\Xi^-\Xi^0$ \\
\hline
$f_1$ &$ f_1(0)$ & 1.00 &0 &0& 1.41&--1.41&--1.00\\
&$ \Lambda_1$ [GeV] &0.69&--0.32$^{\rm a}$&--0.32$^{\rm a}$&0.60&0.60&0.56\\
&$ \Lambda_2$ [GeV] &0.96&--1.72$^{\rm a}$&--1.72$^{\rm a}$&0.81&0.81&0.71\\
&&&&&&&\\
$g_1$ &$ g_1(0)$ & 1.25&0.60&0.60&0.69&--0.69&0.24\\
& $\Lambda_1$ [GeV] &0.76&0.77&0.77&0.77&0.77&0.76\\
& $\Lambda_2$ [GeV] &1.04&1.05&1.05&1.04&1.04&1.04\\
&&&&&&&\\
$g_1/f_1 $& Theor. &1.252&0.736$^{\rm b}$&0.736$^{\rm b}$&0.491&0.491&--0.244\\
& Expt. &1.261&0.742$^{\rm b}$&--&--&--&$<2\times 10^3$\\
&&$\pm 0.004$&$\pm 0.018$&&&&\\
&&&&&&&\\
${f_2 \over M} $ [GeV$^{-1}$]& Theor. &1.81&1.04&1.04&0.76&--0.76&0.73\\
& CVC &1.85&1.17&1.17&0.60&--0.60&1.00\\
&&&&&&&\\
${g_2 \over M}$ [GeV$^{-1}$] & &0&0&0&0&0&0\\
&&&&&&&\\
Rate [$10^6 s^{-1}$]  & Theor. &$1.152\times 10^{-9}$&0.24&0.389&1.47
$^{\rm c}$&3.65$^{\rm d}$&1.55$^{\rm c}$\\
$e$-mode & Expt. &1.125$\times 10^{-9}$&0.25&0.387&--&--&--\\
&&$\pm 0.004$&$\pm 0.06$&$\pm 0.018$&&&\\
&&&&&&&\\
$\alpha_{e\nu}$ & Theor. &--0.101&--0.404&--0.412&0.436&0.438&0.793\\
& Expt. &--0.102&--0.35&--0.404&&&\\
&&$\pm 0.005$&$\pm 0.15$&$\pm 0.044$&&&\\
$\alpha_{e}$ & Theor. &--0.112&--0.701&--0.704&0.287&0.288&--0.514\\
& Expt. &--0.083&&&&&\\
&&$\pm 0.002$&&&&&\\
$\alpha_{\nu}$ & Theor. &0.989&0.647&0.645&0.850&0.850&--0.314\\
& Expt. &0.998&&&&&\\
&&$\pm 0.025$&&&&&\\
$\alpha_{B}$ & Theor. &--0.548&0.070&0.077&--0.710&--0.711&0.518\\
& Expt. &&&&&&\\
\hline\hline
\end{tabular}\end{center}
{$^{\rm a}$ Instead of $\Lambda_i$ we list $f_1^{(i)}$.
$^{\rm b}$ Instead of $g_1/f_1$ we list $\sqrt{3/2} g_1$.
$^{\rm c}$ $\times 10^{-6}$.
$^{\rm d}$ $\times 10^{-8}$.}
\end{table}
\normalsize

\begin{table}
\caption[Results for $\Delta S = 1$ weak beta decay.]
{Results for $\Delta S = 1$ weak beta decay with parameter set~6.
Experimental data are from PDG \protect\cite{part90}.}
\label{tab:strange}
\begin{center}\small\begin{tabular}{cccccccc}
\hline\hline
 & & $\Lambda p $& $\Sigma^0 p $& $\Sigma^- n $& $\Xi^-\Lambda $&
$\Xi^-\Sigma^0 $& $\Xi^0\Sigma^+ $ \\
\hline
$f_1$ &$ f_1(0)$ & --1.19&--0.69&--0.97&1.19&0.69&0.98\\
&$ \Lambda_1$ [GeV] &0.71&0.64&0.64&0.68&0.75&0.75\\
&$ \Lambda_2 $[GeV] &0.98&0.84&0.90&0.89&1.05&1.05\\
&&&&&&&\\
$g_1$ &$ g_1(0)$ & --0.99&0.19&0.27&0.33&0.94&1.33\\
& $\Lambda_1$ [GeV] &0.81&0.83&0.83&0.81&0.81&0.81\\
& $\Lambda_2$ [GeV] &1.12&1.16&1.16&1.10&1.12&1.12\\
&&&&&&&\\
$g_1/f_1 $& Theor. &0.826&--0.275&--0.275&0.272&1.362&1.362\\
& Expt. &0.718&--&--0.340&0.250&1.287&$< 2.93$\\
&&$\pm 0.015$&&$\pm 0.017$&$\pm 0.042$&$\pm 0.158$&\\
&&&&&&&\\
${f_2 \over M} $[GeV$^{-1}$]& Theor. &--0.85&0.44&0.62&0.070&0.98&1.38\\
& CVC &--1.19&--&1.12&--0.080&1.38&1.95\\
&&&&&&&\\
${g_2 \over M}$ [GeV$^{-1}$] & &--0.025&0.0043&0.0061&--$^{\rm a}$&--$^{\rm a}$
&--$^{\rm a}$\\
&&&&&&&\\
Rate [$10^6 s^{-1}$] & Theor. &3.51&2.72&5.74&2.96&0.549&0.942\\
$e$-mode & Expt. &3.170&--&6.88&3.36&0.53&--\\
&&$\pm 0.058$&&$\pm 0.26$&$\pm 0.19$&$\pm 0.10$&\\
&&&&&&&\\
Rate [$10^6 s^{-1}$] & Theor. &0.58&1.18&2.54&0.80&$7.47\times 10^{-3}$&
$7.74\times 10^{-3}$\\
$\mu$-mode & Expt. &0.60&--&3.04&2.1&--&--\\
&&$\pm 0.13$&&$\pm 0.27$&$\pm 2.1$&&\\
&&&&&&&\\
$\alpha_{e\nu}$ & Theor. &--0.100&0.443&0.437&0.531&--0.252&--0.248\\
& Expt. &--0.017$^{\rm b}$&&0.279&0.53&&\\
&&$\pm 0.023$&&$\pm 0.026$&$\pm 0.1$&&\\
$\alpha_{e}$ & Theor. &--0.021&--0.536&--0.537&0.236&--0.226&--0.223\\
& Expt. &0.125&&--0.519$^{\rm c}$&&&\\
&&$\pm 0.066$&&$\pm 0.104$&&&\\
$\alpha_{\nu}$ & Theor. &0.992&--0.318&--0.318&0.592&0.973&0.973\\
& Expt. &0.821&&--0.230$^{\rm c}$&&&\\
&&$\pm 0.066$&&$\pm 0.061$&&&\\
$\alpha_{B}$ & Theor. &--0.582&0.568&0.569&--0.519&--0.437&--0.439\\
& Expt. &--0.508&&0.509$^{\rm c}$&&&\\
&&$\pm 0.065$&&$\pm 0.102$&&&\\
\hline\hline
\end{tabular}\end{center}
{$^{\rm a} {g_2 \over g_1M} \simeq 0.023$ since ${g_2 \over
g_1}\simeq$ constant. $^{\rm b}$ From Ref. \cite{dwor90}. $^{\rm c}$
{}From Ref. \cite{hsue88}.}
\end{table}
\normalsize

\subsection{The rates and $f_1(0), g_1(0)$}
We fit our remaining
parameters $m_s$ and $\beta_\Sigma =\beta_\Lambda $ to the rate $R$
and the ratio $g_1/f_1$ for the processes $\Lambda\to pe^-\bar\nu_e$ and
$\Sigma^-\to ne^-\bar\nu_e$, and $\beta_\Xi$ is fitted to the rate for
$\Xi^-\to \Lambda e^-\bar\nu_e$. We find parameter set~5 of
Table \ref{tab:para} and get
\begin{eqnarray}
R(\Lambda\to pe^-\bar\nu_e) &=& 3.38\times 10^6 {\rm s}^{-1}\;,\nonumber\\
R(\Sigma^-\to ne^-\bar\nu_e) &=& 5.79\times 10^6 {\rm s}^{-1}\;,\nonumber\\
g_1/f_1(\Lambda\to pe^-\bar\nu_e) &=& 0.782\;,\nonumber\\
g_1/f_1(\Sigma^-\to ne^-\bar\nu_e) &=& -0.261\;.\nonumber
\end{eqnarray}
But the value for $m_s$ seems to be too small if we compare it with the
well confined value in the meson sector of the same model \cite{jaus90}.
By considering the constraints $\Delta m = m_s - m_{u/d} \simeq
140$ MeV and $m_s/m_{u/d}\simeq 1.4$ \cite{scad81} we choose
$m_s = 0.40$ GeV (set~6). The results with set~6 have been collected
in Tables \ref{tab:nstrange} and \ref{tab:strange}.

The largest discrepancy between theory and experiments comes from the rates
and $g_1/f_1$ for the processes $\Lambda\to pe^-\bar\nu_e$ and
$\Sigma^-\to ne^-\bar\nu_e$. By changing the axial couplings of the
quarks, i.e. $g_{1us} \simeq 0.9$, we could
improve the rates of both reactions, but the ratios $g_1/f_1$ clearly
force us to use $g_{1us} =1$. Another modification could be the
$\Lambda-\Sigma^0$-mixing. Let us write
\begin{equation}
\Lambda_{\rm phys} =A\;\Lambda + B\;\Sigma^0\;, \quad
\Sigma^0_{\rm phys} = - B\;\Lambda + A\;\Sigma^0\;,\quad A^2+B^2=1\;.
\end{equation}
{}From the measurement $f_1/g_1(\Sigma^-\to \Lambda e^-\bar\nu_e)$ we
get a constraint $A \ge 0.9961, \ B\le 0.0078$ which gives
\begin{eqnarray}
f_1/g_1(\Lambda\to p e^-\bar\nu_e) \ge 0.773 \quad (\ge 0.73 \hbox{ for }
\Delta m = 63 \hbox{ MeV})\;,
\end{eqnarray}
and reduces the rate by about 1 \%. Therefore, $\Lambda-\Sigma^0$-mixing
only improves one of the four values for which theory and experiment differ.
Actually, this inconsistency
of our values is a general feature of quark models with a SU(6)
flavor-spin symmetry.\footnote{The bag model calculation
\cite{dono82} gives similar results:
$g_1/f_1(\Lambda\to pe^-\bar\nu_e) = 0.84$, and
$g_1/f_1(\Sigma^-\to ne^-\bar\nu_e) = -0.28$.} The ratio
$g_1/f_1$ can generally be written as
\begin{equation}
{g_1 \over f_1} = \rho\eta\left({g_1 \over f_1}\right)_{\rm non-rel}\;,
\end{equation}
where $(g_1/f_1)_{\rm non-rel}$ is the non-relativistic value.
The quantity $\rho$ is a relativistic suppression factor due to the
`` small '' components in the quark spinors (in the bag-model)
\index{bag model} or
due to the Melosh-transformation \index{Melosh rotation}
(in our model). The quantity $\eta$ is an
enhancing factor due to SU(3) \mi{symmetry breaking} in $\Delta S = 1$
transitions. From Tables \ref{tab:nstrange} and \ref{tab:strange} we see that
$\rho \simeq 0.73 - 0.76$ depending on the strangeness content of
the wave functions and $\eta \simeq 1.11$. This simple estimate
shows that every quark model is {\it a priori} constrained to
\begin{equation}
{g_1/f_1(\Lambda\to pe^-\bar\nu_e) \over g_1/f_1(\Sigma^-\to ne^-\bar\nu_e)}
= -3
\label{eq:ratio}
\end{equation}
in contrast to the experimental value $-2.11\pm 0.15$ for $g_2 = 0$. For
$g_2 \ne 0$ it is measured that \cite{jens83}
\begin{equation}
\left| {g_1 \over f_1}\right|_{\Lambda p} = 0.715 + 0.28 {g_2 \over f_1}\;,
\label{eq:5.31}
\end{equation}
and \cite{hsue88}
\begin{equation}
\left| {g_1 \over f_1} - 0.237{g_2 \over f_1} \right|_{\Sigma^-n} =
0.34 \pm 0.017\;,
\end{equation}
which will bring the data closer to $-3$, but in our model $g_2/g_1 \simeq
0.025$ which is much too small to remove the discrepancy.

\subsection{$f_2(0)$ and $g_2(0)$}
Our model agrees with the \mi{conserved vector current} (CVC)
 hypothesis.
\index{CVC|see{conserved vector current}}
The deviations have the same origin as the too small neutron magnetic
moment since $f_2$ and the \mi{magnetic moments} have similar
analytic forms. If we
take $\mu_p$ as in Sec.~\ref{sec:mm}, CVC will reproduce our values. The
experimental situation is not yet clear, some experiments confirm \cite{hsue88}
and some disprove \cite{dwor90} CVC.

For $\Delta S = 1$ transitions the prediction of $g_2/g_1$ for
nonrelativistic quark models is $\sim$~0.37 and for the bag model
\index{bag model} $\sim$~0.15
\cite{dono82}. Our model gives also a constant value
\footnote{$(g_2)_{\Xi\Lambda}$ and $(g_2)_{\Xi\Sigma}$ could only
be calculated for $\Delta m=63$ MeV and we get $(g_2/g_1)_{\Delta S=1}
\simeq 0.11$.\label{note}}
\begin{equation}
\left({g_2 \over g_1}\right)_{\Delta S =1}\simeq 0.025\;.
\end{equation}
For $\Delta S =0$ transitions we get
\begin{equation}
\left({g_2 \over g_1}\right)_{\Delta S =0}\simeq 0.0033\;,
\end{equation}
if we put $m_d-m_u = 7$ MeV.
This confirms the viewpoint of the PDG \cite{part90} which fixes $g_2=0$.
Experiments also find a vanishing or small $g_2$ \cite{garc85}.

With CVC and the absence of $g_2$ we reach the same conclusion that was
reached in nuclear physics.

\subsection{$K^2$-dependence of the form factors}
\begin{figure}
\centerline{\psfig{figure=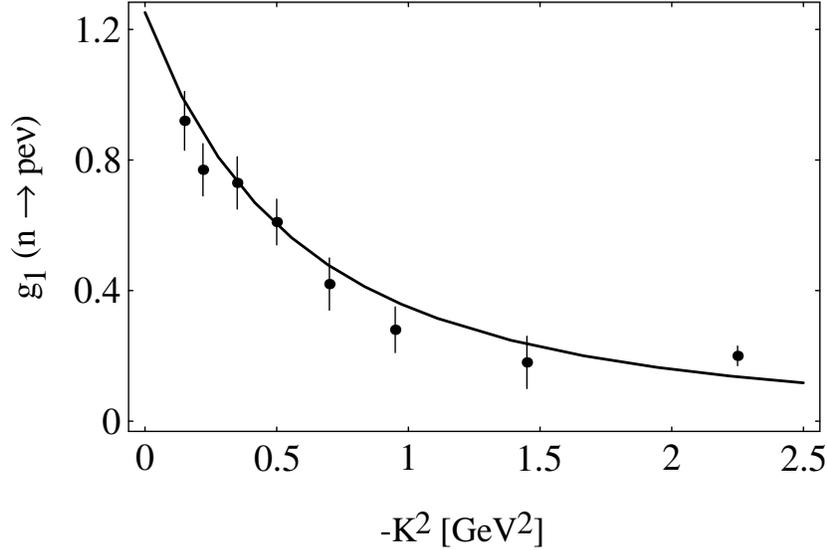,height=216pt}}
\caption[The axial vector form factor $g_1(K^2)$ for the
$np$-transition.]
{The axial vector form factor $g_1(K^2)$ for the $np$-transition.
The dipole formula is compared with the experimental data
taken from Ref.~\protect\cite{beli85}.}
\label{fig:ga}
\end{figure}

Tables \ref{tab:nstrange} and \ref{tab:strange} suggest that the form
factor of Eq.~(\ref{eq:pol}) can be approximated by the dipole form
\begin{equation}
f(K^2) \simeq {f(0) \over \left( 1-K^2 / \Lambda_2^2\right)^2} \;.
\end{equation}
The axial vector form factor $g_1$ for the neutron decay
\index{neutron beta decay} gives a
value $M_A = \Lambda_2 = 1.04$ GeV compared to the experimental
value $M_A = (1.00 \pm 0.04)$ GeV \cite{beli85,brun90}.
Figure \ref{fig:ga} compares the dipole-fit with
experimental points \cite{beli85}.

If we take the dipole Ansatz we can compare our values for $M_V$ and $M_A$
with the results of other work (see Table \ref{tab:dipole}).

\begin{table}
\caption{The parameters $M_V$ and $M_A$ for various models.}
\label{tab:dipole}
\begin{center}\begin{tabular}{ccccccccc}
\hline\hline
&\multicolumn{2}{c}{This work}&\multicolumn{2}{c}{Gaillard et al.\cite
{gail84}}&\multicolumn{2}{c}{Garcia et al. \cite{garc85}}
&\multicolumn{2}{c}{Gensini \cite{gens90}}\\
&$M_V$&$M_A$&$M_V$&$M_A$&$M_V$&$M_A$&$M_V$&$M_A$\\
\hline
$np$&0.96&1.04&0.84&1.08&0.84&0.96&0.84&1.08\\
$\Sigma\Lambda$&-&1.05&-&1.08&-&0.96&-&1.08\\
$\Sigma\Sigma$&0.81&1.04&0.84&1.08&0.84&0.96&0.84&1.08\\
$\Xi\Xi$&0.71&1.04&0.84&1.08&0.84&0.96&0.84&1.08\\
$\Lambda p$&0.98&1.12&0.98&1.25&0.97&1.11&0.94&1.16\\
$\Sigma p$&0.84&1.16&0.98&1.25&0.97&1.11&0.94&1.16\\
$\Sigma n$&0.90&1.16&0.98&1.25&0.97&1.11&0.94&1.16\\
$\Xi\Lambda$&0.89&1.10&0.98&1.25&0.97&1.11&0.94&1.16\\
$\Xi\Sigma$&1.05&1.12&0.98&1.25&0.97&1.11&0.94&1.16\\
\hline\hline
\end{tabular}\end{center}
\end{table}

The contribution of $M_V$ and $M_A$ to the rate and to $x = g_1/f_1$
to first order is
\begin{eqnarray}
{\Delta R \over R} &=& {8 \over 7}{\beta^2M^2 \over (1+3x^2)}
\left({1 \over M_V^2}+{5x^2 \over M_A^2}\right)\nonumber\\
&&\\
{\Delta x^2 \over x^2} &=&-{8 \over 7}\beta^2M^2\left[{(1 -
\alpha_{e\nu})\alpha_{e\nu}
 \over M_V^2} + {6+5\alpha_{e\nu} \over M_A^2}\right] \nonumber
\end{eqnarray}
which shows that our parameters give for the decay
$\Sigma^-\to n e^-\bar\nu_e$ a 0.3\% larger rate and a
5\% smaller $g_1/f_1$ than with the parameters of \mi{Gaillard} et al. that
are often used for the experimental analysis. Although this does not
explain the inconsistency of the data with our calculation, it shows that
future high-statistics experiments should pay more attention to
$M_V$ and $M_A$ in analyzing $g_1/f_1$.

\newpage
\section{Cabibbo fit and $V_{us}$}\label{sec:cabibbo}
\index{Cabibbo model|(}
\index{symmetry breaking}
\index{flavor SU(3) breaking|see{symmetry breaking}}
There are at present some questions concerning flavor SU(3) breaking in
semileptonic weak hyperon decay \cite{roos90,ratc90,gens90}. This symmetry
breaking is included to all orders in our approach. We can extract this
effect from our model and fit the experimental values within the Cabibbo
model. This has been done for the bag model \index{bag model} in 1987
\cite{dono82,dono87}, but since then some of the experimental data have
changed. The main difficulty is the rate for $\Xi^-\to\Lambda
e^-\bar\nu_e$, which comes out too small and is made even worse by their
symmetry breaking scheme.

The reason for using the Cabibbo model is the bad fit for the relevant
data in Table \ref{tab:strange}. The experimental deviations are as large
as 17\% and do not permit a precise determination of the
Kobayashi-Maskawa matrix element $V_{us}$.

The Cabibbo model analyses the experimental results in terms of
three parameters $V_{us}$, $F$, and $D$. The form factor $f_1(0)$ is
given by SU(3) symmetry, $f_2(0)$ by its CVC-values and
$g_1(0)$ as follows:

\begin{center}\begin{tabular}{cccccc}
\hline\hline
$np$&$\Lambda p$&$\Sigma n$&$\Sigma^\pm \Lambda$&$\Xi \Lambda$&
$\Xi^-\Sigma$\\
\hline
$F+D$&$-\sqrt{3/2}(F+D/3)$&$-F+D$&$\sqrt{2/3}D$&$\sqrt{3/2}(F-D/3)$
&$1/\sqrt{2}(F+D)$\\
\hline\hline
\end{tabular}\end{center}

The purpose of a Cabibbo fit is to determine the values of $V_{us}$,
$F$, and $D$ corresponding to the best agreement between experiment
and theory.

We choose the value $V_{ud}=0.9735$ of Ref.~\cite{rasc90}, and perform
a least-squares $\chi^2$-fit. Radiative corrections are taken into account.

Table \ref{tab:cabibbo} gives the residuals \index{residual}
$R_i=\left[x_i({\rm fit})-x_i({\rm meas})\right]/\delta x_i({\rm meas})$
for each reaction, $\chi^2=\sum_i R_i^2$ for 11 degrees of freedom,
and the fit variables.

\begin{table}
\caption[Fits to the Cabibbo model.]{Fits to the Cabibbo model.
The decay rates are given in units of $10^6 s^{-1}$, except for the
neutron decay with $10^3 s^{-1}$. The
residuals are given by $\left[x_i({\rm fit})-x_i({\rm meas})\right]/\delta
x_i({\rm meas})$.}
\begin{center}\begin{tabular}{lcrrrrrrr}
\hline\hline
&Expt.&\multicolumn{7}{c}{Residuals of fit no.}\\
&&1&2&3&4&5&6&7\\
\hline
$R(n \to pe\nu)$&$1.125\pm 0.013$&$-1.1$&$-1.1$&$-1.2$&$-1.5$
&$-1.2$&$-1.2$&$-1.1$\\
$R(\Lambda \to pe\nu)$&$3.170\pm0.058$&0.9&0.8&1.0&1.5&1.2&0.3&0.4\\
$R(\Sigma \to ne\nu)$&$6.88\pm 0.26$&--0.5&--0.3&--0.3&--1.1&--0.5&--0.5
&--0.7\\
$R(\Sigma^- \to \Lambda e\nu)$&$0.387\pm 0.018$&2.7&2.7&1.1&2.2&0.4&1.2&1.1\\
$R(\Sigma^+ \to \Lambda e\nu)$&$0.25\pm 0.06$&0.2&0.2&--0.1&0.1&--0.2&
--0.1&--0.1\\
$R(\Xi^- \to \Lambda e\nu)$&$3.36\pm 0.19$&--2.2&--2.1&--2.8&--3.5&--3.3
&1.1$^{\rm a}$&1.1$^{\rm a}$\\
$R(\Xi^- \to \Sigma e\nu)$&$0.53\pm
0.10$&0.0&--0.1&--0.2&0.2&--0.3&--0.3&--0.3\\
$R(\Lambda \to p\mu\nu)$&$0.60\pm 0.13$&--0.3&--0.3&--0.3&--0.3&--0.3
&--0.4&--0.4\\
$R(\Sigma \to n\mu\nu)$&$3.04\pm 0.27$&--0.2&--0.1&--0.1&--0.4&--0.2&--0.2
&--0.3\\
$R(\Xi^- \to \Lambda \mu\nu)$&$2.1\pm 2.1$&--0.6&--0.6&--0.6&--0.6&--0.6&--0.6
&--0.6\\
$g_1(n \to pe\nu)$&$1.261\pm 0.004$&--0.1&--0.1&--0.2&--0.9&--0.3&--0.1&0.1\\
${g_1 \over f_1}(\Lambda \to pe\nu)$&$0.718\pm 0.015$&0.9&1.0&2.8&5.2&3.5&
2.7&1.1$^{\rm b}$\\
${g_1 \over f_1}(\Sigma \to ne\nu)$&$-0.340\pm 0.017$&0.9&1.0&--0.6&--1.8
&--1.4&--0.8&--0.5\\
${g_1 \over f_1}(\Xi^- \to \Lambda e\nu)$&$0.25\pm 0.05$&--0.9&--0.9&
--1.1&--0.8&--1.1&--1.1&--1.1\\
\hline
$F$&&0.468&0.469&0.461&0.458&0.457&0.460&0.462\\
$D$&&0.793&0.792&0.799&0.800&0.802&0.801&0.799\\
$V_{us}[\pm 0.003]$&&0.227&0.226&0.226&0.222&0.223&0.225&0.225\\
$\chi^2$&&17.5&16.9&21.4&54.8&30.3&14.3&7.7\\
\hline\hline
\end{tabular}\end{center}
{$^{\rm a}$We take the
experimental value $1.83\pm 0.79$ \cite{thom80}. $^{\rm b}$The effective value
$0.744\pm 0.015$ is taken.}
\label{tab:cabibbo}
\end{table}

\begin{table}
\caption{Symmetry breaking for $f_1$. The ratio $f_1/f_1^{\rm SU(3)}$
is shown.}
\label{tab:breakf}
\begin{center}\begin{tabular}{ccccc}
\hline\hline
&\multicolumn{2}{c}{This work}&\multicolumn{1}{c}{Donoghue}
&\multicolumn{1}{c}{Krause}\\
&$\Delta m=63$MeV&$\Delta m=133$MeV&&\\
\hline
$\Delta S = 0$&1&1&1&1\\
$\Lambda p$&0.978&0.976&0.987&0.943\\
$\Sigma p$&0.979&0.975&0.987&-\\
$\Sigma n$&0.978&0.975&0.987&0.987\\
$\Xi\Lambda$&0.981&0.976&0.987&0.957\\
$\Xi\Sigma$&0.982&0.976&0.987&0.943\\
\hline\hline
\end{tabular}\end{center}
\label{tab:XI}
\end{table}

\begin{table}
\caption{Symmetry breaking for $g_1$. The ratio $g_1/g_1^{\rm SU(3)}$
is shown.}
\label{tab:breakg}
\begin{center}\begin{tabular}{cccc}
\hline\hline
&\multicolumn{2}{c}{This work}&\multicolumn{1}{c}{Donoghue}\\
&$\Delta m=63$MeV&$\Delta m=133$MeV&\\
\hline
$np$&1.000&1.000&1.000\\
$\Sigma\Lambda$&0.959&0.981&0.9383/0.9390\\
$\Sigma\Sigma$&0.955&0.982&-\\
$\Xi\Xi$&0.916&0.977&-\\
$\Lambda p$&1.021&1.072&1.050\\
$\Sigma p$&1.011&1.051&-\\
$\Sigma n$&1.012&1.056&1.040\\
$\Xi\Lambda$&0.987&1.072&1.003\\
$\Xi\Sigma$&0.981&1.061&0.9954\\
\hline\hline
\end{tabular}\end{center}
\label{tab:XII}
\end{table}

\index{fit}
Two fits have been made without symmetry breaking. Fit 1 uses the
commonly used values of $M_{V/A}$ given in Ref. \cite{gail84}
and fit 2 uses our masses, which gives
a slightly improved $\chi^2$. Therefore, we use our values for all the
other fits. The rate for $\Sigma^- \to \Lambda e^-\bar\nu$ produces the
largest deviation, indicating SU(3) symmetry breaking. Using Tables
\ref{tab:XI} and \ref{tab:XII} we fit using $\Delta m=63$ MeV (fit 3),
$\Delta m=133$ MeV (fit 4), and the results of \mi{Donoghue} (fit 5).
The symmetry breaking
scheme for $\Delta m=133$ MeV cannot be correct. We improved the above
deviation at the cost of introducing a large contribution to
$\chi^2$ from the rate $\Xi^- \to
\Lambda e^-\bar\nu$ and the ratio $g_1/f_1$ for $\Lambda \to p e^-\bar\nu$.
Fit 3 performs best, but is still worse than no symmetry breaking. Since
there are doubts about the experimental rate for $\Xi^- \to \Lambda
e^-\bar\nu$ \cite{roos90} we use $(1.83\pm 0.79)\times 10^6 s^{-1}$
\cite{thom80} for fit 6 with $\Delta m=63$ MeV. This case is nearly as good
as no symmetry breaking, which gives $\chi^2=13.8$. We have to remember
Eq.~(\ref{eq:5.31}) and footnote~\ref{note} on page~\pageref{note}, and the
nonvanishing $g_2$ gives an effective $g_1/f_1=0.744\pm 0.015$. With these
data, fit 7 is in excellent agreement with experiment ($\chi^2=7.7$ for 11
DF). We observe that $V_{us}$ lowers its value from fit 1 to 7, so that we get
closer to the one derived from the meson sector. The analysis of $K_{e3}$
decays yields
\begin{equation}
V_{us}=\cases{0.2196\pm 0.0023&(Ref. \cite{leut84}) ,\cr
              0.2199\pm 0.0017&(Refs. \cite{jaus91,bark92}).}
\end{equation}
Other values from hyperon beta decay are
\begin{equation}
V_{us}=\cases{0.2258\pm 0.0027&(Ref. \cite{garc92}) ,\cr
              0.222\pm 0.003&(Ref. \cite{part90}) .}
\end{equation}
The last value is derived from WA2 data \cite{bour83} with the symmetry
breaking scheme from Ref.~\cite{dono87}. Our result is in excellent
agreement with the recent value from Ref.~\cite{garc92}, and it is also
consistent with the other baryonic value \cite{part90}.
But the discrepancy with the
meson sector still remains.

In conclusion,  there is evidence for symmetry breaking from the rate
$\Sigma^- \to \Lambda e^-\bar\nu$. But symmetry breaking alone
makes the fit even worse. We also have to use a new value of $M_{V/A}$
and a small second-class axial coupling as given by our model. In this
Section, we considered symmetry breaking due to the mass difference
$m_s-m_{u/d}$ alone, but an asymmetric wave function as treated in
Chapter~\ref{ch:Asymmetric} breaks SU(3) symmetry as well.
\index{hyperon decay|)}
\index{Cabibbo model|)}

\newpage
\section{Conclusions}
Parameter set~6 yields the best fit for the semileptonic weak decays
in the MQM. It agrees with the data within 17\%, except for some of the
correlations and asymmetries.
The main discrepancy with experiments lies with the decays
$\Lambda\to pe^-\bar\nu_e$, and  $\Sigma^-\to ne^-\bar\nu_e$, which are
the main sources for determining the Kobayashi-Maskawa mixing matrix
element $V_{us}$. This is the reason why we switched over to the Cabibbo
model in calculating $V_{us}$.

The parameter set~6 is quite different from the set~4 found by a
comparison with the baryon magnetic moments. This result shows that the
representation of the model discussed up to now does not give a
consistent picture of the $u$, $d$, and $s$ sector of the baryons.
A consistent picture can be achieved with the help of asymmetric wave
functions as shown in the next Chapter. Even the fit within the weak
sector improves dramatically.

\chapter{Asymmetric wave functions}\label{ch:Asymmetric}

\section{The diquark model} 
\index{wave function!asymmetric}

In this Chapter we investigate the effects of an asymmetric wave function.
Since the early days of the quark model \mi{diquark} clustering has been
studied. In Gell-Mann's \index{Gell-Mann} original paper on quarks
\cite{gell64}, he mentions the term diquark in a footnote. Refs.
\cite{ida66,lich67} first took the idea seriously, and in both papers, a
baryon is described as a \mi{bound state} of a quark and a diquark. A
recent review with many references can be found in Ref. \cite{tori88}. It
has even been seen that many gluon effects can be simulated by diquarks
\cite{fred88}.

The concept of diquarks is also useful in treating deep-inelastic electron
scattering. Feynman \cite{feyn72} observed that the experimental ratio of
the neutron to proton structure function can be qualitatively explained if
nearly all the momentum is carried by a leading quark, which is a $u$ quark
in the case of a proton and a $d$ quark in the case of a neutron. In both
cases a scalar diquark with isospin zero remains. Close \cite{clos73}
includes both scalar and vector diquarks, which have the same probability
within SU(6) symmetry. The conclusion from the analysis \cite{clos73} is that a
scalar diquark is more probable than a vector one at large momentum
transfer. This is in agreement with the fact that the \mi{QCD spin-spin
force}, originally introduced into the quark model in Ref. \cite{deru70},
is attractive and strongest in the spin-0 quark-quark state.

Most generally the proton can be composed of the four combinations for
spin and isospin either equal to zero or one:
\bea
\vert p \rangle &=& {\rm A}\; u(ud)_0\phi + {\rm B}\; d(uu)_0\phi
+{\rm C}\;u(ud)_1\phi +{\rm D}\; d(uu)_1\phi +\; \hbox{perm.}\;,\\
&& {\rm A}^2+{\rm B}^2+{\rm C}^2+{\rm D}^2=1 \;,\nonumber
\eea
where the parentheses indicate the diquark clustering and the spin of the
quark pair is given as a subscript. A preliminary calculation for the
magnetic moments of the baryon octet suggests that C and D are small or
zero. The fit shows that if we put B = --0.2 the parameters
$\beta_q$ and $\beta_Q$, discussed below, can be
chosen to be equal. Note that SU(6) is still broken in this case.

Considering the number of the parameters and because of the picture from
deep-inelastic electron scattering we only keep the spin-isospin-0 part.
To implement
this feature we have to modify the wave functions in Section
\ref{sec:wavefunction}. In order to get a spin-isospin-0 clustering we
write the proton wave function as
\begin{equation} \label{eq:proton}
	\frac{-1}{\sqrt{18}}\left[
	-uud\left( \phi_1\chi^{\rho1}+\phi_2\chi^{\rho2}\right)
	+udu\left( \phi_1\chi^{\rho1}-\phi_3\chi^{\rho3}\right)
	+duu\left( \phi_2\chi^{\rho2}+\phi_3\chi^{\rho3}\right)
	\right] \; .
\end{equation}

\index{wave function!baryon octet}
The wave functions for the $\Sigma$s and $\Xi$s are obtained by changing the
flavor wave function accordingly. The $\Lambda$ wave function is given by

\begin{equation} \label{eq:lambda}
	\frac{-1}{\sqrt{12}}\left[
	\phi_3\chi^{\rho3}\left( uds-dus \right)
	+\phi_2\chi^{\rho2}\left( usd-dsu \right)
	+\phi_1\chi^{\rho1}\left( sud-sdu \right)
	\right] \; .
\end{equation}

The function $\phi_i$ is the momentum wave function with symmetry in the quarks
different from $i$. We choose
\begin{equation} \label{eq:phii}
	\phi_i=N_ie^{-X_i} \; ,
\end{equation}
where the normalization factor $N_i$ is given below and the $X_i$ are the
generalized forms of $M^2/2\beta$:
\begin{eqnarray} \label{eq:difunction}
X_3 & = &
\frac{Q_\perp^2}{2\eta(1-\eta)\beta_Q^2}+\frac{q_\perp^2}
{2\eta\xi(1-\xi)\beta_q^2}+\frac{m_1^2}{2\eta\xi\beta_q^2}+\frac{m_2^2}
{2\eta(1-\xi)\beta_q^2}+\frac{m_3^2}{2(1-\eta)\beta_Q^2}
\; ,
\nonumber \\
 X_2 & = &
q_\perp^2\frac{(1-\eta)(1-\xi)\beta_Q^2+\xi\beta_q^2}{2\beta_Q^2\beta_q^2
\eta\xi(1-\xi)(1-\eta+\xi\eta)}+
Q_\perp^2\frac{(1-\xi)(1-\eta)\beta_q^2+\xi\beta_Q^2}{2\beta_Q^2\beta_q^2\eta
(1-\eta)(1-\eta+\xi\eta)}
\nonumber \\
 &  &+
q_\perp Q_\perp \frac{\beta_Q^2-\beta_q^2}{\beta_Q^2
\beta_q^2\eta(1-\eta+\xi\eta)}+
\frac{m_1^2}{2\eta\xi\beta_q^2}+\frac{m_2^2}{2\eta(1-\xi)\beta_Q^2}+
\frac{m_3^2}{2(1-\eta)\beta_q^2} \; ,
\nonumber \\
 X_1 & = &
q_\perp^2\frac{(1-\xi)\beta_q^2+\xi(1-\eta)\beta_Q^2}{2\beta_Q^2\beta_q^2
\eta\xi(1-\xi)(1-\xi\eta)}+Q_\perp^2\frac{(1-\xi)\beta_Q^2+\xi(1-\eta)
\beta_q^2}{2\beta_Q^2\beta_q^2\eta(1-\eta)(1-\xi\eta)}
\nonumber \\
 &  &-
q_\perp Q_\perp \frac{\beta_Q^2-\beta_q^2}
{\beta_Q^2\beta_q^2\eta(1-\xi\eta)}+
\frac{m_1^2}{2\eta\xi\beta_Q^2}+\frac{m_2^2}{2\eta(1-\xi)\beta_q^2}+
\frac{m_3^2}{2(1-\eta)\beta_q^2} \; .
\end{eqnarray}

There is a special form of $N_i$ with $N_1=N_2=N_3$:

	\begin{equation} 
		N_i^2=\left[ \int e^{-2X_3}\frac{d\xi dq_\perp d\eta dQ_\perp}
		{\xi\eta(1-\xi)(1-\eta)} \right]^{-1}  \; .
	\end{equation}

However, this normalization gives too small rates for the semileptonic
decays. We use instead our usual normalization

	\begin{equation} 
		\frac{N_c}{(2\pi)^6} \int d^3qd^3Q |\phi_i|^2 = 1 \; .
	\end{equation}

\newpage
\section{Results and discussions} 

The calculations of the matrix elements are similar to those in
the symmetric case, but they are more involved because
the wave function is no longer symmetric. We give an example for the
vector current ($K^2=0$) [see Eq.~\ref{eq:asy1}]
\begin{eqnarray} \label{eq:dispin}
	\lefteqn{
	\langle \uparrow\uparrow\downarrow | \uparrow\uparrow\downarrow
	\rangle_{\rm asym}	 =  \langle \uparrow\uparrow\downarrow |
	\uparrow\uparrow\downarrow \rangle_{\rm sym}} \nonumber \\
		&	&  - \frac{2(a-a')^2(bb'+Q_\perp)cd(q_L Q_R)^2}
	{(a'^2+Q^2_\perp)(a^2+Q^2_\perp)\sqrt{b'^2+Q^2_\perp}\sqrt{b^2+Q^2_\perp}
	(c^2+q^2_\perp)(d^2+q^2_\perp)}
	\; .
\end{eqnarray}
For the numerical calculation it is important to simplify the additional
term to reduce the six-dimensional integral to a
five-dimensional one (see Appendix \ref{sec:numerical}).

Before presenting the results we give some general considerations. To take
an asymmetric wave function is the most natural extension to the minimal
model. Since we do not have this possibility in the meson sector, this
extension does not contradict the minimal meson model
\cite{jaus90,jaus91}. Another question concerns the ratio in
Eq.~(\ref{eq:ratio}). It is in principle possible to improve this value
drastically to give $-2.26$ ($m_{u/d}=0.26$ GeV, $m_s=0.39$ GeV,
$\beta_{QN}=0.3$ GeV, $\beta_{qN}=0.7$ GeV, $\beta_{Q\Sigma}=0.7$ GeV,
$\beta_{q\Sigma}=0.3$ GeV), which shows great flexibility compared to other
models.

\begin{table}[b]
\caption[The parameters of the asymmetric constituent quark
model.]{The parameters of the asymmetric constituent quark model.
All numbers are given in GeV.}
\begin{center}\begin{tabular}{ccccccccccc}
\hline\hline
&$m_{u/d}$&$m_s$&$\beta_{QN}$&$\beta_{qN}$&$\beta_{Q\Sigma/\Lambda}$&
$\beta_{q\Sigma/\Lambda}$&$\beta_{Q\Xi}$&$\beta_{q\Xi}$&$f_{1us}$&$g_{1us}$\\
\hline
Set 8&0.26&0.395&0.55&0.55&0.35&0.65&0.34&0.65&1.32&1.17\\
\hline\hline
\end{tabular}\end{center}
\label{tab:asympara}
\end{table}

We minimize the function $\Delta = \sum_i | x_i({\rm fit})
-x_i({\rm meas})|$ for the fit. The details of the fit are described in
Appendix~\ref{sec:fit}. Figure~\ref{fig:fit} shows $\Delta$ as a function
of the various parameters. The black areas in the density plots indicate the
minimum, the white areas the maximum of the function $\Delta$. The fixed
values of the parameters in the plots are the ones in
Table~\ref{tab:asympara}. The figures (a) and (b) show the fit for the
magnetic moments of the proton and neutron, and the weak axial form
factor $g_1(n\rightarrow p e^-\bar\nu_e)$. We can see that there is no large
diquark clustering in the nucleon sector ($\beta_{QN} = \beta_{qN}$).
The magnetic moments of the hyperons are fitted in figures (c), (d) and
(e). There is a strong diquark clustering in the strange baryon sector
($\beta_q\sim 2\beta_Q$). The mass of the $u$ and $d$ quarks is
$(0.26\pm 0.20)$ GeV for both the nucleons and hyperons. The strange
quark mass is fixed to be 1.5 times $m_{u/d}$. Figure (f) gives the fit
to the rates and the axial form factors for the decays $\Lambda \rightarrow
p e^-\bar\nu_e$ and $\Sigma \rightarrow n e^-\bar\nu_e$ as a function of
the weak quark form factors.

\begin{figure}
\centerline{\psfig{figure=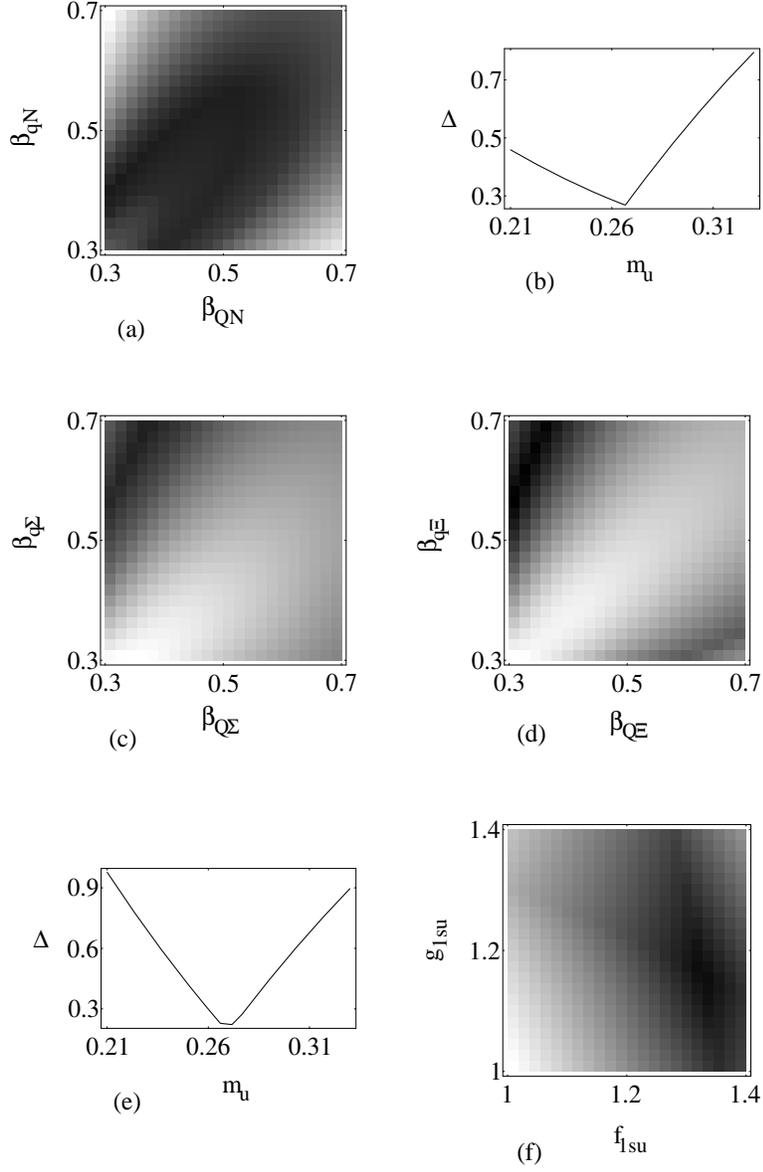,height=500pt}}
\caption[Parameter space of the asymmetric wave function fit.] {Parameter
space of the asymmetric wave function fit. The deviation from the
experimental data is plotted against the various parameters. The black
areas in the density plot show the minimum, the white areas the maximum of
the difference between the experimental values and those given by the fit.
The $\beta$s and the masses
are given in units of GeV.(a), (b) nucleon sector; (c), (d), (e) magnetic
moments of the hyperons; (f) semileptonic, strangeness changing, weak
decay.}
\label{fig:fit}
\end{figure}

We summarize the results for the asymmetric parameter set in Table
\ref{tab:diquark}. There is a considerable improvement for the magnetic
moments of the neutron, the $\Sigma$s and $\Xi$s, and for the rates and the
ratios $g_1/f_1$ of the semileptonic decays. Non-zero quark anomalous magnetic
form factors could even give better results for the magnetic moments
of the neutron and the $\Xi^-$ (see Table \ref{tab:moment}). Parameter
set~8 fits both electromagnetic and weak properties of the baryon octet.
Weak form factors of the quarks are needed for the strangeness changing
transition $s \rightarrow u$. Since the rate is proportional to
$|V_{us}|^2f_{1us}^2$ and we have to fit both values, we cannot
predict an accurate value for $V_{us}$.

\begin{table}
\caption[Electroweak properties of the baryon octet.]
{Electroweak properties of the baryon octet. The calculations with
parameter sets~4, 6 and 8 are compared. We see that set~4 is only able to
fit the magnetic moments (see Section~\protect\ref{sec:mm}), set~6 fits the
weak sector (see Chapter~\protect\ref{ch:decay}), and set~8 can fit both
sectors simultaneously. The magnetic moments are given in units of the
nuclear magneton, the decay rates are given in $10^6 s^{-1}$.}
\begin{center}\begin{tabular}{lcrrr}
\hline\hline
Particle&Experiment \cite{part90}&Set 4&Set 6&Set 8\\
\hline
$\mu (p)$&2.79 $\pm 10^{-7}$&2.85&2.78&2.82\\
$\mu (n)$&--1.91 $\pm 10^{-7}$&--1.83&--1.62&--1.66\\
$\mu (\Sigma^+)$&2.42 $\pm$ 0.05 &2.59&3.23&2.48\\
$\mu (\Sigma^-)$&--1.157 $\pm$ 0.025 &--1.30&--1.36&--1.09\\
$\mu (\Lambda)$&--0.613 $\pm$ 0.004 &--0.48&--0.72&--0.64\\
$\mu (\Xi^0)$&--1.250 $\pm$ 0.014 &--1.25&--1.87&--1.28\\
$\mu (\Xi^-)$& --0.679 $\pm$ 0.031&--0.99&--0.96&--0.78\\
$\frac{g_1}{f_1}(np)$&1.261 $\pm$ 0.004&1.63&1.252&1.25\\
$\frac{g_1}{f_1}(\Lambda p)$&0.718 $\pm$ 0.015&0.957&0.826&0.760\\
$\frac{g_1}{f_1}(\Sigma n)$&--0.340 $\pm$ 0.017&--0.319&--0.275&--0.238\\
$\frac{g_1}{f_1}(\Xi^- \Lambda)$&0.250 $\pm$ 0.042&0.319&0.272&0.190\\
$\frac{g_1}{f_1}(\Xi^- \Sigma^0)$&1.287 $\pm$ 0.158&1.594&1.362&1.13\\
$R (\Lambda p)$&3.17 $\pm$ 0.058&0.14&3.51&3.20\\
$R (\Sigma n)$&6.88 $\pm$ 0.26&0.16&5.74&6.68\\
$R (\Xi^- \Lambda)$&3.36 $\pm$ 0.19&0.10&2.96&3.61\\
$R (\Xi^- \Sigma^0)$&0.53 $\pm$ 0.10&0.02&0.55&0.48\\
\hline\hline
\end{tabular}\end{center}
\label{tab:diquark}
\end{table}

General features exhibited by these results are
\begin{itemize}
\item In the nucleon sector, there is no diquark clustering ($\beta_{QN}
  = \beta_{qN}$).
  \index{nucleons!diquark clustering}
\item There is a strong diquark clustering in the strange baryon sector
($2\beta_Q\sim\beta_q$).
\item The momentum scale parameter for the diquark pair is about the
same for all baryons ($\beta_{QN} =
\beta_{qN}\sim\beta_{q\Sigma/\Lambda}=\beta_{q\Xi}$).
\end{itemize}

Table~\ref{tab:comparison} compares the results of this work with
other models. The nonrelativistic quark model yields accurate
magnetic moments, but fails in the weak decay sector. The results derived
from QCD sum rules \index{QCD sum rule technique} and the \mi{bag model} are
comparable to those from our model.
Nevertheless,  some of the data are reproduced
best within our model. The fits from the \mi{lattice simulation} and the
\mi{Skyrme model} are too small.

\begin{table}
\caption[Comparison with other models for the electroweak properties
of the baryon octet.]
{Electroweak properties of the baryon octet. The calculations of
the present work with parameter set~8 are compared with the static
nonrelativistic quark model (NQM), QCD sum rule (SR) \protect\cite{chiu87},
lattice simulations (Latt.) \protect\cite{lein91}, bag model (Bag)
\protect\cite{bari90}, and \mi{Skyrme model} (Skyr.)
\protect\cite{kunz90,adki83}. The magnetic moments are given in units of the
nuclear magneton. \index{lattice simulation}}
\begin{center}\begin{tabular}{lcrrrrrr}
\hline\hline
Particle&Experiment \cite{part90}&Set 8&NQM&SR&Latt.&Bag&Skyr.\\
\hline
$\mu (p)$&2.79 $\pm 10^{-7}$&2.82&2.82&3.04&2.3&2.78&1.97\\
$\mu (n)$&--1.91 $\pm 10^{-7}$&--1.66&--1.88&--1.79&
--1.3&--1.83&--1.24\\
$\mu (\Sigma^+)$&2.42 $\pm$ 0.05 &2.48&2.70&2.73&1.9&2.65&2.25\\
$\mu (\Sigma^-)$&--1.157 $\pm$ 0.025 &--1.09&--1.05&--1.26&
--0.88&--1.40&--0.88\\
$\mu (\Lambda)$&--0.613 $\pm$ 0.004 &--0.64&--0.60&--0.50&
--0.41&--0.60&--0.59\\
$\mu (\Xi^0)$&--1.250 $\pm$ 0.014 &--1.28&--1.43&--1.32&--0.96&
--1.40&--1.42\\
$\mu (\Xi^-)$& --0.679 $\pm$ 0.031&--0.78&--0.49&--0.93&--0.42&
--0.53&--0.40\\
$\frac{g_1}{f_1}(np)$&1.261 $\pm$ 0.004&1.25&1.67& & &1.224&0.61\\
$\frac{g_1}{f_1}(\Lambda p)$&0.718 $\pm$ 0.015&0.760&1.00& & &0.757& \\
$\frac{g_1}{f_1}(\Sigma n)$&--0.340 $\pm$ 0.017&--0.238&
--0.33& & &--0.252& \\
$\frac{g_1}{f_1}(\Xi^- \Lambda)$&0.250 $\pm$ 0.042&0.190&0.33&
 & &0.167& \\
$\frac{g_1}{f_1}(\Xi^- \Sigma^0)$&1.287 $\pm$ 0.158&1.13&1.67&
 & &1.256& \\
\hline\hline
\end{tabular}\end{center}
\label{tab:comparison}
\end{table}

\chapter{Discussion and conclusions}\label{ch:conclusion}

We have considered a relativistic model for the three-quark core of the
baryons.
A field-theory calculation of matrix elements between bound states
\index{bound state} is given
with the help of the quasipotential reduction of the Bethe-Salpeter
equation. \index{Bethe-Salpeter equation}

The input to our model fits is the constituent quark masses and the
momentum range
parameters. They are essentially free parameters in the framework of
spectroscopic models, but they are kept fixed for the entire set of
reactions. This is not an easy task because the minimal model will not do
the job. The physical picture that emerges in our analysis is an
asymmetric three quark state with a spin-isospin-0 diquark
for the hyperons and a symmetric wave function for the nucleons, which is
Lorentz shaped in the momentum distribution.
Only for the strangeness-changing weak decay do we need
nontrivial form factors. With this input we can fit almost every
measurement within $6\%$; many fits are even more accurate (see
Table~\ref{tab:diquark}).

There is no need for quark form factors in the electromagnetic sector
except for fine tuning. Wave functions can be described without admixtures
of mixed permutation symmetry. In fact, a slight asymmetry in the wave
function does correspond to a nonvanishing mixed permutation symmetry,
but we need a large \mi{diquark} clustering, as we have seen in the previous
Chapter.

The symmetry breaking scheme and the $M_V$ and $M_A$ of our model can
explain all present data for the weak beta decay of the hyperons within the
Cabibbo theory (see Sec.~\ref{sec:cabibbo}). \index{Cabibbo model}
The value for $V_{us} = 0.225 \pm 0.003$ that we get has
recently been confirmed in a different analysis \cite{garc92}. But a
discrepancy with the meson sector still remains.

We list some features of our model, which have not been looked at before
in this framework:

\begin{enumerate}
\item Asymmetric wave functions.

\item  Wave functions with admixture of mixed permutation symmetry.

\item  Wave functions with realistic high-momentum features up to more than
30 GeV$^2$.

\item Comprehensive calculations for both the electromagnetic and the
weak sector.

\item  Consistent symmetry breaking scheme for the Cabibbo theory.

\item Derivatives of the weak form factors.
\end{enumerate}

It is interesting to note that the mass parameters in this work are
similar to those in the meson sector of the same model \cite{jaus91}.
This gives us confidence concerning the generality of the model.

We conclude with some remarks on how to improve the model described in
this thesis.

\begin{enumerate}
\item  In addition to the three valence quarks, one should consider the
effects of higher Fock states (valence quarks, gluons).

\item  The momentum wave function should be derived from a potential.

\item  Gluon corrections should be calculated.

\item  A different diquark clustering should be investigated.

\end{enumerate}

Points 1 and 3 have been considered to some extent in our model because
diquark clustering simulates gluon effects and these also give rise to higher
Fock states. Point 2 should justify the choice of our wave function.

\appendix
\chapter{Computational methods}\label{ch:methods}

The main programs for this thesis are written in {\sc form} \cite{verm89},
{\sc Mathematica} \cite{wolf91}, {\sc FORTRAN} \cite{fort78}, and {\sc C}
\cite{kern88}, together with the {\sc NAG library} \cite{nag14}. They run
on an Alliant FX/80, a Sun SPARC station, a Macintosh Classic, and a
Macintosh IIsi. The final version of the programs are written in {\sc
Mathematica} and {\sc C} to meet two goals:
\index{form}
\index{Mathematica}
\index{FORTRAN}
\index{NAG}
\index{C}
\index{Alliant}
\index{Macintosh}
\index{Sun SPARC}

\begin{enumerate}
\item  The \mi{software} should be easy to modify and to maintain since we want
to check many different flavors of the quark model.
\item Because the formulae are large, there should be a way to check
their correctness easily.
\end{enumerate}

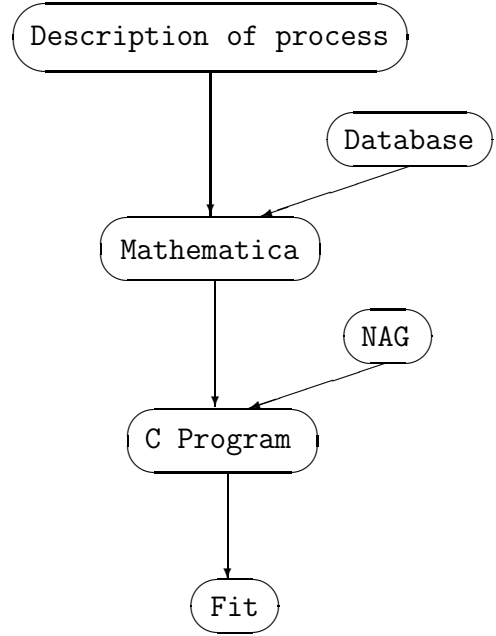
\begin{figure}
\centerline{{\tt    \setlength{\unitlength}{0.92pt}
\begin{picture}(217,279)
\thinlines    \put(173,202){\vector(-3,-1){61}}
              \put(17,252){Description of process}
              \put(54,164){Mathematica}
              \put(64,86){C Program}
              \put(91,17){Fit}
              \put(146,209){Database}
              \put(153,128){NAG}
              \put(173,213){\oval(68,22)}
              \put(91,168){\oval(90,26)}
              \put(164,132){\oval(36,22)}
              \put(96,89){\oval(78,26)}
              \put(101,21){\oval(36,22)}
              \put(91,255){\oval(162,28)}
              \put(91,241){\vector(0,-1){59}}
              \put(93,155){\vector(0,-1){52}}
              \put(98,76){\vector(0,-1){44}}
              \put(163,121){\vector(-3,-1){56}}
\end{picture}}}
\caption[Flow chart of the basic steps for this thesis.]
{Flow chart of the basic steps for this thesis. The aim was to keep the
input for the programs as small and simple as possible to make sure to get
correct results. The steps in between have been automated to minimize
typographical errors.}
\label{fig:flow}
\end{figure}

The general design philosophy is given in Fig. \ref{fig:flow}. The input,
the description of the process, should be small and similar to the physical
notation. The properties of the Melosh transform
\index{Melosh rotation}
and of the various wave
functions should be kept in a database, so that they have to be typed only
once and can be thoroughly checked. From this starting point,
the symbolic program should produce the large formulae in an automated way
and splice them into the {\sc C} program. With this method it is easy to
achieve the two above mentioned goals, because the input is short and can
be given in a physical language.

The rest of Appendix \ref{ch:methods} is organized as follows: In
Sec.~\ref{sec:symbolic} we give details of the symbolic implementation.
Numerical questions are treated in Sec.~\ref{sec:numerical}, in
Sec.~\ref{sec:fit} we show our procedure for the multidimensional fit,
and in the last Section some parts of the program are listed.

\newpage
\section{Symbolic calculation} \label{sec:symbolic}

In implementing the wave functions of Eqs.~(\ref{eq:wavefunction}),
(\ref{eq:proton}), and (\ref{eq:lambda}) we use some simple relations between
them. For instance:
\bea
| n \rangle &=& - | p \rangle \left( u \leftrightarrow d \right)\;,\nonumber\\
| \Sigma^+ \rangle &=& - | p \rangle \left( d \rightarrow s \right)\;,\\
| p \downarrow \rangle &=& - | p \uparrow\rangle
\left( \uparrow \leftrightarrow \downarrow \right)\;.\nonumber
\eea
The wave functions in the {\sc Mathematica} program are written in an
obvious notation. The flavors up, down, and strange are denoted by their
initials {\tt u}, {\tt d}, {\tt s}. The functions in Eq.~(\ref{eq:phii}) are
{\tt p1}, {\tt p2}, and {\tt p3}, respectively. Spin up and spin down are
denoted by {\tt a} and {\tt b} respectively. This part of the {\sc Mathematica}
program reads:

\begin{verbatim}
prot = -(uud (p1(aba-aab)+p2(baa-aab))
        +udu (p1(aab-aba)+p3(baa-aba))
        +duu (p2(aab-baa)+p3(aba-baa)))/Sqrt[18];
lam = (sud p1(aba-aab)+usd p2(baa-aab)
      +sdu p1(aab-aba)+uds p3(baa-aba)
      +dsu p2(aab-baa)+dus p3(aba-baa))/Sqrt[12];

ab = {aab -> -bba,aba -> -bab,baa -> -abb};
ud = {uud -> -ddu,udu -> -dud,duu -> -udd};
us = {uud -> -ssd,udu -> -sds,duu -> -dss};
ds = {uud -> -uus,udu -> -usu,duu -> -suu};
udds = {uud -> dds,udu -> dsd,duu -> sdd};
usdu = {uud -> ssu,udu -> sus,duu -> uss};
sigma = {uud->(uds+dus)/Sqrt[2],udu->(usd+dsu)/Sqrt[2],
       duu->(sud+sdu)/Sqrt[2]};

proton[spin_] := If[spin==1/2,prot,prot /. ab];
neutron[spin_] := If[spin==1/2,prot /. ud,
                               prot /. ud /.ab];
sigmaP[spin_] := If[spin==1/2,prot /. ds,
                               prot /. ds /.ab];
sigma0[spin_] := If[spin==1/2,prot /. sigma,
                               prot /. sigma /.ab];
sigmaM[spin_] := If[spin==1/2,prot /. udds,
                               prot /. udds /.ab];
xi0[spin_] := If[spin==1/2,prot /. usdu,
                               prot /. usdu /.ab];
xiM[spin_] := If[spin==1/2,prot /. us,
                               prot /. us /.ab];
lambda[spin_] := If[spin==1/2,lam,lam /.ab];
\end{verbatim}

The wave function of the $| \Xi^0\downarrow\rangle$ baryon for example can
now be called by {\tt xi0[-1/2]}. The transition matrix elements for the
weak decay in Eq.~(\ref{eq:formfactors}) can be defined by a function {\tt
transition[.,.]}. \index{transition[.,.]}
\begin{verbatim}
transition[b2_,b1_] :=
     Block[{b},b=b2 /. bar;
         Expand[3 b b1] /. flavourWeak /. spin
     ]
\end{verbatim}
This function needs some additional definitions for the spin and flavor
part.
\begin{verbatim}
flavourWeak ={udu udd -> 1,dud duu -> 1,ssd ssu -> 1,
              dsd dsu -> 1,sdd sdu -> 1,usd usu -> 1,
              sud suu -> 1,usu usd -> 1,
              uds udu -> 1,dus duu -> 1,dds ddu -> 1,
              sds sdu -> 1,dss dsu -> 1,sus suu -> 1,
              uss usu -> 1,sdu sdd -> 1,dsu dsd -> 1,
              uud->0,udu->0,duu->0,ddu->0,dud->0,udd->0,
              uds->0,dus->0,usd->0,dsu->0,sud->0,sdu->0,
              uus->0,usu->0,suu->0,dds->0,dsd->0,sdd->0,
              ssu->0,sus->0,uss->0,ssd->0,sds->0,dss->0};

spin = {aabs aab->aabaab,aabs aba->aababa,aabs baa->
        aabbaa,abas aab->abaaab,abas aba->abaaba,
        abas baa->ababaa,baas aab->baaaab,
        baas aba->baaaba,baas baa->baabaa};
bar = {aab->aabs,aba->abas,baa->baas,p1->p1s,
       p2->p2s,p3->p3s};
\end{verbatim}

We now turn to the spin part of the bracket and write the Melosh
transform in Eq.~(\ref{eq:melosh}) as: \index{Melosh rotation}
\begin{verbatim}
Melosh={{{a c-qr Ql,-a ql-c Ql},{c Qr +a qr,a c-ql Qr}},
        {{a d+qr Ql,a ql-d Ql},{d Qr-a qr,a d+ql Qr}},
        {{b,Ql},{-Qr,b}}} ;

aab = {Melosh[[1]].{1,0},Melosh[[2]].{1,0},
       Melosh[[3]].{0,1}};
aba = {Melosh[[1]].{1,0},Melosh[[2]].{0,1},
       Melosh[[3]].{1,0}};
baa = {Melosh[[1]].{0,1},Melosh[[2]].{1,0},
       Melosh[[3]].{1,0}};
abb = {Melosh[[1]].{1,0},Melosh[[2]].{0,1},
       Melosh[[3]].{0,1}};
bab = {Melosh[[1]].{0,1},Melosh[[2]].{1,0},
       Melosh[[3]].{0,1}};
bba = {Melosh[[1]].{0,1},Melosh[[2]].{0,1},
       Melosh[[3]].{1,0}};
\end{verbatim}
For calculating the spin bracket we define the function {\tt
bracket[.,.]}, e.g. the transition $\langle \uparrow\downarrow\uparrow
| A^+ | \uparrow\downarrow\uparrow \rangle$ can be computed by {\tt
bracket[aba,aba]}. \label{page:o} \index{bracket[.,.]}
\begin{verbatim}
o = {{{1,0},{0,1}},{{1,0},{0,1}},{{1,0},{0,-1}}};
qQrule1={qr->qt2/ql,Qr->Qt2/Ql};
qQrule2={ql^n_Integer?Negative->(qt2/qr)^n,
         Ql^n_Integer?Negative->(Qt2/Qr)^n}
qQrule3={ql Qr->qtQt,(ql Qr)^2->qlQr2,(ql Qr)^3->qlQr3,
        qr Ql->qtQt,(qr Ql)^2->qlQr2,(qr Ql)^3->qlQr3};
conjugate={qr->ql,ql->qr,Qr->Ql,Ql->Qr,a->as,b->bs};
bracket[x_,y_] :=
   Block[{z},z = x /. conjugate;
      Factor[ReplaceAll[ReplaceAll[ReplaceAll[Expand[
         Product[z[[i]].o[[i]].y[[i]],{i,1,3}]],qQrule1],
         qQrule2],qQrule3]]
        ]
\end{verbatim}
The computation of {\tt bracket[aba,aba]} takes 40 seconds on a
Macintosh~IIsi for $K^2=0$:
\begin{verbatim}
In = bracket[aba,aba]//Timing

Out = {39.23333333333333333*Second,
   (b*bs - Qt2)*(a^2*as^2*c^2*d^2 +
      2*a^2*c*d*qlQr2 - 4*a*as*c*d*qlQr2 +
      2*as^2*c*d*qlQr2 + a^2*as^2*c^2*qt2 +
      a^2*as^2*d^2*qt2 + a^2*as^2*qt2^2 +
      2*a*as*c^2*d^2*Qt2 + 2*a*as*c^2*qt2*Qt2 -
      2*a^2*c*d*qt2*Qt2 + 4*a*as*c*d*qt2*Qt2 -
      2*as^2*c*d*qt2*Qt2 + 2*a*as*d^2*qt2*Qt2 +
      2*a*as*qt2^2*Qt2 + c^2*d^2*Qt2^2 +
      c^2*qt2*Qt2^2 + d^2*qt2*Qt2^2 + qt2^2*Qt2^2
      )}
\end{verbatim}

In order to obtain a convenient form for the numerical implementation in
Sec.~\ref{sec:numerical} we define the function {\tt list}, which gives
a list of the coefficients of the spin parts.
\begin{verbatim}
coeff[a_,b_]:=Block[{x},x=Coefficient[a,b];
      {Coefficient[x,p1*p1s],Coefficient[x,p1*p2s],
      Coefficient[x,p1*p3s],Coefficient[x,p2*p1s],
      Coefficient[x,p2*p2s],Coefficient[x,p2*p3s],
      Coefficient[x,p3*p1s],Coefficient[x,p3*p2s],
      Coefficient[x,p3*p3s]}]
list[a_]:=Block[{x},x=Expand[N[a]];
  Flatten[{coeff[x,aabaab],coeff[x,aababa],coeff[x,aabbaa],
           coeff[x,abaaab],coeff[x,abaaba],coeff[x,ababaa],
           coeff[x,baaaab],coeff[x,baaaba],coeff[x,baabaa]}]]
\end{verbatim}

\newpage
\section{Numerical calculation} \label{sec:numerical}

The integrals to be handled are normally six dimensional. In some special
cases, such as for $K^2=0$ and for a symmetric wave function, we have only four
dimensional integrals to do. But it is obvious that a fast and accurate
integration
routine is crucial for our analysis. After testing various library and
self-written routines we use the routine {\tt d01fcf} \index{d01fcf}
 from the {\sc Nag
library} \cite{nag14}. {\tt d01fcf} is based on the {\tt half} subroutine
\cite{door76} and uses the basic rule described by Ref. \cite{genz80}.
The routine attempts to evaluate a multidimensional integral, with
constant and finite limits, to a specified relative accuracy, using a
repeated subdivision of the hyper-rectangular region into smaller
hyper-rectangles. In each subregion, the integral is estimated using a
seventh-degree rule, and an error estimate is obtained by comparison with
a fifth-degree rule, which uses a subset of the same points. The error
estimate is therefore less time consuming than for the Gauss
integration. The fourth differences of the integrand along each
coordinate axis are evaluated, and the subregion is marked for possible
future subdivision in half along that coordinate axis which has the
largest absolute fourth difference.

For a \mi{multidimensional integral} it is important to keep the
dimensionality of the
integrals as small as possible. By introducing cylindrical coordinates we
can easily reduce the six dimensions to five. Defining new variables by
\bea
&q_1 = q \cos\theta\;,   &Q_1 = Q \cos\phi\;, \nonumber\\
&q_2 = q \sin\theta\;,   &Q_2 = Q \sin\phi\;, \nonumber\\
&q_3 = q_3\;,\qquad   &Q_3 = Q_3\;, \nonumber\\
&0 < \theta < 2\pi\;,  &0 < \phi < 2\pi\;,
\eea
we can write the difficult parts of the formulae [e.g.
(\ref{eq:difunction}) and (\ref{eq:dispin})] as follows:
\bea
q_\perp Q_\perp &=& q Q \cos{(\theta-\phi)}\;, \nonumber\\
(q_L q_R)^2 &=& q^2Q^2 \cos{[2(\theta-\phi)]}\;,\\
(q_L q_R)^3 &=& q^3Q^3 \cos{(\theta-\phi)} \left\{ 2\cos{[2(\theta-\phi)]}-1
\right\} \;.\nonumber
\eea
Since the integrand depends only on $(\theta-\phi)$, one integral is
trivial. This is crucial because the routine {\tt d01fcf} looses much
efficiency for more than five integrations. In some cases the
$(\theta-\phi)$-integration can also be done yielding modified
Bessel functions \index{modified Bessel function}
of the first kind, but the speed of the routine {\tt
d01fcf} would not increase.

\newpage
\section{Fitting procedures} \label{sec:fit}

The procedure to \mi{fit} our variables to the experimental data is rather
involved since the number of variables is high for some versions of our
model. One way to go is to fit some subset of the experiments, e.g. the
nucleon sector, and fix the values of the parameters so obtained when
fitting the rest of the data.
 The other way is to define a function that measures
the error of the fit. The {\sc Mathematica} function for the magnetic
moments of the octet could be written as:
\index{Mathematica}
\begin{verbatim}
f[mu_,bQs_,bqs_,bQx_,bqx_]:=
    Abs[1.42-sP[mu,bQs,bqs]]+
    Abs[-0.16-sM[mu,bQs,bqs]]+
    Abs[-0.61-ml[mu,bQs,bqs]]+
    Abs[-1.25-x0[mu,bQx,bqx]]+
    Abs[0.32-xM[mu,bQx,bqx]];
\end{verbatim}
The calls {\tt sP}, {\tt sM}, {\tt ml}, {\tt x0}, and {\tt xM} give the
\mi{magnetic moments} of the $\Sigma^+$, $\Sigma^0$, $\Lambda$, $\Xi^0$,
and $\Xi^-$ respectively as a function of the mass of the $u$ quark
({\tt mu}), and the scale factors $\beta_{q\Sigma}$ ({\tt bqs}),
$\beta_{Q\Sigma}$ ({\tt bQs}), $\beta_{q\Xi}$ ({\tt bqx}), and
$\beta_{Q\Xi}$ ({\tt bQx}).
Because of speed considerations we do not directly compute the magnetic
moments at runtime, but we build a lookup table, which we interpolate
with the {\sc Mathematica} function {\tt Interpolate}, e.g.
\index{Interpolate[.]}
\begin{verbatim}
sP=Interpolation[data_sP,InterpolationOrder->3].
\end{verbatim}
A local minimum of the function {\tt f} can be found by the {\sc
Mathematica} command
\index{FindMinimum}
\begin{verbatim}
FindMinimum[f[mu,bQs,bqs,bQx,bqx],
    {mu,{.25,.26},.21,.33},{bQs,{.5,.52},.3,.7},
    {bqs,{.5,.52},.3,.7},{bQx,{.5,.52},.3,.7},
    {bqx,{.5,.52},.3,.7}]
\end{verbatim}

\begin{figure}
\centerline{\psfig{figure=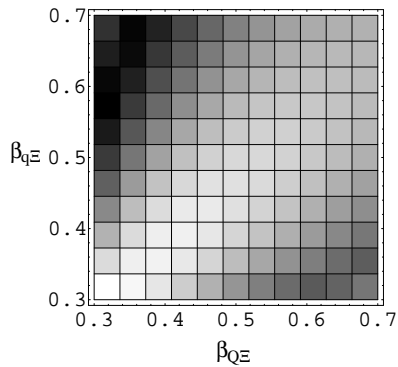,height=144pt}}
\caption[Example of a density plot for the magnetic moments of the baryon
octet.]{Example of a density plot for the magnetic moments of the baryon
octet. The black areas show the minimum, the white areas the maximum of the
deviation from the experimental data. The $\beta$s are given in units of
GeV. One can easily see that an asymmetric wave function is strongly
favored.}
\label{fig:fitexample}
\end{figure}

To search for a global minimum we use the graphics capability of
{\sc Mathematica}. The command
\index{DensityPlot}
\begin{verbatim}
DensityPlot[f[.25,.35,.65,bQx,bqx],{bQx,.3,.7},{bqx,.3,.7}]
\end{verbatim}
produces the plot in Fig.~\ref{fig:fitexample}. The black areas show the
minimum, the white areas the maximum of the function {\tt f} mentioned
above. We can easily see in Fig.~\ref{fig:fitexample} that an asymmetric
wave function is strongly favored.

\newpage
\section{Program listing} \label{listing}

In this Section we present some parts of the listing of the {\sc C} program
for the strangeness-changing weak semileptonic decay with no momentum
transfer ($K^2=0$). The programs for $K^2 \not = 0$ are larger, and the
one for electromagnetic properties is shorter. We will not present the
entire program, but simply comment on some of its parts. The aim was to keep
the program modular and flexible.

The {\sc FORTRAN} routine {\tt d01fcf} \index{d01fcf} can be called from
within a {\sc C} program in the following way:
\begin{verbatim}
  d01fcf_(&ndim,a,b,&minpts,&maxpts,g1sn_,&eps,&acc,&lenwrk,
          wrkstr,&sum_g1sn,&ifail);
\end{verbatim}
The routine names must be extended by an underscore, and the variables
must be called by addresses. In the above call we evaluate the form
factor $g_1$ for the process $\Sigma^- \rightarrow n$ and store its value
in the variable \verb|sum_g1sn|. The function {\tt g1sn} is given below,
and the parameters are fixed by experience as follows:
\begin{verbatim}
#define LENWRK 100000
#define N      5
#define PI2    6.283185308
...
  int ndim=N,ifail=1,minpts=0,maxpts=(int)(93*LENWRK/7),lenwrk=LENWRK;
  static double a[N]={0.,-3.,0.,-3.,0.},b[N]={3.,3.,3.,3.,PI2};
  double eps,acc,wrkstr[LENWRK], ...
\end{verbatim}
The succeeding parts of the {\sc C} program are directly produced by the
{\sc Mathematica} programs explained in Sec.~\ref{sec:symbolic}. The
{\sc Mathematica} command
\begin{verbatim}
FactorTerms[transition[sigmaM[.5],neutron[.5]]]//list
\end{verbatim}
produces the array in the function {\tt g1sn}.
\begin{verbatim}
double g1sn_(ndim,z)
int *ndim;
double *z;
{
	double trans[81]={-0.1666666666666666667, -0.1666666666666666667, 0,
  -0.1666666666666666667, -0.1666666666666666667, 0, 0, 0,
  0, 0.1666666666666666667, 0.1666666666666666667, 0, 0, 0,
  0, 0, 0, 0, 0, 0, 0, 0.1666666666666666667,
  0.1666666666666666667, 0, 0, 0, 0, 0.1666666666666666667,
  0, 0, 0.1666666666666666667, 0, 0, 0, 0, 0,
  -0.1666666666666666667, 0, 0, 0, 0, 0, 0, 0, 0, 0, 0, 0,
  -0.1666666666666666667, 0, 0, 0, 0, 0, 0,
  0.1666666666666666667, 0, 0, 0.1666666666666666667, 0, 0,
  0, 0, 0, -0.1666666666666666667, 0, 0, 0, 0, 0, 0, 0, 0,
  0, 0, 0, -0.1666666666666666667, 0, 0, 0, 0},axial();
	return(axial(z,trans));
}
\end{verbatim}
The function {\tt axial}, called on the last line, contains the entire
implementation of the Melosh transform, the kinematics, and the variable
transform discussed in the proceeding Section. The wave functions are taken
from Eq.~(\ref{eq:difunction}). The spin transitions {\tt aabaab,
aababa, ...} are calculated with {\sc Mathematica} as shown in
Sec.~\ref{sec:symbolic}:
\begin{verbatim}
aabaab=bracket[aab,aab];
aababa=bracket[aab,aba];
...
Splice["weak.mc"]
\end{verbatim}
The last command {\tt Splice} \index{Splice[.]} splices {\sc Mathematica}
output into an external file. The output looks as follows:
\begin{verbatim}
double axial(z,t)
double *z,*t;
{
  double q,qt2,Q,Qt2,theta,M,M3,xi,eta,p1,p2,p3,p1s,p2s,p3s,
         alpha2,beta2,alpha2s,beta2s,m12,m22,m32,m32s,aabaab,aababa,
         aabbaa,abaaab,abaaba,ababaa,baaaab,baaaba,baabaa,res,qtQt,
         qlQr2,qlQr3,Ms,Q3,Q3s,Q_2,Q_2s,e3,e3s,e12,e12s,a,as,b,bs,c,d,
         X1,X2,X3,X1s,X2s,X3s,q_2,e1,e2,q3;
  q=z[0]; q3=z[1]; Q=z[2]; Q3=z[3]; theta=z[4];

  qt2 = q*q; Qt2 = Q*Q;
  q_2 = qt2 + q3 * q3; Q_2 = Qt2 + Q3 * Q3;
  qtQt = q*Q*cos(theta);
  qlQr2 = q*q*Q*Q*cos(2.*theta);
  qlQr3 = q*q*q*Q*Q*Q*cos(theta)*(2.*cos(2.*theta)-1);

  e1 = sqrt(q_2 + m1 * m1); e2 = sqrt(q_2 + m2 * m2); M3 = e1 + e2;
  e3 = sqrt(Q_2 + m3 * m3); e12 = sqrt(Q_2 + M3 * M3);M = e3 + e12;
  xi = (e1 + q3)/M3; eta = (e12 + Q3)/M;
  Ms = sqrt((Qt2+(1-eta)*M3*M3+eta*m3s*m3s)/eta/(1-eta));
  Q3s = (eta-0.5)*Ms-(M3*M3-m3s*m3s)/2.0/Ms;
  Q_2s = Qt2 + Q3s * Q3s;
  e3s = sqrt(Q_2s + m3s * m3s); e12s = sqrt(Q_2s + M3 * M3);

  a = M3 + eta * M;       as = M3 + eta * Ms;
  b = m3 + (1 - eta) * M; bs = m3s + (1 - eta) * Ms;
  c = m1 + xi * M3;
  d = m2 + (1 - xi) * M3;

  alpha2 = alpha*alpha; beta2 = beta*beta;
  alpha2s = alphas*alphas; beta2s = betas*betas;
  m12=m1*m1; m22=m2*m2; m32=m3*m3; m32s=m3s*m3s;
  X3=(Qt2/eta/(1-eta)/alpha2+qt2/eta/xi/(1-xi)/beta2+
      m12/eta/xi/beta2+m22/eta/(1-xi)/beta2+m32/(1-eta)/alpha2);
  X2=qt2*((1-eta)*(1-xi)*alpha2+xi*beta2)/alpha2/beta2
     /eta/xi/(1-xi)/(1-eta+xi*eta)+Qt2*((1-xi)
     *(1-eta)*beta2+xi*alpha2)/alpha2/beta2/eta
     /(1-eta)/(1-eta+xi*eta)+2.*(alpha2-beta2)/alpha2
     /beta2/eta/(1-eta+xi*eta)*q*Q*cos(theta)+
     m12/eta/xi/beta2+m22/eta/(1-xi)/alpha2+m32/(1-eta)/beta2;
  X1=qt2*((1-xi)*beta2+xi*(1-eta)*alpha2)/alpha2/beta2
     /eta/xi/(1-xi)/(1-xi*eta)+Qt2*((1-xi)*alpha2+xi*(1-eta)*
     beta2)/alpha2/beta2/eta/(1-eta)/(1-xi*eta)-2.*(alpha2-beta2)
     /alpha2/beta2/eta/(1-xi*eta)*q*Q*cos(theta)+
     m12/eta/xi/alpha2+m22/eta/(1-xi)/beta2+m32/(1-eta)/beta2;
#ifdef GAUSS
  p3=N3*exp(-X3/2.); p2=N2*exp(-X2/2.); p1=N1*exp(-X1/2.);
#else
  p3=N3*pow((X3+1),-4); p2=N2*pow((X2+1),-4); p1=N1*pow((X1+1),-4);
#endif
  X3s=(Qt2/eta/(1-eta)/alpha2s+qt2/eta/xi/(1-xi)/beta2s+
      m12/eta/xi/beta2s+m22/eta/(1-xi)/beta2s+m32s/(1-eta)/alpha2s);
  X2s=qt2*((1-eta)*(1-xi)*alpha2s+xi*beta2s)/alpha2s/beta2s
     /eta/xi/(1-xi)/(1-eta+xi*eta)+Qt2*((1-xi)
     *(1-eta)*beta2s+xi*alpha2s)/alpha2s/beta2s/eta
     /(1-eta)/(1-eta+xi*eta)+2.*(alpha2s-beta2s)/alpha2s
     /beta2s/eta/(1-eta+xi*eta)*q*Q*cos(theta)+
     m12/eta/xi/beta2s+m22/eta/(1-xi)/alpha2s+m32s/(1-eta)/beta2s;
  X1s=qt2*((1-xi)*beta2s+xi*(1-eta)*alpha2s)/alpha2s/beta2s
     /eta/xi/(1-xi)/(1-xi*eta)+Qt2*((1-xi)*alpha2s+xi*(1-eta)*
     beta2s)/alpha2s/beta2s/eta/(1-eta)/(1-xi*eta)-2.*(alpha2s-beta2s)
     /alpha2s/beta2s/eta/(1-xi*eta)*q*Q*cos(theta)+
     m12/eta/xi/alpha2s+m22/eta/(1-xi)/beta2s+m32s/(1-eta)/beta2s;
#ifdef GAUSS
  p3=N3*exp(-X3s/2.); p2=N2*exp(-X2s/2.); p1=N1*exp(-X1s/2.);
#else
  p3=N3*pow((X3s+1),-4); p2=N2*pow((X2s+1),-4); p1=N1*pow((X1s+1),-4);
#endif

  aabaab=(Qt2 - b*bs)*(Qt2*Qt2*c*c*d*d + 2*Qt2*a*as*c*c*d*d +
    a*a*as*as*c*c*d*d - 2*a*a*c*d*qlQr2 + 4*a*as*c*d*qlQr2 -
    2*as*as*c*d*qlQr2 + Qt2*Qt2*c*c*qt2 + 2*Qt2*a*as*c*c*qt2 +
    a*a*as*as*c*c*qt2 + 2*Qt2*a*a*c*d*qt2 -
    4*Qt2*a*as*c*d*qt2 + 2*Qt2*as*as*c*d*qt2 + Qt2*Qt2*d*d*qt2 +
    2*Qt2*a*as*d*d*qt2 + a*a*as*as*d*d*qt2 + Qt2*Qt2*qt2*qt2 +
    2*Qt2*a*as*qt2*qt2 + a*a*as*as*qt2*qt2);
  aababa=(a - as)*(b + bs)*(Qt2*Qt2*c*c*d*d + Qt2*a*as*c*c*d*d +
    ...
    28 lines omitted
    These formulae are given in Eqs. (B.41) - (B.57)
    ...
  baabaa = -((Qt2 - b*bs)*(Qt2*Qt2*c*c*d*d + 2*Qt2*a*as*c*c*d*d +
      a*a*as*as*c*c*d*d + 2*a*a*c*d*qlQr2 - 4*a*as*c*d*qlQr2 +
      2*as*as*c*d*qlQr2 + Qt2*Qt2*c*c*qt2 + 2*Qt2*a*as*c*c*qt2 +
      a*a*as*as*c*c*qt2 - 2*Qt2*a*a*c*d*qt2 +
      4*Qt2*a*as*c*d*qt2 - 2*Qt2*as*as*c*d*qt2 +
      Qt2*Qt2*d*d*qt2 + 2*Qt2*a*as*d*d*qt2 + a*a*as*as*d*d*qt2 +
      Qt2*Qt2*qt2*qt2 + 2*Qt2*a*as*qt2*qt2 + a*a*as*as*qt2*qt2));

  res = t[0]*aabaab*p1*p1s + t[1]*aabaab*p1*p2s + t[2]*aabaab*p1*p3s +
        t[3]*aabaab*p2*p1s + t[4]*aabaab*p2*p2s + t[5]*aabaab*p2*p3s +
        t[6]*aabaab*p3*p1s + t[7]*aabaab*p3*p2s + t[8]*aabaab*p3*p3s +
        ...
        22 lines omitted
        ...
        t[75]*baabaa*p2*p1s + t[76]*baabaa*p2*p2s + t[77]*baabaa*p2*p3s +
        t[78]*baabaa*p3*p1s + t[79]*baabaa*p3*p2s + t[80]*baabaa*p3*p3s;
  res /= (a*a+Qt2)*(d*d+qt2)*(as*as+Qt2)*(c*c+qt2)*
         sqrt(b*b+Qt2)*sqrt(bs*bs+Qt2);
  res *= sqrt(e3s*e12s*M/e3/e12/Ms);
  return(q*Q*res);
}
\end{verbatim}
The {\sc C} preprocessor directives \verb|#ifdef GAUSS, #else, #endif|
give us the possibility of choosing either the Gauss shaped wave function
or the Lorentz shaped one by simply typing the line
\begin{verbatim}
#define GAUSS
\end{verbatim}
or by commenting it out.

The weak vector form factor is calculated in an analogous way; the line
\begin{verbatim}
o = {{{1,0},{0,1}},{{1,0},{0,1}},{{1,0},{0,-1}}};
\end{verbatim}
on page~\pageref{page:o} has just to be replaced by
\begin{verbatim}
o = {{{1,0},{0,1}},{{1,0},{0,1}},{{1,0},{0,1}}};
\end{verbatim}
The listings for the calculation for $K^2 \not =0$ are longer, so we
omit them here.


\pagestyle{myheadings}
\markboth{}{}
\chapter*{Acknowledgments}

It is a pleasure to acknowledge the kind help of the following persons:

\begin{itemize}

\item Especially, I would like to thank my advisor {\em Prof. Dr. W.~Jaus\/}
for suggesting the problem of this thesis, for many stimulating
discussions, and for carefully reading the manuscript. His advice and
criticism proved to be indispensable.

\item Furthermore, I am indebted to {\em Prof. Dr. G.~Rasche\/} for his
continuous interest in this thesis and his kind help.

\item I would like to thank {\em Kurt Sonnenmoser\/} for many helpful
discussions about computer problems.

\item I am grateful to the members of the Institute for Theoretical Physics
of the University of Zurich for the friendly atmosphere.

\item Finally, I would like to express my profound gratitude towards my
dear {\em wife\/} for her encouragement and her interest in my thesis.

\end{itemize}
This work was supported by the Canton of Zurich and the Swiss National
Science Foundation.

\addcontentsline{toc}{chapter}{Acknowledgments}
\pagestyle{headings}

\cleardoublepage

\cleardoublepage
\addcontentsline{toc}{chapter}{Index}
\begin{theindex}

  \item Ademollo-Gatto theorem
    \subitem extension of, 38
  \item Alliant, 57
  \item angular correlation, 32

  \indexspace

  \item bag model, 2, 31, 42, 44, 53
  \item Berestetskii, 1
  \item Bethe-Salpeter equation, 2, 3, 6, 55
  \item bound state, 2, 6, 10, 21, 49, 55
  \item bracket[.,.], 60
  \item bremsstrahlung, 33

  \indexspace

  \item C, 57
  \item Cabibbo model, 44--46, 55
  \item center of mass motion, 2, 5, 10
  \item charge radius, 19
  \item Chernyak, 16
  \item chiral perturbation theory, 31
  \item Chung, 1, 13
  \item Coester, 1, 13
  \item color, 9
  \item confinement scale, 11, 13
  \item conserved vector current, 42
  \item current-current interaction Hamiltonian, 31

  \indexspace

  \item d01fcf, 62, 65
  \item DensityPlot, 64
  \item diquark, 3, 15, 49, 55
  \item Donoghue, 46
  \item Dziembowski, 1

  \indexspace

  \item Faddeev equation, 2, 6, 7
  \item FindMinimum, 63
  \item fit, 44, 63
  \item form, 57
  \item FORTRAN, 57

  \indexspace

  \item Gaillard, 43
  \item Gell-Mann, 1, 49

  \indexspace

  \item Heitler, 2
  \item hyperon decay, 1, 3, 30--46
    \subitem angular correlation, 32
    \subitem decay rate, 32

  \indexspace

  \item Interpolate[.], 63

  \indexspace

  \item Jaus, 1

  \indexspace

  \item kinematic subgroup, 5
  \item Kobayashi-Maskawa matrix, 1, 3, 31

  \indexspace

  \item ladder approximation, 6
  \item lattice simulation, 53, 54
  \item light front, 1--3, 9, 15, 16
    \subitem variables, 10

  \indexspace

  \item Macintosh, 57
  \item magnetic moments, 1, 13, 21, 27--30, 42, 63
  \item many-particle system, 5
  \item mass operator, 5, 7, 10
  \item Mathematica, 57, 63
  \item Melosh rotation, 1, 10, 42, 57, 60
  \item modified Bessel function, 62
  \item multidimensional integral, 62

  \indexspace

  \item NAG, 57
  \item neutron beta decay, 21, 43
  \item nucleons
    \subitem diquark clustering, 53
    \subitem electromagnetic form factors, 19--24

  \indexspace

  \item Poincar{\accent 19 e} group, 2, 5

  \indexspace

  \item QCD spin-spin force, 49
  \item QCD sum rule technique, 16, 53
  \item quark structure, 13, 20, 21, 30, 36

  \indexspace

  \item radiative corrections, 33
  \item residual, 44

  \indexspace

  \item Sachs form factors, 19
  \item Skyrme model, 53, 54
  \item software, 57
  \item Splice[.], 66
  \item Sun SPARC, 57
  \item symmetry breaking, 31, 38, 42, 44

  \indexspace

  \item Terent'ev, 1
  \item transition[.,.], 59

  \indexspace

  \item uncertainty principle, 1

  \indexspace

  \item vector
    \subitem four-vector, 5
    \subitem light-front vectors, 5
  \item vertex function, 6, 7, 9

  \indexspace

  \item wave function, 2, 7, 12--17
    \subitem asymmetric, 15, 49
    \subitem baryon octet, 11, 50
    \subitem configuration mixing, 14
    \subitem from vertex functions to, 9
  \item weak magnetism, 32
  \item Weber, 1

  \indexspace

  \item Zhitnitsky, 16

\end{theindex}


\end{document}